\newcommand{\Jybeam}  {\mbox{Jy}~\mbox{beam}^{-1}}
\documentclass{aa}  

\usepackage{graphicx}
\usepackage{txfonts}
\usepackage{hyperref}
\usepackage{xcolor}
\newcommand{\RomanNumeralCaps}[1]
    {\MakeUppercase{\romannumeral #1}}
\begin{document} 

   \title{Near-Infrared Observations of Outflows and YSOs in the Massive Star-Forming Region AFGL 5180}

   \author{S. Crowe
          \inst{1,2}
          \and
          R. Fedriani\inst{1}
          \and
          J. C. Tan\inst{2,3}
          \and
          M. Whittle\inst{2}
          \and
          Y. Zhang\inst{4}
          \and
          A. Caratti o Garatti\inst{5}
          \and
          J.P. Farias\inst{6}
          \and
          A. Gautam\inst{2}
          \and
          Z. Telkamp\inst{2}
          \and
          B. Rothberg\inst{7,8}
          \and
          M. Grudi\'c\inst{9}
          \and
          M. Andersen\inst{10}
          \and
          G. Cosentino\inst{3}
          \and
          R. Garcia-Lopez\inst{11}
          \and
          V. Rosero\inst{12}
          \and
          K. Tanaka\inst{13}
          \and
          E. Pinna\inst{14,15}
          \and
          F. Rossi\inst{14,15}
          \and
          D. Miller\inst{16}
          \and
          G. Agapito\inst{14,15}
          \and
          C. Plantet\inst{14,15}
          \and
          E. Ghose\inst{14,15}
          \and
          J. Christou\inst{16,17}
          \and
          J. Power\inst{16}
          \and
          A. Puglisi\inst{14,15}
          \and
          R. Briguglio\inst{14,15}
          \and
          G. Brusa\inst{16}
          \and
          G. Taylor\inst{16}
          \and
          X. Zhang\inst{16}
          \and
          T. Mazzoni\inst{14,15}
          \and
          M. Bonaglia\inst{14,15}
          \and
          S. Esposito\inst{14,15}
          \and
          C. Veillet\inst{16}
          }

   \institute{Instituto de Astrof\'isica de Andaluc\'ia, CSIC, Glorieta de la Astronom\'ia s/n, E-18008 Granada, Spain 
        \and
            Dept. of Astronomy, University of Virginia, Charlottesville, Virginia 22904, USA 
        \and
            Dept. of Space, Earth \& Environment, Chalmers University of Technology, 412 93 Gothenburg, Sweden 
        \and
            Dept. of Astronomy, Shanghai Jiao Tong University, 800 Dongchuan RD. Minhang District, Shanghai, China 
        \and
            INAF-Osservatorio Astronomico di Capodimonte, Salita Moiariello
            16, I-80131 Napoli, Italy 
        \and
            Department of Astronomy, University of Texas at Austin, TX 78712 
        \and
            U.S. Naval Observatory, 3450 Massachusetts Avenue NW, Washington, DC 20392, USA 
        \and
            Department of Physics \& Astronomy, George Mason University, 4400 University Drive, MS 3F3, Fairfax, VA 22030, USA 
        \and
            Carnegie Observatories, 813 Santa Barbara St, Pasadena, CA 91101, USA 
        \and
            Gemini Observatory, NSFs National Optical-Infrared Astronomy Research Laboratory, Casilla 603, La Serena, Chile 
        \and
            University College Dublin, School of Physics, Belfield, Dublin 4, Ireland 
        \and
            National Radio Astronomy Observatory, 1003 Lopezville Road, Socorro, NM 87801, USA 
        \and
            Department of Earth and Planetary Sciences, Tokyo Institute of Technology, Meguro, Tokyo, 152-8551, Japan 
        \and
            INAF - Osservatorio Astrofisico di Arcetri, largo E. Fermi 5, 50125 Firenze, Italy 
        \and
            ADONI - ADaptive Optics National lab in Italy, Italy 
        \and
            Large Binocular Telescope Observatory, 933 N. Cherry Ave 552, Tucson, AZ 85721, USA 
        \and
            NOIRLab/Gemini Observatory, 950 N Cherry Ave, Tucson, AZ 85719, USA 
             }

  \abstract
   {Massive stars play important roles throughout the universe, however their formation remains poorly understood. Observations of jets and outflows in high-mass star-forming regions, as well as surveys of young stellar object (YSO) content, can help test theoretical models of massive star formation.}
   {We aim at characterizing the massive star-forming region AFGL 5180 in the near-infrared (NIR), identifying outflows and relating these to sub-mm/mm sources, as well as surveying the overall YSO surface number density to compare to massive star formation models.}
   {Broad- and narrow-band imaging of AFGL 5180 was made in the NIR with the Large Binocular Telescope, in both seeing-limited ($\sim0.5\arcsec$) and high angular resolution ($\sim0.09\arcsec$) Adaptive Optics (AO) modes, as well as with the Hubble Space Telescope. 
   Archival continuum data from the Atacama Millimeter/Submillimeter Array (ALMA) was also utilized.}
   {At least 40 jet knots were identified via NIR emission from H$_2$ and [FeII] tracing shocked gas. 
   Bright jet knots outflowing from the central most massive protostar, S4 (estimated mass $\sim11\:M_{\odot}$, via SED fitting), are detected towards the east of the source and are resolved in fine detail with the AO imaging. Additional knots are distributed throughout the field, likely indicating the presence of multiple driving sources. Sub-millimeter sources detected by ALMA are shown to be grouped in two main complexes, AFGL 5180 M and a small cluster $\sim15\arcsec$ (0.15 pc in projection) to the south, AFGL 5180 S. From our NIR continuum images we identify YSO candidates down to masses of $\sim 0.1\:M_\odot$. Combined with the sub-mm sources, this yields a surface number density of such YSOs of $N_* \sim 10^3 {\rm pc}^{-2}$ within a projected radius of about 0.1 pc. Such a value is similar to those predicted by models of both Core Accretion from a turbulent clump environment and Competitive Accretion. The radial profile of $N_*$ is relatively flat on scales out to 0.2~pc, with only modest enhancement around the massive protostar inside 0.05~pc, which provides additional constraints on these massive star formation models.
   %
   %
   }
   {This study demonstrates the utility of high-resolution NIR imaging, in particular with AO, for detecting outflow activity and YSOs in distant regions. The presented images reveal the complex morphology of outflow-shocked gas within the large-scale bipolar flow of a massive protostar, as well as clear evidence for several other outflow driving sources in the region. 
   Finally, this work presents a novel approach to compare the observed YSO surface number density from our study against different models of massive star formation. 
   }


   \maketitle
%

\section{Introduction}\label{sec:introduction}
    
Massive stars play a significant role throughout the Universe, especially via the regulation of the physical and chemical evolution of galaxies that ultimately influences all star and planet formation activity in these systems. However, the formation mechanism of massive stars remains actively debated, especially between Core Accretion models, i.e., scaled-up versions of low-mass star formation theory \citep[e.g.,][]{mckee03}, and Competitive Accretion models \citep[e.g.,][]{bonnell01,wang10,grudic22} \citep[see, e.g.,][for a review]{tan14}. Detailed high-resolution observational studies of regions of massive star formation, where individual sources can be resolved and characterized, are needed to test theoretical models.

One such observational campaign is the SOFIA Massive (SOMA) Star Formation Survey \citep[see][and Telkamp et al. in prep.]{zhang13,debuizer17,liu19,liu20,fedriani23a}, which utilizes mid- and far-infrared (MIR \& FIR) observations from the Stratospheric Observatory for Infrared Astronomy (SOFIA) FORCAST instrument from 7 to 37~$\rm \mu m$, complemented by archival Spitzer and Herschel MIR to FIR imaging to place constraints on properties of massive protostars via fitting of their spectral energy distributions (SEDs) to radiative transfer models \citep{zhang18} based on the Turbulent Core Accretion (TCA) theory \citep{mckee03}. This paper is part of the near-infrared (NIR) follow-up, SOMA-NIR survey, on these massive star-forming regions that have previously been observed in the MIR \& FIR, in particular focusing on the SOMA region AFGL 5180 (Telkamp et al. in prep). The objective of this follow-up with the SOMA-NIR survey is to establish additional constraints on the protostellar properties derived from the SEDs. This will be achieved, for example, by characterizing the outflow geometry driven by the massive protostars and investigating the stellar content in the vicinity of these protostars.

The NIR follow-up of these star-forming regions is important to help identify young stellar objects (YSOs) across a range of masses, study continuum emission that often defines outflow cavities created and illuminated by the massive protostars, and detect emission lines from shock-heated gas in the outflows from both high- and low-mass sources.
Fast outflowing gas is ubiquitously seen from protostellar systems \citep[e.g.,][]{bally16,ray21} and this is expected to be a natural consequence of the accretion process during star formation \citep[e.g.,][]{shu87,beltran16}. Shocks in these flows heat the gas, creating conditions that can dissociate and ionize and/or excite $\rm H_2$ molecules leading to emission of atomic recombination lines, such as [FeII], and $\rm H_2$ ro-vibrational transitions that are detectable in the NIR.
Protostellar jets have been seen to extend over several parsecs from their driving source \citep[e.g.,][]{davis04,varricatt10,caratti15,fedriani18}. 



Here we present NIR observations of the massive star-forming region AFGL 5180, also known as IRAS 06058+2138 or G188.949+0.915 (peak pixel in our near-infrared image: RA(J2000)=06:08:53.38, Dec(J2000)=+21:38:28.38; see §\ref{sec:observations}), which is part of the regions covered by the SOMA survey (Telkamp et al., in prep.). 
Previous studies have noted the presence of a prominent east-west $^{12}$CO outflow in the region, with a blueshifted lobe to the east and a redshifted lobe to the west \citep{snell88}. \citet{tamura91} and \citet{hodapp94} also noted the high level of nebulosity associated with the source, particularly in the infrared.
\citet{tofani95} reported detection of several water masers towards the region, clustering both towards the main nebulosity and towards the south of it, providing evidence for two main sites of star formation. \citet{davis98} identified multiple knots of shocked molecular hydrogen emission,
suggesting the presence of multiple outflows. \citet{tamura91} identified a separate NW-SE outflow using NIR K-band polarization maps, and \citet{yao00} associated this outflow with an embedded source identified through NIR polarimetric lines. \citet{longmore06} identified six sources in the mid-infrared (MIR) which they associate with star formation activity. \citet{mutie21} used archival ALMA band 6 \& 7 data to identify gas cores associated with YSO candidates, finding 8 cores in the region, MM1-8, of which one, MM1, appears to be associated with massive star formation activity.
    
AFGL 5180 has also been well-studied over several decades as a site of a strong Class II 6.7 GHz methanol maser associated with the main outflowing region \citep{szymczak00,goedhart14,mutie21}. Such detections are well known to be associated with sites of star formation \citep{caswell95,walsh98}, and almost exclusively massive star formation \citep{breen13}. Additionally, the maser emission in AFGL 5180 has been known to be highly variable \citep{goedhart04,mutie21}, with some indication of cyclic or sinusoidal variation in early epochs followed by linear or exponential decay in later periods \citep{goedhart04,goedhart07,goedhart14}; this may indicate variable accretion and/or episodic events. 
    
\citet{carpenter95b} identified the region within the wider Gemini OB1 molecular cloud complex, hypothesizing that star formation in the association has been triggered by swept-up shells of gas, in the case of AFGL 5180 likely powered by the O9.5 star \textit{LS V +21 27} \citep{romanlopes19}.
However, this claim has been challenged in later papers which propose that star formation in AFGL 5180 and the wider dust ridge potentially associated with \textit{LS V +21 27} is rather the result of a cloud-cloud collision, on the basis of the ages and distributions of star-forming regions and the morphology of the redshifted and blueshifted components of the clouds \citep{vasyunina10,shimoikura13,maity23}.


There are several distance estimates to AFGL 5180 in the literature, with most between 1.5-2 kpc \citep{humphreys78,oh10,niinuma11,reid09}. Notably, a recent compendium of distances towards nearby molecular clouds by \citet{zucker20}, calculated using Bayesian statistics in combination with \textit{Gaia} parallax measurements to determine distances to $\sim5\%$ accuracy, places the Gemini OB1 Association at $\sim$1.8-2 kpc \citep[see Table A.1 of][]{zucker20}. In particular, the closest region in Gemini OB1 to AFGL 5180 is noted at a distance of 2 kpc, which is the distance we adopt in this study.


The structure of our paper is as follows. We present the observations and data reduction used in this study in §\ref{sec:observations}. Results and characterization of the region are presented in §\ref{sec:Results}. Discussion of the findings is made in §\ref{sec:Discussion}, and a summary and conclusion are given in §\ref{sec:Conclusions}.
\section{Observations and Data Reduction}\label{sec:observations}
    \begin{table*}[t]
        \centering
        \caption{Summary of observations for AFGL 5180. Total exposure time refers to the time on the source; in the case of the ground-based LBT observations, half of this time was dedicated to taking sky frames.}      
        \label{obs_table}             
        \begin{tabular}{c c c c c c c }     
        \hline\hline       
        Date & Telescope/ & Filter & Total exp. time & FOV$^\dagger$ & Angular res.\\ 
        (yyyy-mm-dd) & Instrument & (Name) & (s) &  & \\ 
        \hline
           2020-10-10 & LBT/LUCI1\&2 & K & 300 & $4\arcmin\times4\arcmin$ & $0.5\arcsec$\\  
           2020-10-10 & LBT/LUCI1 & H2 & 600 & $4\arcmin\times4\arcmin$ & $0.5\arcsec$\\ 
           2020-10-10 & LBT/LUCI2 & Br\_gam & 600 & $4\arcmin\times4\arcmin$ & $0.5\arcsec$\\ 
           2020-11-03 & LBT/SOUL & K & 300 & $0.5\arcmin\times0.5\arcmin$ & $0.09\arcsec$\\ 
           2020-11-03 & LBT/SOUL & H2 & 600 & $0.5\arcmin\times0.5\arcmin$ & $0.09\arcsec$\\ 
           2020-11-03 & LBT/SOUL & Br\_gam & 600 & $0.5\arcmin\times0.5\arcmin$ & $0.09\arcsec$\\ 
           2016-02-13 & HST/WFC3 & F110W & 400 & $2\arcmin\times2\arcmin$ & $0.12\arcsec$\\ 
           2016-02-13 & HST/WFC3 & F128N & 900 & $2\arcmin\times2\arcmin$ & $0.13\arcsec$\\ 
           2016-02-13 & HST/WFC3 & F160W & 350 & $2\arcmin\times2\arcmin$ & $0.16\arcsec$\\
           2016-02-13 & HST/WFC3 & F164N & 900 & $2\arcmin\times2\arcmin$ & $0.17\arcsec$\\
           2016-04-23 & ALMA & Band 6 & 600 & $44\arcsec$ & $0.32\arcsec\times0.25\arcsec$\\
           2017-12-02 & ALMA & Band 7 & 600 & $30\arcsec$ & $0.068\arcsec\times0.067\arcsec$\\
           \hline    
           \multicolumn{6}{l}{$^\dagger$ Listed FOV sizes for the ALMA images are for regions with primary beam response  $>0.1$.}\\
        \end{tabular}
    \end{table*}

\subsection{Large Binocular Telescope}\label{sec:LBT_obs}
    
Observations were taken on the 10th of October, 2020 with the Large Binocular Telescope (LBT) in binocular mode, i.e. utilizing both mirrors simultaneously, with the LBT Utility Camera in the Infrared \citep[LUCI;][]{seifert03} instrument in seeing limited mode (UV-2020B-04; PI: J. C. Tan). The N3.75 camera with a pixel scale of 0.12\arcsec and FOV of $4\arcmin\times4\arcmin$ was used. The filters K and Br$\gamma$, which are centered at the wavelengths 2.194 and 2.170~$\mu{\rm m}$, respectively, were employed on the LUCI1 instrument mounted on the left (SX) mirror, and the filters K and H$_2$, which are centered at the wavelengths  2.194 and 2.124~$\mu{\rm m}$, respectively, were employed on the LUCI2 instrument mounted on the right (DX) mirror. The central coordinates of the image are RA(J2000)=06:08:53.60, Dec(J2000)=+21:38:15.61. The images have a position angle (PA) of 0$^{\circ}$.
            
Observations were also taken on the 3rd of November, 2020 with the LUCI-1 instrument (see previous paragraph) in the Adaptive Optics (AO) assisted mode with the Single conjugated adaptive Optics Upgrade for the LBT \citep[SOUL;][]{pinna16} as part of the SOUL Commissioning Science Run scheduled for November 1-6, 2020 (UV-2020B-501; PI: J. C. Tan). The N30 camera with a pixel scale of 0.015\arcsec and FOV of $30\arcsec\times30\arcsec$ was used. The filters K, H$_2$, and Br$\gamma$ which are centred at the wavelengths  2.194,  2.124,  2.170\,$\mu{\rm m}$, respectively, were employed on the LUCI-1 insrument mounted on the left (SX) mirror. The central coordinates of the image were RA(J2000)=06:08:54.042, Dec(J2000)=+21:38:26.920. The AO guide star used, 082-000146 (RA(J2000)=06:08:54.271, Dec(J2000)=+21:38:24.324, R=15.41\,mag), was located at 4.2$\arcsec$ from the center of the image. The images were taken with a PA of 150$^{\circ}$, though the final images used for analysis were rotated to a PA=0$^{\circ}$.
            
The data were reduced with custom python scripts using the python packages ccdproc \citep{ccdproc}, astropy \citep{astropy} and photutils \citep{photutils} in the standard way, i.e. via subtraction of sky frames and flat field correction. The images were registered to one another and astrometrically calibrated by matching stars to the images from Hubble Space Telescope (HST), see §\ref{sec:HST_obs}, with an estimated residual of $0.13\arcsec$ for the seeing-limited images and $0.02\arcsec$ for the AO images. The Strehl ratio of the AO images were 0.15. The angular resolution of the seeing-limited and AO images were $\sim0.5\arcsec$ and $\sim0.09\arcsec$, respectively, derived by determining the Full Width at Half Maximum (FWHM) of Moffat profiles of isolated sources identified in each of the images \citep[see Appendix A of][for the details]{fedriani23b}.
            
\subsection{Hubble Space Telescope}\label{sec:HST_obs}
    
Observations were taken with the HST on the 13th of February, 2016 (program ID: 14465; PI: J. C. Tan). The Wide Field Camera 3 (WFC3) instrument was used with a pixel scale of 0.13\arcsec and FOV of $2\arcmin\times2\arcmin$. The filters F110W (J-band), F128N (Pa$\beta$), F160W (H-band), and F164N ([FeII]), which have the mean wavelengths 1.180, 1.284, 1.544, and 1.645 $\mu{\rm m}$, and therefore diffraction-limited angular resolutions of $0.12\arcsec$, $0.13\arcsec$, $0.16\arcsec$, and $0.17\arcsec$, respectively, were employed. The data was downloaded from the Mikulski Archive for Space Science (MAST) as a Hubble Advanced Product (HAP), which are reduced and calibrated using the standard pipeline. The images were astrometrically corrected by matching stars to the Gaia DR3 catalogue \citep{gaiadr3}, with an estimated residual of 0.0323\arcsec, determined by measuring the mean separation between isolated sources identified using the algorithm \textit{DAOFIND} \citep{stetson87} and their Gaia counterparts. The final drizzle images have a position angle (PA) of 0$^{\circ}$.
        
\begin{figure*}
\centering
\includegraphics[width=1\textwidth,height=1\textheight,keepaspectratio]{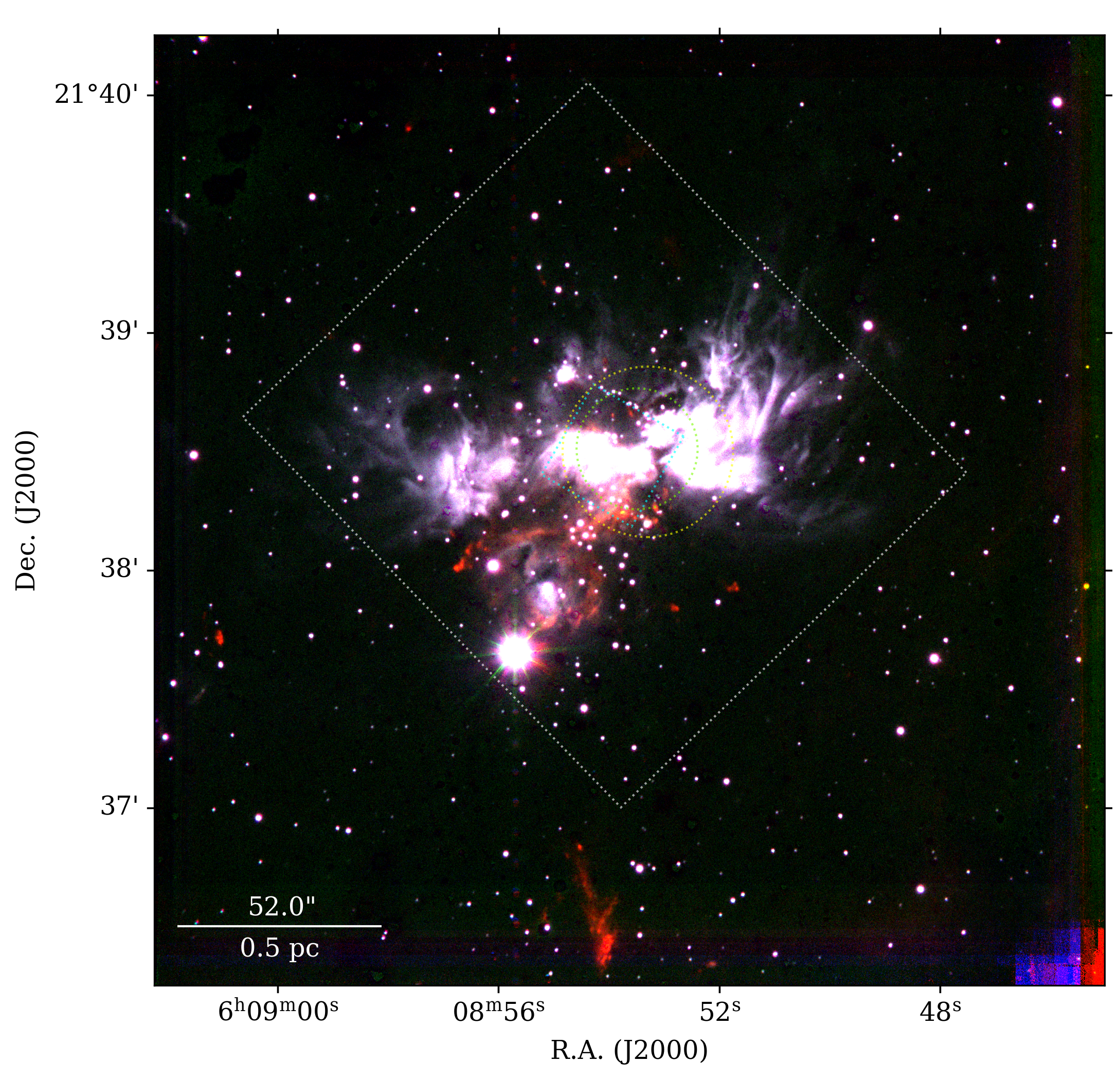}
\caption{\label{RGB_LBT_limited} RGB image of the LBT LUCI seeing-limited data. H$_2$ emission is shown in red, K-band in green, and Br$\gamma$ in blue. The FOV of the LUCI-1 SOUL AO observations (Fig. \ref{RGB_LBT_AO}) is indicated by the dotted cyan box, and the FOV of the HST observations (Fig. \ref{RGB_HST}) is represented by the dotted white box. The FOV of the ALMA Bands 6 and 7 observations are indicated by the yellow and green circles, respectively.}
\end{figure*}
    
\begin{figure*}
\centering
\includegraphics[width=1\textwidth,height=1\textheight,keepaspectratio]{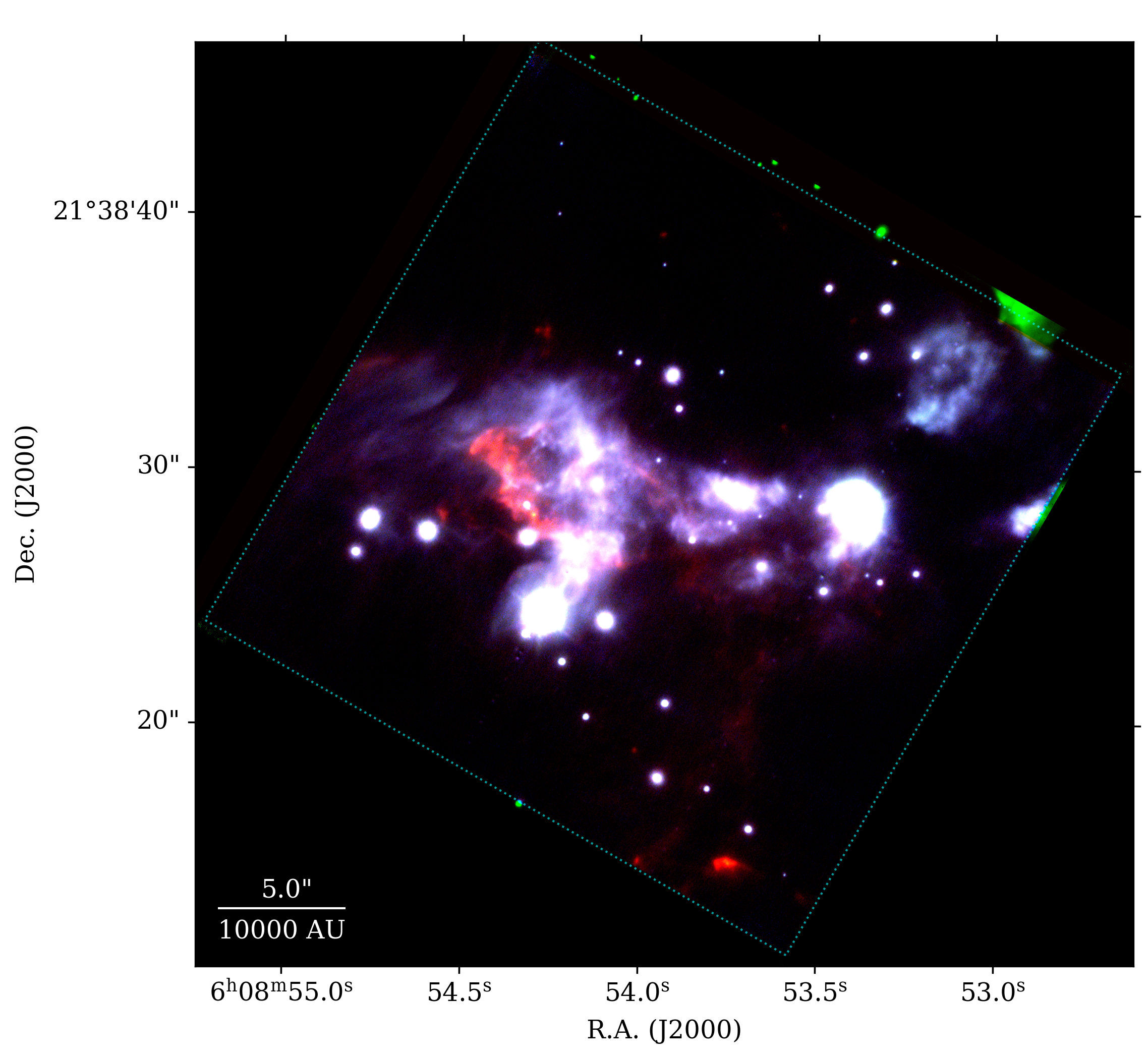}
\caption{\label{RGB_LBT_AO} RGB image of the LUCI-1 SOUL Adaptive Optics data. H$_2$ emission is shown in red, K-band in green, and Br$\gamma$ in blue. The FOV of the AO observations is indicated by the dotted cyan box.}
\end{figure*}
    
\begin{figure*}
\centering
\includegraphics[width=1\textwidth,height=1\textheight,keepaspectratio]{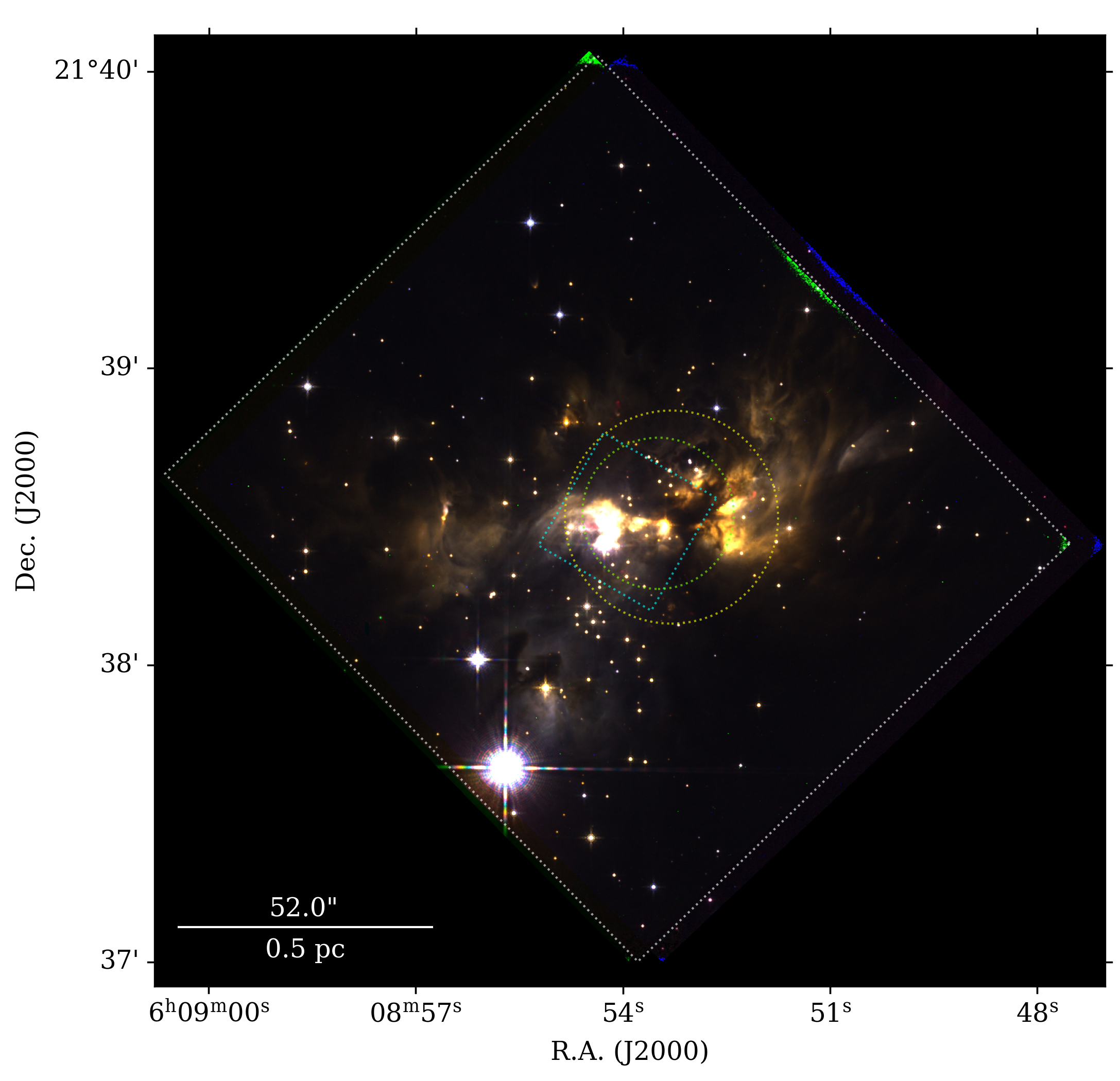}
\caption{\label{RGB_HST} RGBC image of the HST WFC3 data. [FeII] emission is shown in red, H-band in green, J-band in blue, and Pa$\beta$ in cyan. The FOV of the HST observations is indicated by the dotted white box, and the FOV of the LUCI-1 SOUL AO observations (Fig. \ref{RGB_LBT_AO}) is indicated by the dotted cyan box. The FOV of the ALMA Bands 6 and 7 observations are indicated by the yellow and green circles, respectively.}
\end{figure*}

\subsection{Atacama Large Millimeter/submillimeter Array}\label{sec:ALMA_obs}

We also used Atacama Large Millimeter/submillimeter Array (ALMA) 1.3 mm (Band 6) and 0.9 mm (Band 7) continuum data. The ALMA Band 6 observations were carried out on 2016 April 23 with the C36-3 configuration, and on 2016 September 8 with the C36-6 configuration (ALMA project ID: 2015.1.01454.S, PI: Y. Zhang). The total integration times were 3.5 and 6.5 minutes in the two configurations, respectively. Forty-one antennas were used with baselines ranging from 15–462 m in the C36-3 configuration, and 38 antennas were used with baselines ranging from 15 m to 3.2 km in the C36-6 configuration. J0750+1231 and J0510+1800 were used for bandpass calibration and flux calibration. J0603+2159 was used as a phase calibrator. The ALMA Band 7 observations were carried out on 2017 December 2 using the C43-7 configuration with 48 antennas and baselines ranging from 41.4 m to 6.9 km (ALMA project ID: 2017.1.00178.S, PI: T. Hirota). The total integration time was 10 minutes. J0510+1800 was used for bandpass calibration and flux calibration and J0613+2604 was used as a phase calibrator.

The data were calibrated and imaged in CASA \citep{casa}. After calibration, we performed self-calibration for each band and configuration using the continuum data combining the line-free channels. We first performed two phase-only self-calibration iterations with solution intervals of 30 and 6s, and then one iteration of amplitude self-calibration with the solution interval equal to the scan intervals. The effective bandwidth of line-free channels was 1.3 GHz and 3.7 GHz for the 1.3 and 0.9 mm continuum, respectively. The CASA tclean task was used to image the data, using Briggs weighting with the robust parameter set to 0.5. The 1.3 mm continuum image has a synthetic beam of $0\farcs32\times0\farcs25$ and rms noise level of $3.1\times 10^{-4}~\Jybeam$, and the 0.9 mm continuum image has a synthetic beam of $0\farcs068\times0\farcs067$ and rms noise of $1.3\times 10^{-4}~\Jybeam$. The details of all observations are summarized in Table \ref{obs_table}.
        
\section{Results}\label{sec:Results}\label{sec:data_presentation}

Here we present the NIR data obtained from LBT and HST on the AFGL 5180 complex, featuring broad coverage of the entire region from LBT seeing-limited and HST data as well as high-resolution observations of the inner main outflowing region covered by LUCI-1 SOUL. We also present the ALMA Bands 6 and 7 data which reveal the presence of several sub-mm cores believed to contain the driving sources of the jet knots seen in the NIR.

\subsection{NIR Imaging of the AFGL 5180 Complex}\label{sec:NIR_presentation}
    
Figure \ref{RGB_LBT_limited} shows an RGB image of the fullest extent of the AFGL 5180 complex covered by the LBT seeing-limited data ($4\arcmin\times4\arcmin$); red is used for the H$_2$ filter, green for K, and blue for Br$\gamma$. Readily apparent is the prominent east-west outflow, which is captured in all three LBT LUCI filters used; however, the narrow-band H$_2$ image reveals a large number of additional features, including an extended emission protruding from the main complex to the southeast, as well as several additional knots, particularly the bright bow shock to the south, extending over several arcminutes from the central region. These numerous H$_2$ knots, which appear to emanate in a wide variety of directions away from the central region, provide evidence of multiple outflows in the complex (see §\ref{sec:knot_ID}).
    
Figure \ref{RGB_LBT_AO} shows an RGB image, with the same color scheme as the previous image, of the inner outflowing region of AFGL 5180 as seen with LUCI-1 SOUL AO, in particular capturing in high resolution the eastern blue-shifted jet. The emission in the NIR has a cone-like morphology, although the complex structure makes it difficult to determine a precise value of the opening angle of the cone.
Nevertheless, the cone appears to trace directly back to a source located near the bright emission on the right in the image, however this driving protostar may be obscured in the NIR image. As in the seeing-limited image, a number of prominent H$_2$ features are seen, both within the main cone and also in more widespread locations, e.g., to the S and NE, suggesting the possibility of multiple outflow driving sources in the region.
            
Figure \ref{RGB_HST} shows an RGBC image from the HST of the AFGL 5180 complex. Here red is used for F164N, green is used for F160W, blue is used for F110W, and cyan is used for F128N. The prominent east-west outflow is again clearly seen, with a few [FeII] knots contained within the eastern jet and a large reflection nebula to the west. A few [FeII] knots can also be seen to the north, west, and south, which are noticeably off-axis with respect to the main outflow, corroborating the LBT LUCI H$_2$ data and again indicating the presence of multiple driving sources. See \S\ref{sec:knot_ID} for more details on jet knot identification.
    
\subsection{Continuum-Subtracted Images}\label{sec:continuum-sub}
        
In order to more easily identify line-emitting outflow structures,
continuum-subtracted images were created following the procedures outlined in \citet{long20}. In short, fluxes from several isolated point sources in the continuum- and narrow-band images were compared to determine their flux ratio, which was then used to subtract the continuum from the narrow-band images pixel-by-pixel. This process was performed on all of the narrow-band data. The LBT H$_2$ and Br$\gamma$ images (both seeing-limited and AO) were subtracted using the K-band image, while the HST [FeII] and Pa$\beta$ images were subtracted using the H- and J-band images, respectively.
    

We note the presence of residuals in Figs. \ref{big_diagram} and \ref{knot16}, where the continuum-subtraction was performed imperfectly on the stars in the image, resulting in a combination of bright and dark pixels. As an example, see the bottom-left star in the bottom panels of 
Fig. \ref{knot16}. We note also that some imperfections in the continuum-subtraction arise from color differences of the regions of H- and K-band nebulosity compared to those of the stars.
Nonetheless, such features are typical of the continuum-subtraction method, as also seen in other studies \citep[e.g.,][]{bally22,bally23}, but they do not interfere significantly with the identification of real features.

    
\subsection{Characterization of the AFGL 5180 Complex}\label{sec:Characterization}

\begin{table*}
\centering
\caption{
Compact cores identified in the ALMA Bands 6 \& 7 data (see §\ref{sec:cores}). Cores which had no emission or are not detected in a given band are denoted by $\cdots$, and those which are outside of the field of view of a given band are denoted by -. S/N values are with respect to the measured RMS noise of the image, and values derived from Band 7 and Band 6, respectively, are separated by a slash (/). Coordinates are given from the Band 7 data, except for S12-14. Associated source names are given from \citet{mutie21}.}
\label{core_table}             
\begin{tabular}{c c c c c c c c c}     
\hline\hline       
Source Name &Associated Source &R.A.(J2000)&Decl.(J2000)& $S_6$ & $S_7$ & $M_{\rm mm}$ & Area & S/N\\
& &(J2000) & (J2000) & (mJy) & (mJy) & ($M_{\odot}$) & ($\mathrm{arcsec^2}$) & \\
\hline     
S1 &$\cdots$&06:08:53.25 & 21:38:28.95 & $\cdots$ & 1.8 & 0.017 & 0.010  & 11\\
S2 &$\cdots$& 06:08:53.26 & 21:38:26.83 & $\cdots$ & 4.2 & 0.040 & 0.018  & 17\\
S3 &$\cdots$& 06:08:53.32 & 21:38:25.59 & $\cdots$ & 1.2 & 0.012 & 0.011  & 8\\
S4 & $^\dagger$MM1 & 06:08:53.33 & 21:38:29.01 & 51.7 & 102.6 & 0.97/1.9 & 0.012/0.156 & 600/200\\
S5 &$\cdots$& 06:08:53.35 & 21:38:27.06 & $\cdots$ & 4.6 & 0.044 & 0.012 & 25\\
S6 &$\cdots$& 06:08:53.36 & 21:38:25.86 & $\cdots$ & 1.8 & 0.017 & 0.012 & 8\\
S7 &$\cdots$& 06:08:53.37 & 21:38:34.46 & $\cdots$ & 1.3 & 0.012 & 0.009 & 8\\
S8 &$\cdots$& 06:08:53.38 & 21:38:29.97 & $\cdots$ & 0.6 & 0.0059 & 0.006 & 5\\
S9a & $^\ddagger$MM2 & 06:08:53.49 & 21:38:30.65 & 18.8 & 13.6 & 0.13/0.68 & 0.018/0.278 & 45/40\\
S9b & $^\ddagger$MM2 & 06:08:53.49 & 21:38:30.72 & 18.8 & 9.4 & 0.089/0.68 & 0.016/0.278 & 40/40\\
S10 &$\cdots$& 06:08:54.00 & 21:38:34.19 & $\cdots$ & 3.6 & 0.034 & 0.016 & 14\\
S11 &$\cdots$& 06:08:54.14 & 21:38:34.47 & $\cdots$ & 2.8 & 0.026 & 0.013 & 11\\
S12 & MM8 & 06:08:52.97 & 21:38:11.14 & 24.0 & - & 0.71 & 0.313 & 11\\
S13 & MM7 & 06:08:53.22 & 21:38:09.84 & 49.4 & - & 1.5 & 0.534 & 14\\
S14 & MM6 & 06:08:53.35 & 21:38:11.69 & 61.3 & - & 1.8 & 0.689 & 11\\
\hline
\multicolumn{9}{l}{$^\dagger$ RA and Dec from band 6 data: 
06:08:53.33 21:38:28.97}\\
\multicolumn{9}{l}{$^\ddagger$ RA and Dec from band 6 data: 
06:08:53.49 21:38:30.63}\\
\end{tabular}
\end{table*}

\begin{figure*}
\centering
\includegraphics[width=0.565\textwidth,height=0.565\textheight,keepaspectratio]{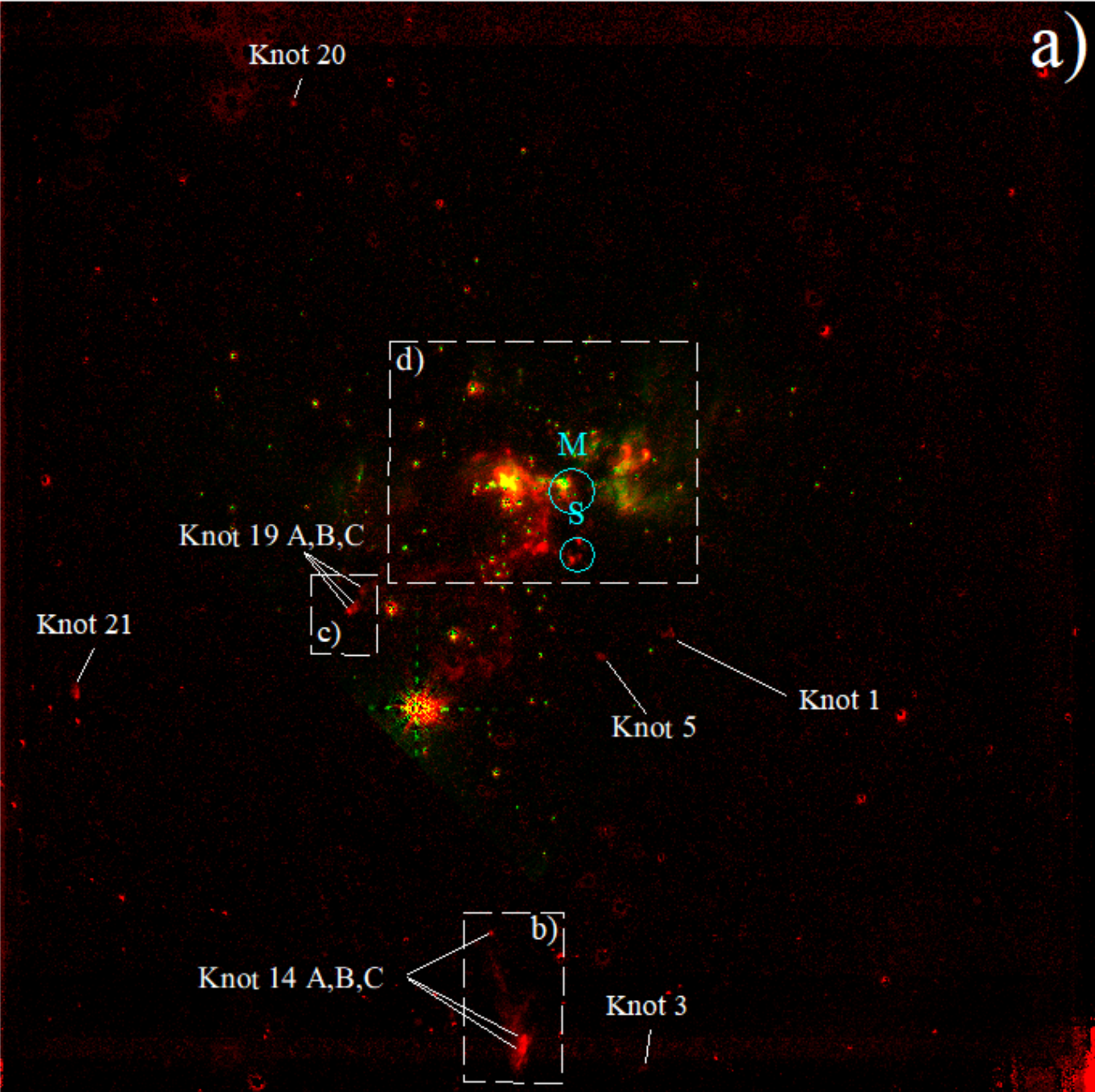}
\includegraphics[width=0.42\textwidth,height=0.42\textheight,keepaspectratio]{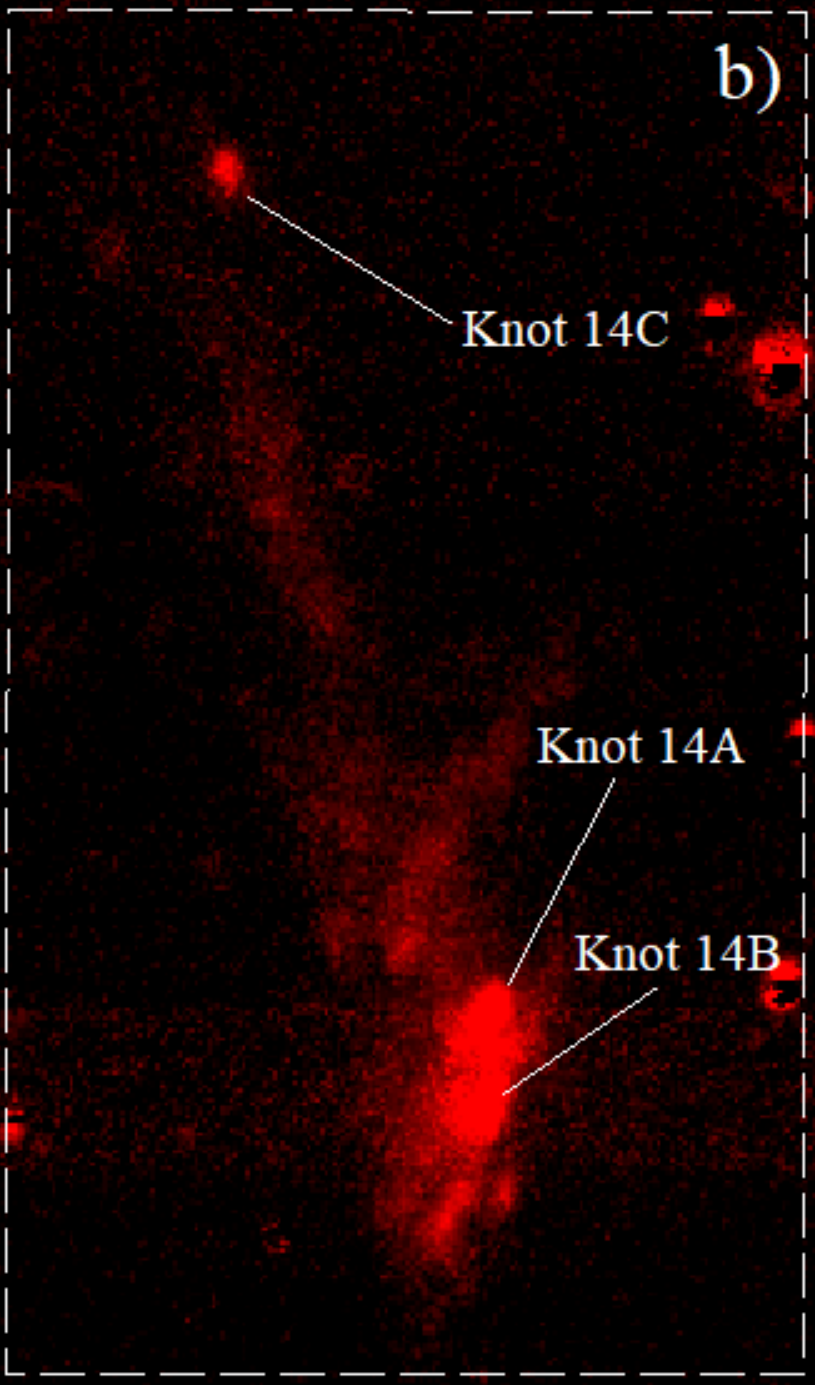}
\includegraphics[width=0.355\textwidth,height=0.355\textheight,keepaspectratio]{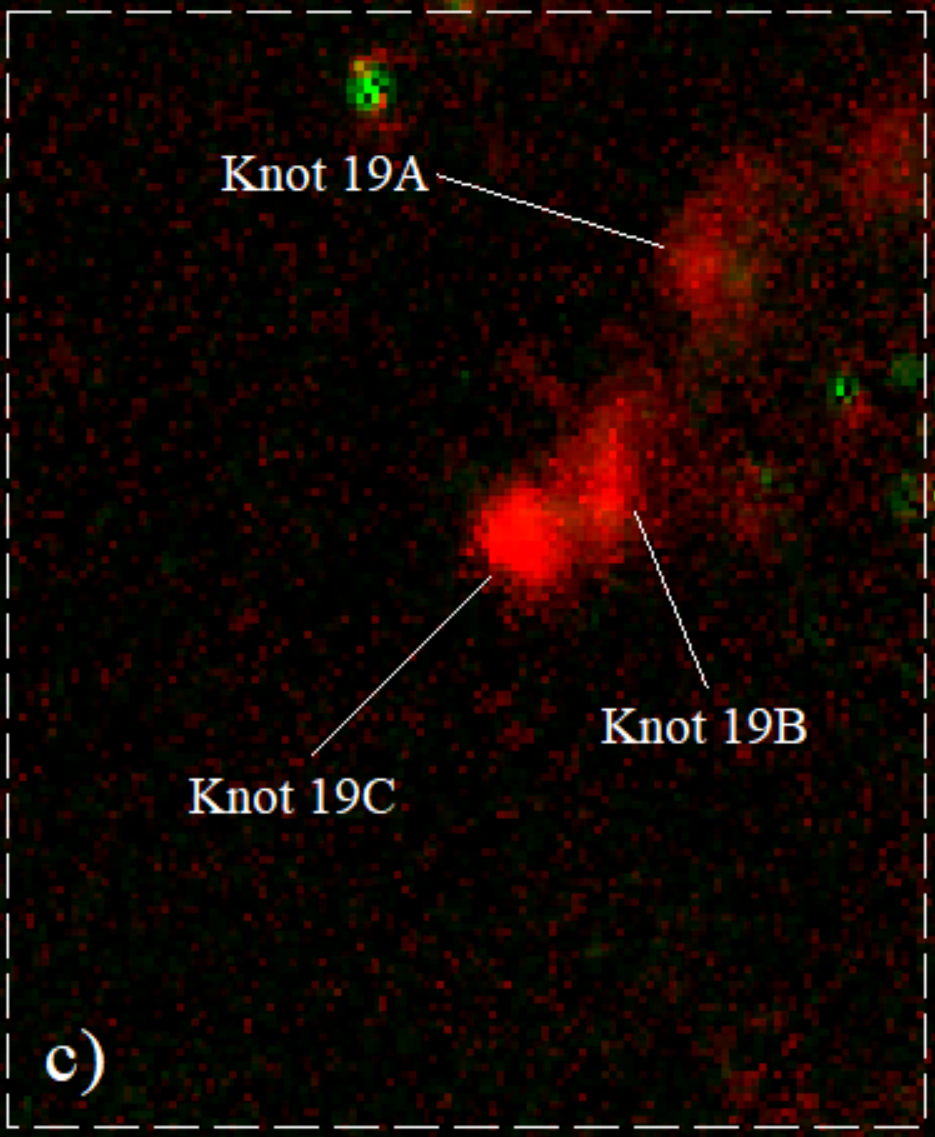}
\includegraphics[width=0.548\textwidth,height=0.548\textheight,keepaspectratio]{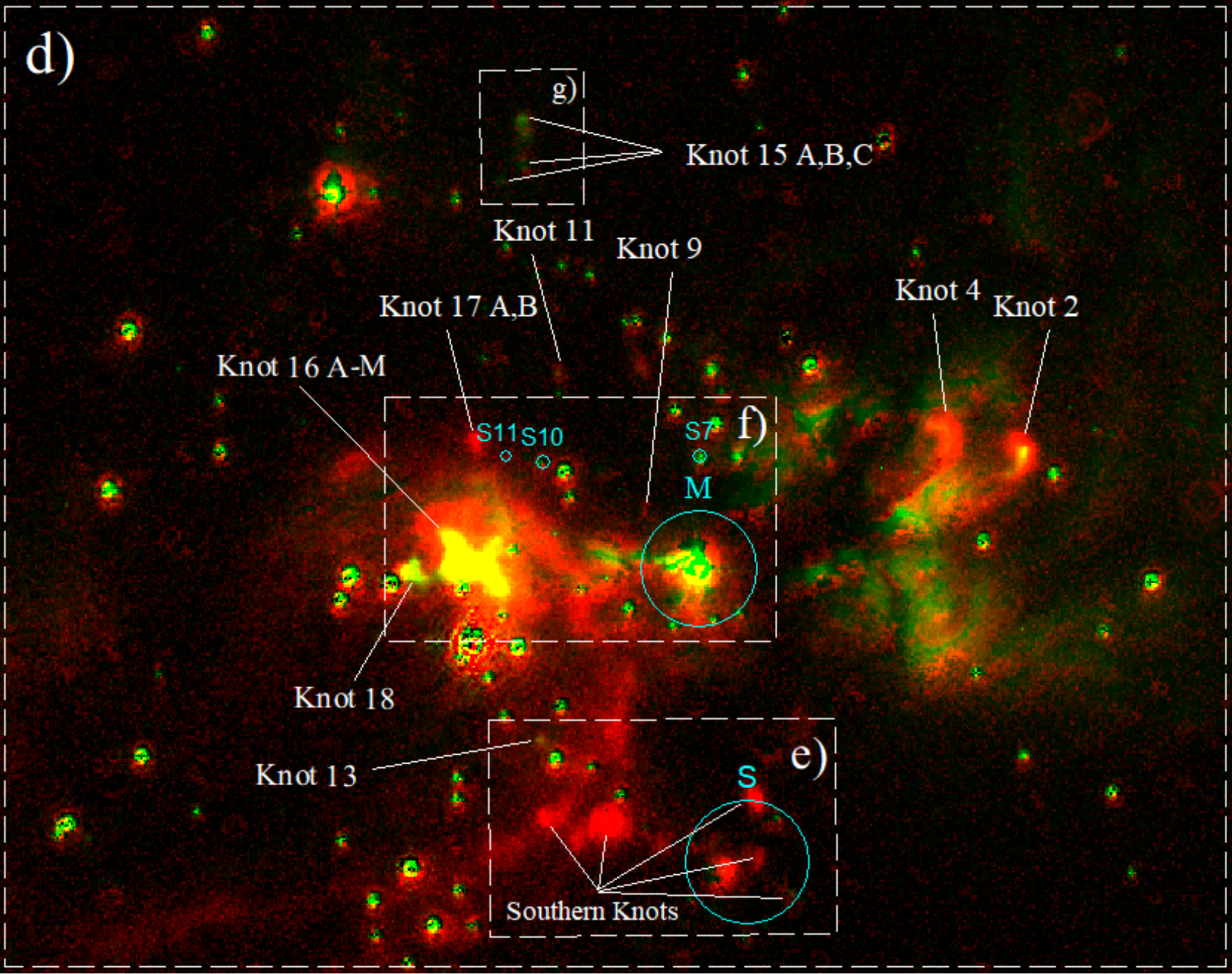}
\caption{\label{big_diagram} Diagram of identified jet features in the continuum-subtracted LBT LUCI H$_2$ data as well as HST [FeII] data; see Table \ref{Knot_table}. North is up and east is to the left. \textit{a)} 2-color image of the AFGL 5180 complex; red is LBT H$_2$ seeing-limited, green is HST [FeII]. The FOV is the same as in Fig. \ref{RGB_LBT_limited}. Knots and the locations of the AFGL 5180 M and S complexes are labeled. \textit{b)} Magnification of the knot 14 complex. \textit{c)} Magnification of the knot 19 complex. \textit{d)} Diagram of the inner outflowing region of AFGL 5180. Knots and cores are labeled in white and cyan, respectively. Note the presence of continuum-subtracted stars in the images that are responsible for the spots of dark and bright pixels; see §\ref{sec:continuum-sub}.}
\end{figure*}
\renewcommand{\thefigure}{\arabic{figure}}
\addtocounter{figure}{-1}
\begin{figure*}
\centering
\includegraphics[width=0.7\textwidth,height=0.7\textheight,keepaspectratio]{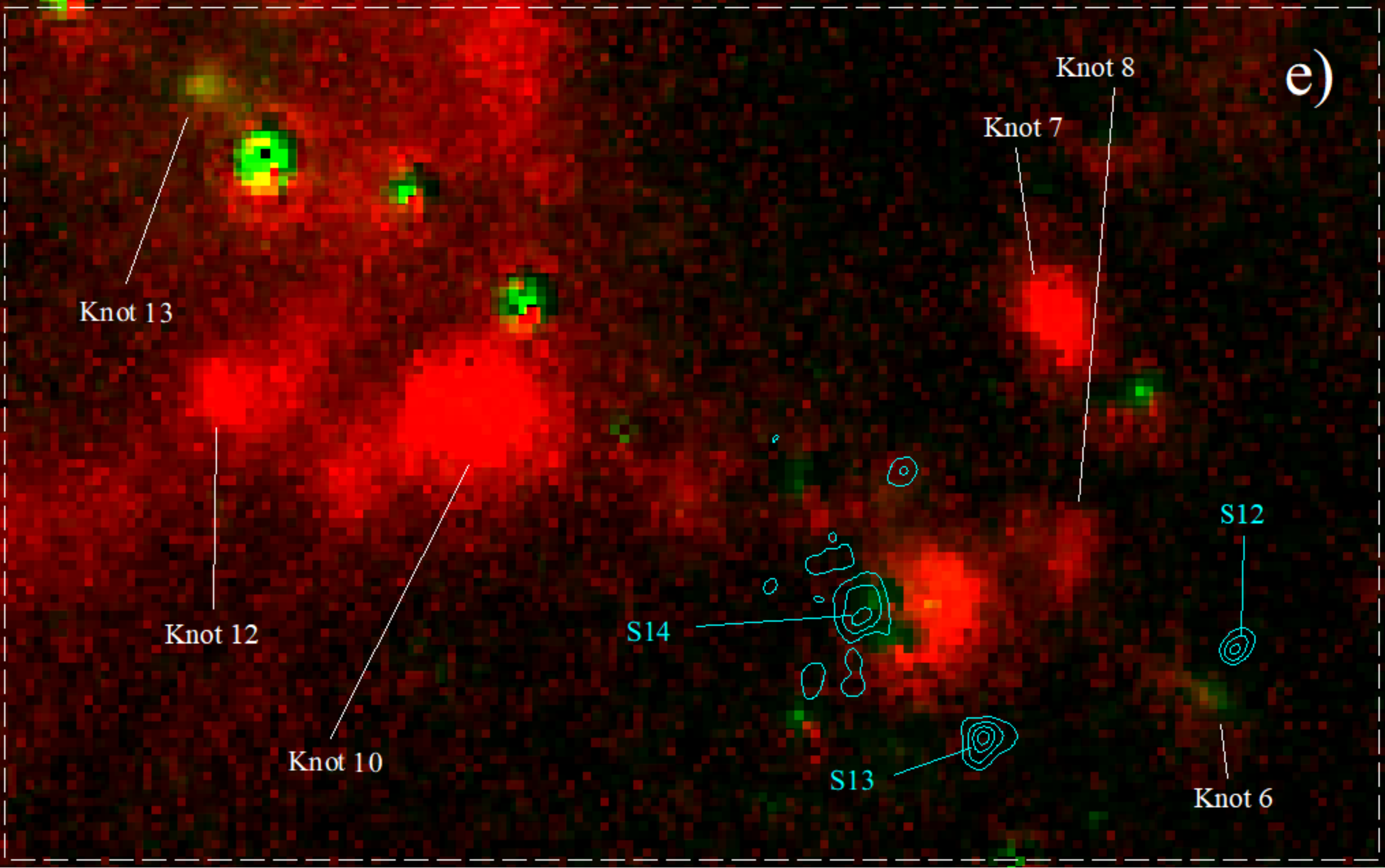}
\includegraphics[width=0.655\textwidth,height=0.655\textheight,keepaspectratio]{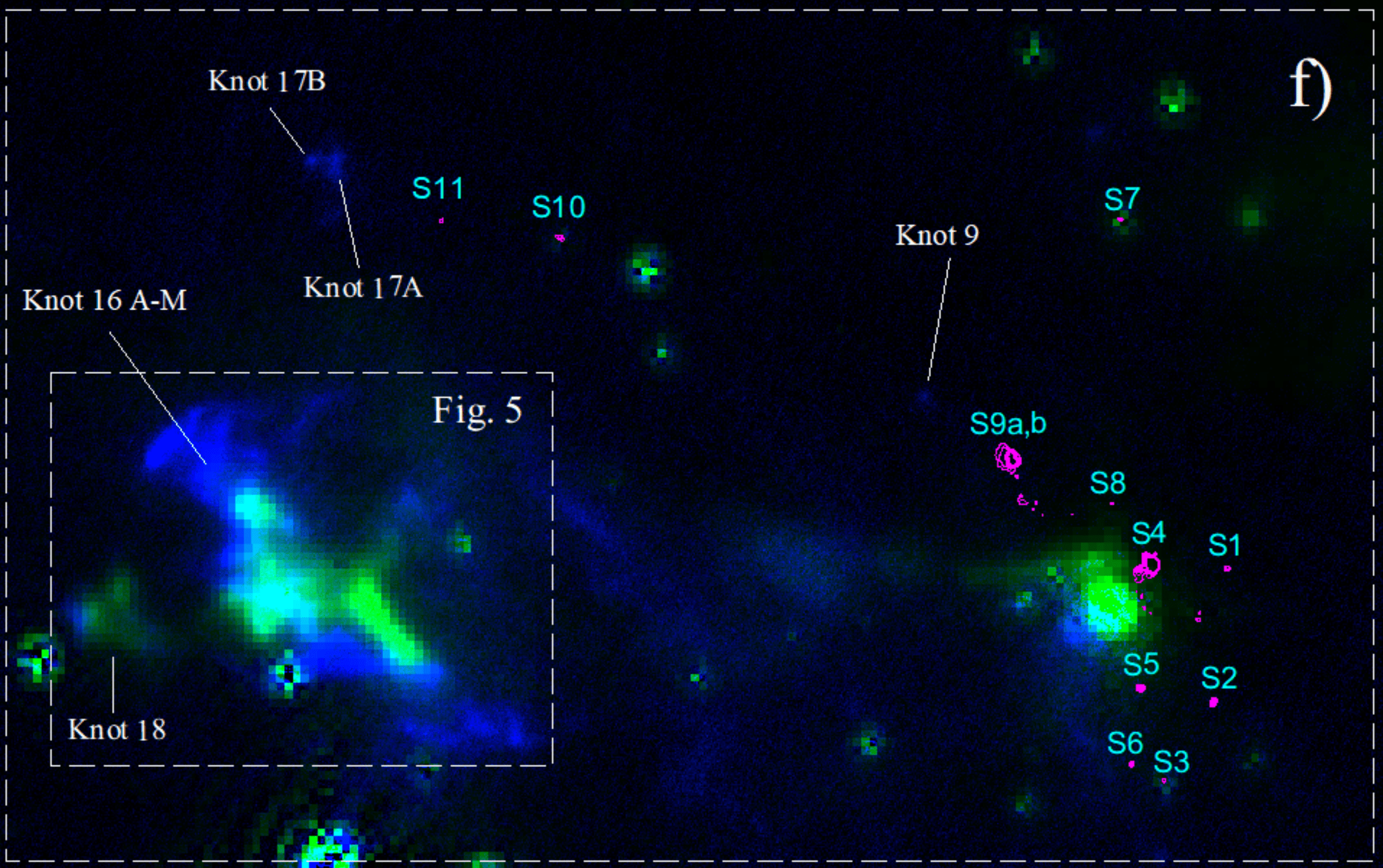}
\includegraphics[width=0.32\textwidth,height=0.32\textheight,keepaspectratio]{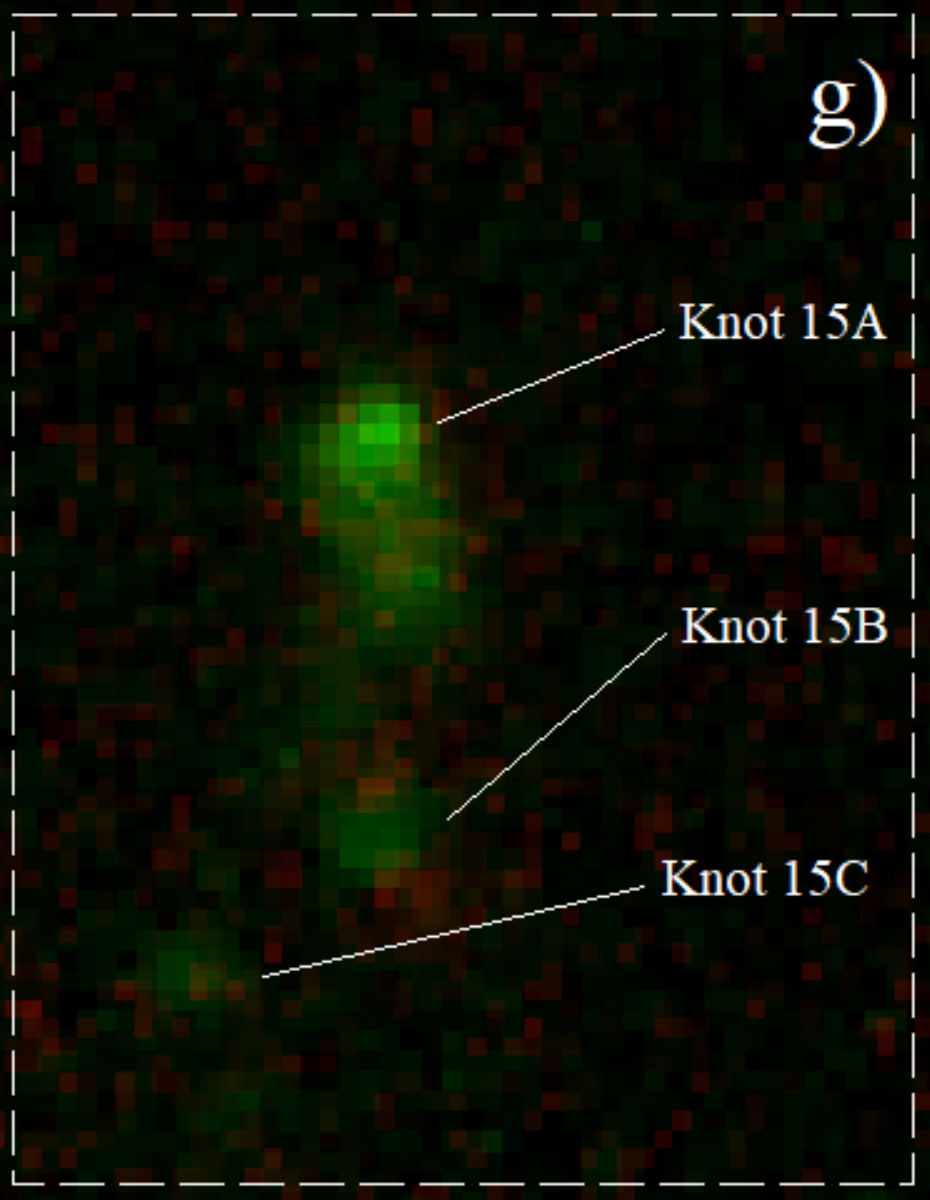}
\caption{cont. \textit{e)} Magnification of the AFGL 5180 S cluster. Knots and cores are labeled in white and cyan, respectively, and ALMA Band 6 contours are shown in cyan from 5 to 20 $\sigma$ in steps of 3. \textit{f)} Magnification of the main outflowing region around AFGL 5180 M. The H$_2$ seeing-limited data (red) is no longer shown, and instead the LUCI-1 SOUL AO data is shown in blue. The FOV is similar to Fig. \ref{RGB_LBT_AO}. Knots and cores are labeled in white and cyan, respectively, and ALMA Band 7 contours are shown in magenta from 5 to 20 $\sigma$ in steps of 3. The FOV of the knot 16 complex shown in Fig. \ref{knot16} is shown as the white dotted box. \textit{g)} Magnification of the knot 15 complex.}
\end{figure*}

\begin{figure*}
\centering
\includegraphics[width=0.6\textwidth,keepaspectratio]{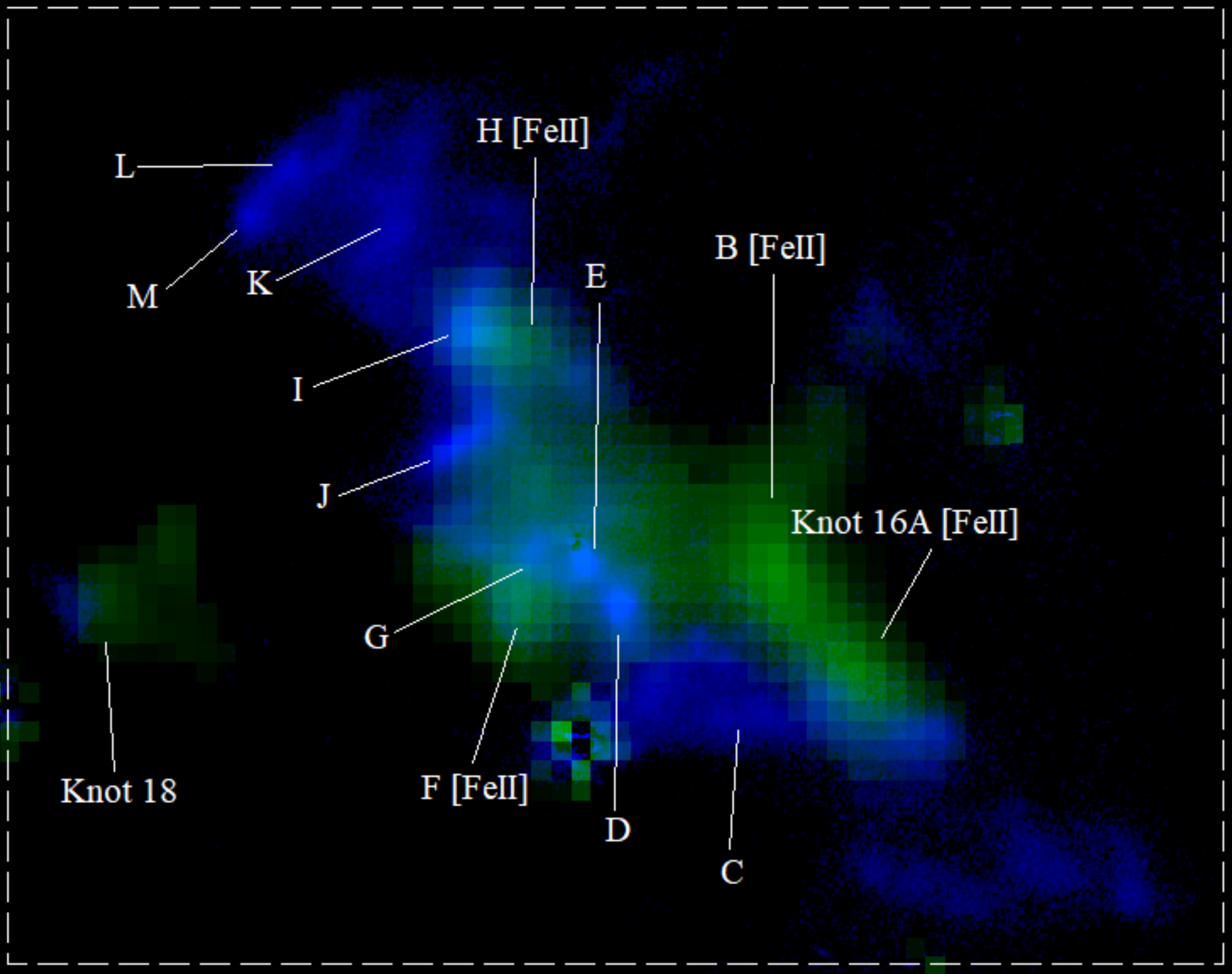}
\includegraphics[width=0.48\textwidth]{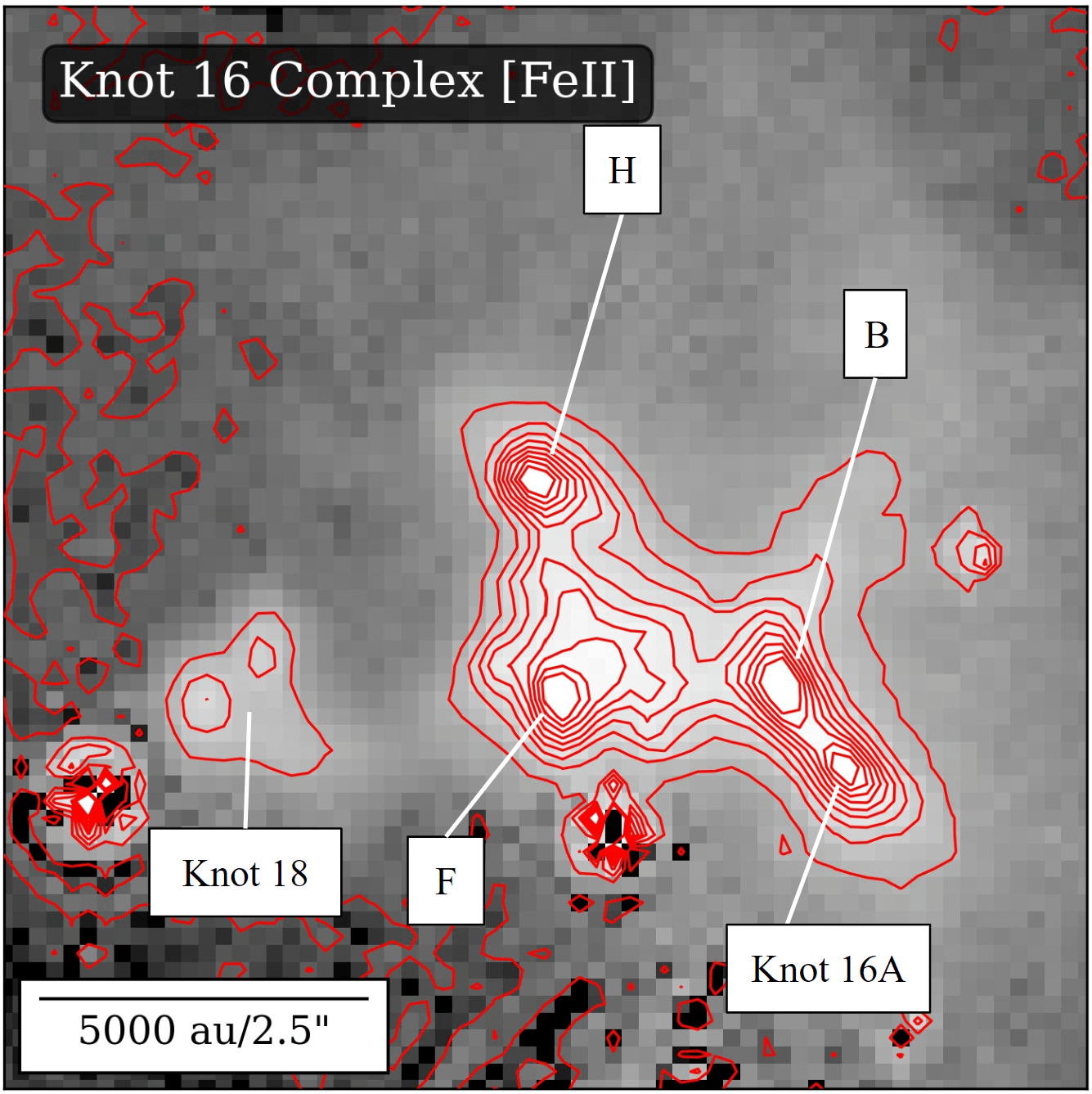}
\includegraphics[width=0.48\textwidth]{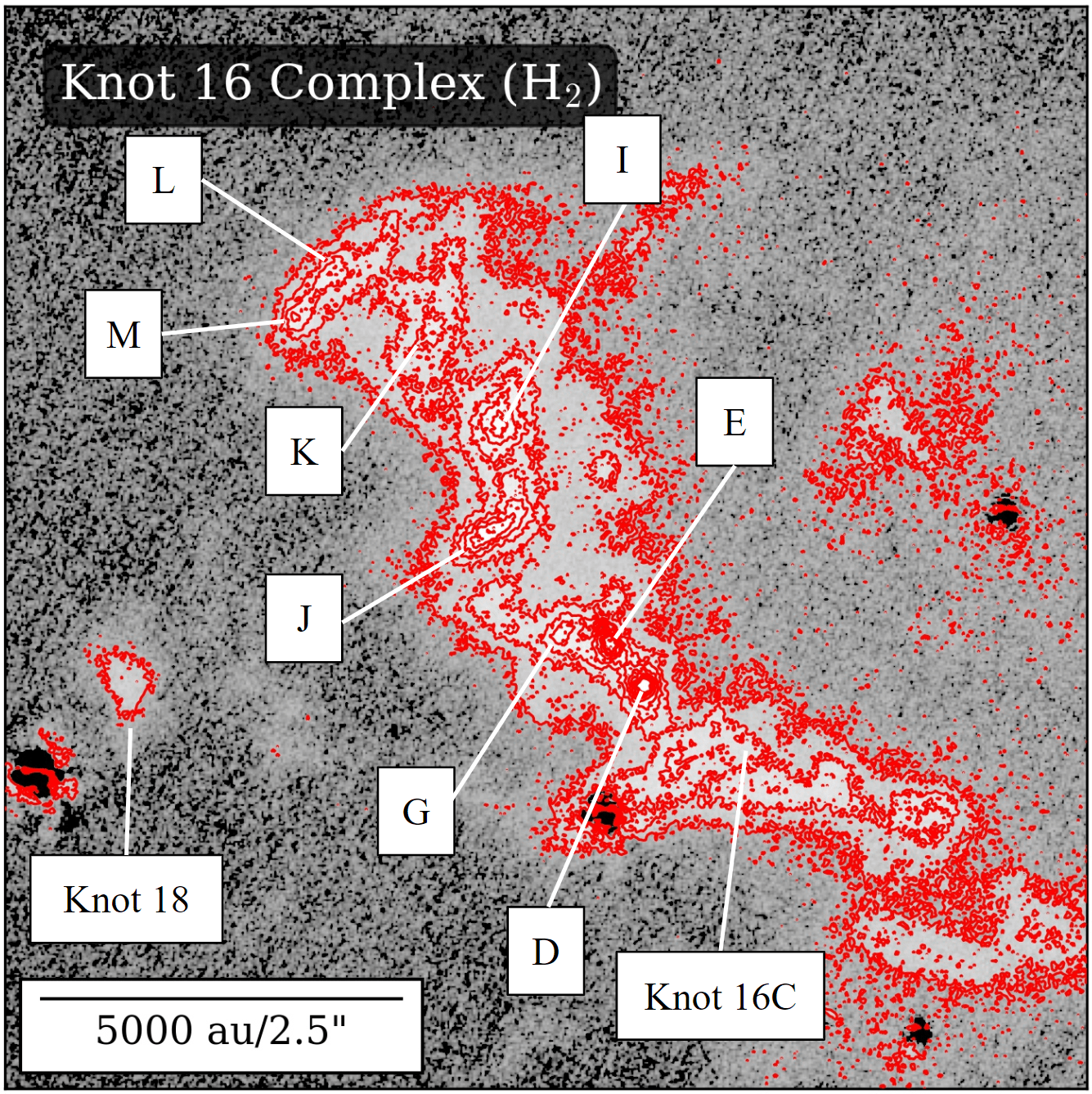}
\caption{\label{knot16} Diagram of the knot 16 complex with all identified features. \textit{Top:} 2-color continuum-subtracted image of the knot 16 complex, with HST [FeII] emission in green and LUCI-1 SOUL AO H$_2$ emission in blue (see Fig. \ref{big_diagram}\textit{f)} for reference). Sub-knots are labeled along with knot 18. Sub-knots of the knot 16 complex are identified first in [FeII] and then in H$_2$. \textit{Bottom left:} Significance level contour map for the sub-knots identified in the knot 16 complex from the continuum-subtracted [FeII] data (gray scale image). The contour levels represent values from 5 to 500$\sigma$, in steps of 55$\sigma$, where $\sigma$ is the standard deviation sampled from the local background (see §\ref{sec:knot_ID}) \textit{Bottom right:} Significance level contour map for the sub-knots identified in the knot 16 complex from the continuum-subtracted H$_2$ data (gray scale image). The contour levels represent values from 5 to 45$\sigma$, in steps of 5$\sigma$, where $\sigma$ is the standard deviation sampled from the local background (see §\ref{sec:knot_ID}). North is up and east is to the left in all panels.}
\end{figure*}

\begin{figure*}
\centering
\includegraphics[width=0.48\textwidth]{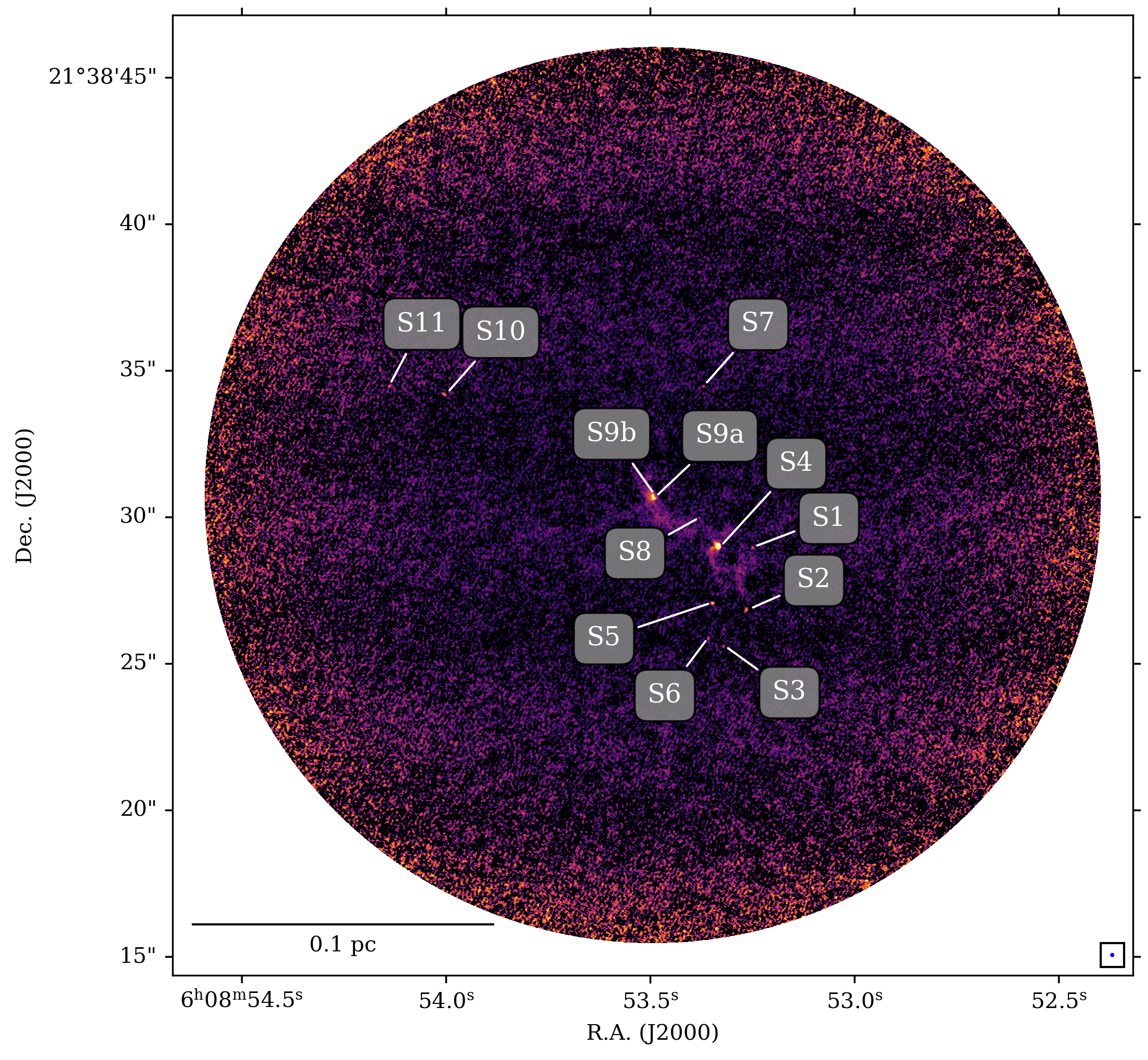}
\includegraphics[width=0.48\textwidth]{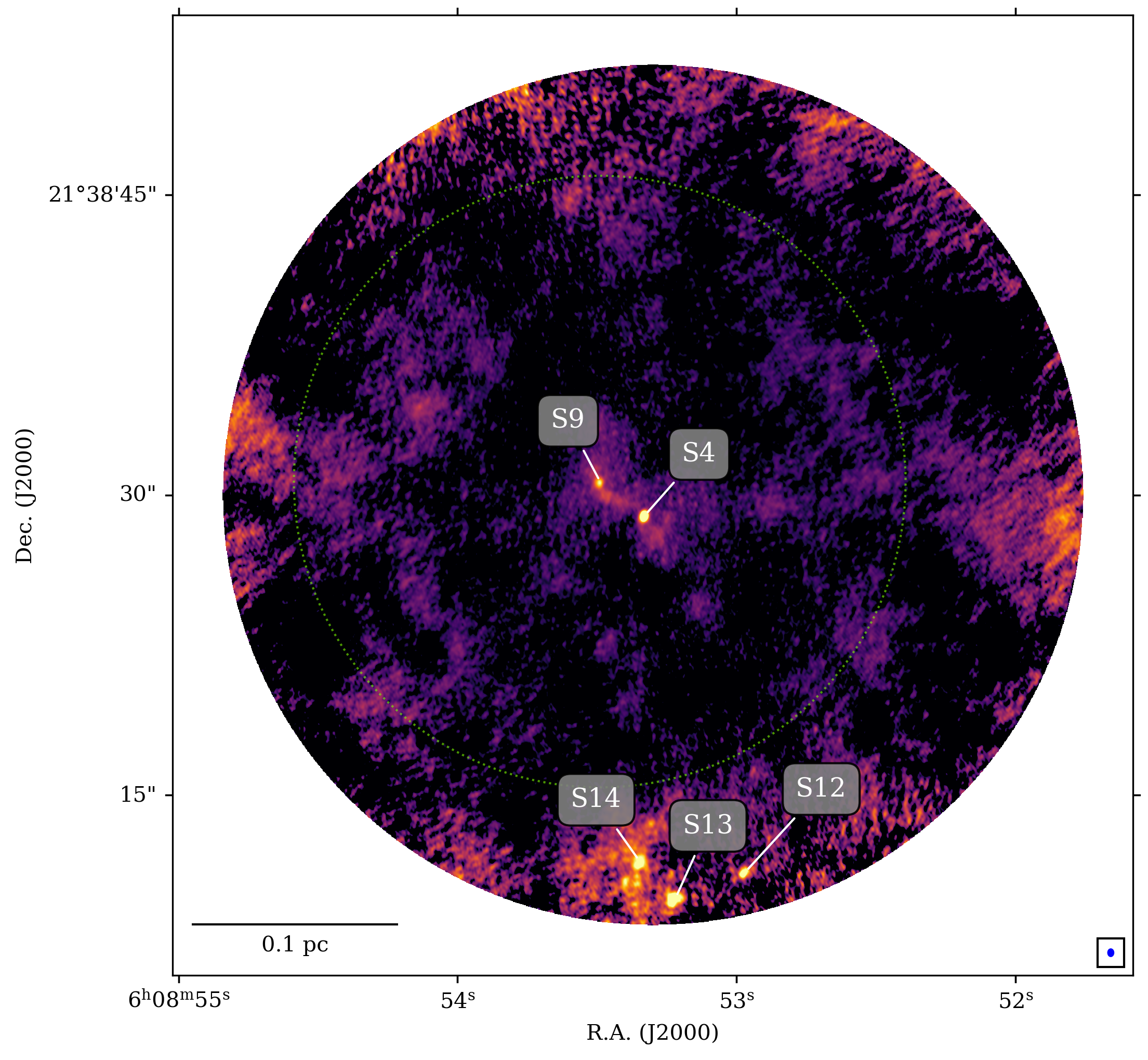}
\caption{\label{alma_diagram} \textit{Left:} ALMA Band 7 image (no primary beam correction applied) with identified cores labeled. The beam size is shown in the bottom-right corner. \textit{Right:} ALMA Band 6 image (no primary beam correction applied) with identified cores labeled. The beam size is shown in the bottom-right corner. The FOV of the ALMA Band 7 observation is shown as the dotted green circle.}
\end{figure*}

\subsubsection{The Sub-mm Population in AFGL 5180}\label{sec:cores}

\citet{mutie21} provide an analysis of ALMA Band 6 data towards AFGL 5180, from which they identify at least 8 dust cores, called MM1$-$8, among which MM1 and MM2 are found to have corresponding compact sources seen in the Band 7 high-resolution image. However, some of the cores identified from the ALMA Band 6 data (see §\ref{sec:ALMA_obs}), such as MM3 and MM4, appear as rather diffuse emission unlike the compact morphology which would be expected of a protostar \citep[see Fig. \ref{alma_diagram} and Fig. 5 of][]{mutie21}. Furthermore, inspection of the Band 7 data reveals many additional compact continuum sources which are unidentified in \citet{mutie21}. For these reasons, we present a re-analysis of the ALMA Band 6 and 7 data, for which we have re-reduced the data (see \S\ref{sec:ALMA_obs}).
        
From the high-resolution Band 7 data, we identify 12 compact continuum sources within $15\arcsec$ from the center, which are readily apparent upon visual inspection (see Fig. \ref{alma_diagram} and Table \ref{core_table}).
Among these sources, S4 and S9a/S9b are reported by \citet{mutie21} but labeled as MM1 and MM2, respectively. The other sources are newly identified. Due to the smaller FOV of the Band 7 image, the southern cores seen in the Band 6 image are not covered by the Band 7 image. Therefore we supplement the source list with these three cores, S12-14 \citep[labeled MM6-8 in][]{mutie21}. However, it is unclear how many compact sources (YSOs) are really harbored in these cores due to the lack of high-resolution observations in these regions.
        
The coordinates of each source are obtained with 2D Gaussian fitting, and the fluxes are determined from primary beam corrected images by integrating within the fitted Gaussian ellipse at $1.5\times\mathrm{FWHM}$. This method primarily retrieves flux from the central compact structure of each source; therefore, the fluxes and masses presented likely represent those of the disk or inner envelope of the sources, rather than the full envelope mass.
The properties of each of these sources are summarized in Table \ref{core_table}.
     
The core masses are estimated from the fluxes following standard assuptions of optically thin thermal dust emission, i.e.,
\begin{equation}
M_{\rm mm} = \frac{S_{\nu}d^2} {\kappa_{\nu}B_{\nu}(T_d)} \,
\end{equation}
where $S_{\nu}$ is the dust continuum flux at the frequency $\nu$ (230 GHz for Band 6; 330 GHz for Band 7), $d$ is the distance to AFGL 5180, which we take to be 2 kpc (see §\ref{sec:introduction}), $\kappa_{\nu}$ is the dust opacity per unit mass, which we take to be 0.899 $\mathrm{cm^{2}\:g^{-1}}$ for Band 6 and 1.77 $\mathrm{cm^{2}\:g^{-1}}$ for Band 7 \citep{ossenkopf94}, and $B_{\nu}$ is the spectral radiance (Planck function) at the frequency $\nu$ and dust temperature $T_d$, which is taken to be 50 K for the southern complex (S12-14) and 42 K for the remaining sources based on Spectral Energy Distribution (SED) fitting results from \citet{minier05}. A gas-to-dust ratio of 100 was assumed. The mass estimates are presented in Table \ref{core_table}. 


Looking at the resulting population of ALMA-identified protostellar YSO candidates, a few trends are noticeable. 
First, all of the sources appear to be contained in relatively low-mass inner envelopes or disks, with the most massive source S4 having an envelope mass of $\sim2\,M_\odot$ as estimated by ALMA Band 6, and the next most massive sources, S14 and S13, having masses of $1.8\,M_\odot$ and $1.5\,M_\odot$, respectively.
The source S4 is also noteworthy as it is associated with a prominent methanol maser, which is well documented by decades of observations \citep{goedhart04, goedhart07, vanderwalt11, goedhart14, mutie21}, a strong indication of massive star-forming activity \citep{breen13}, and which also shows a cyclic pattern in variability, perhaps indicating the past occurrence of accretion bursts
\citep[see also][for a discussion of episodic accretion and ejection processes in massive protostars]{caratti17,cesaroni18,fedriani23b}. S4 appears to be the source primarily responsible for powering the main east-west outflow, being the brightest and most massive source in the AFGL 5180 complex, and positioned at the tip of both the east and west outflow cones. See \S\ref{sec:s4_jet} for further analysis of S4 and the prominent H$_2$ and [FeII] knots it appears to be powering in the eastern outflow. 
        
Second, the sources appear to comprise two main clusters: one containing S1-6, 8, and 9 (hereafter AFGL 5180 M), and another $\sim15\arcsec$ (0.15 pc) to the south of AFGL 5180 M containing S12-14 (hereafter AFGL 5180 S). In addition, there are a few other sources (S7, S10 and S11) distributed to the N and NE (see Fig. \ref{alma_diagram}). Both AFGL 5180 M and S appear to be sites of active star formation, based on the presence of multiple water masers \citep[see, e.g.,][]{tofani95} and dozens of jet knots.
We conclude that these young clusters likely contain many of the driving sources responsible for the jet knots observed in the NIR (see \S\ref{sec:knot_ID}).
                 
\subsubsection{Jet Feature Identification}\label{sec:knot_ID}

Jet knots in the AFGL 5180 complex were identified using the H$_2$ and [FeII] continuum-subtracted images, which trace shocked emission from protostellar jets. The Br$\gamma$ and Pa$\beta$ continuum-subtracted images were also examined, but did not have any identifiable compact knot-like features.
        
Knots are identified as extended emission features in the continuum-subtracted images. Their peak intensity is determined in relation to the local standard deviation sampled in the vicinity of the knot candidate. The minimum size of a knot candidate in each image is determined by the angular resolution of the image, as described below.
        
The pixel scale of the H$_2$ seeing-limited image is 0.12\arcsec/pixel, which, with an angular resolution of $\sim$ 0.5\arcsec, gives a minimum resolvable size of $\sim$ 4 pixels. The pixel scale of the AO H$_2$ image is 0.015\arcsec/pixel, which, with an angular resolution of $\sim$ 0.09\arcsec, gives a minimum resolvable size of $\sim$ 6 pixels. Finally, the pixel scale of the [FeII] image is 0.13\arcsec/pixel, which, with an angular resolution of 0.17\arcsec, gives a minimum resolvable size of $\gtrsim$ 1 pixel.
                     
The minimum intensity peak for knot identification in the H$_2$ (seeing-limited and AO) images was set at $3\sigma$, based on the $\sigma$ level of the dimmest knots apparent in the seeing-limited images, knots 1 and 3 (see Fig. \ref{big_diagram} and Appendix \ref{sec:Knot_significance}). Likewise, the minimum intensity peak for knot identification in the [FeII] image was set at $5\sigma$, based on the $\sigma$ level of the dimmest knot apparent in the image, knot 13 (see Fig. \ref{big_diagram} and Appendix \ref{sec:Knot_significance}). Table \ref{Knot_table} presents a list of all identified knots in the images. See Figures \ref{seeing-limited knots}, \ref{AO_knot_sig}, and \ref{FEII_knot_sig} for a close-in view of each knot. Precise coordinates presented in Table \ref{Knot_table} were determined via 2D Gaussian fitting of each knot, and Table \ref{Knot_table} also provides a prediction of the potential driving core(s) for each knot, and therefore its position angle (relative to N is up for PA=0$^{\circ}$), and angular separation.

        
Associations between knots and driving sources were made by inspection of the jet morphology of the image to identify linear patterns that provide evidence of origin from a particular source/cluster. See Appendix \ref{sec:knot_attribution} for a demonstration of how such associations were made. These associations are, however, tentative, and due to the crowded distribution of the ALMA cores (see §\ref{sec:cores}), in many cases it was challenging to attribute each knot to a single driving core. Therefore, many of the knots presented in Table \ref{Knot_table} are attributed to multiple cores/clusters. Additional data revealing proper motions would allow for knots to be more reliably traced back to specific cores; see §\ref{sec:Discussion} for further discussion. However, in some cases we can still differentiate between knots driven by sources from either of the two main clusters, AFGL 5180 M and S. The results presented in Table \ref{Knot_table} for knots originating from these clusters are calculated assuming S4 and S13 as the centers of the AFGL 5180 M and S clusters, respectively.
        
It is worth noting that many of the knots extending away from the main outflowing structure in AFGL 5180 appear to group towards the southeast (see Fig. \ref{RGB_LBT_limited}). This morphology can be explained if this side of the outflow forms the near-facing cavity, in which case the jet knots would be expected to be blueshifted. If the main protostar is forming via core accretion, then the opposite, far-facing cavity is expected to be present and also include jet knots. However, these would likely suffer from higher levels of extinction and thus appear dimmer and be harder to detect. The prediction of blueshifted knots in the SE group requires confirmation via spectroscopic observations.
        
            
        
It is also worth noting that a few of the identified knots correspond with those identified in the study by \citet{davis98} (see their Fig. 9). In particular: knot C in \citet{davis98} corresponds to knot 10 in our study; their knot D corresponds to our knot 7; their knot A is our knot 19 (split into A, B, and C); and their knot N corresponds with our knot 16 (split into A-M). The knots which they identify as S$_1$ and S$_2$, although visible in the H$_2$ data (see Figs. \ref{RGB_LBT_limited} and \ref{big_diagram}), have not been picked up as significant enough to be considered knots by our definition, i.e., they appear to be diffuse nebulosity rather than concentrated knots. \citet{davis98} discuss the possibility of an east-west outflow connecting their knots A-D, which our results would seem to corroborate with the presence of a clear flow connecting knots 10, 12, 19, and 21 back to AFGL 5180 S. Our results, however, do not support their interpretation of their knot N (our knot 16) being part of a north-south outflow; rather, we provide strong evidence that this knot complex is actually part of an east-west outflow being powered by S4 (see §\ref{sec:s4_jet}). 
\citet{davis98} also identify a collimated jet feature far to the east of the main complex, which is outside the FOV of our data. 
                    
Finally, \citet{davis98} identify an H$_2$ feature to the south of the main AFGL 5180 complex, which is clearly identifiable in our work as the prominent bow-shaped feature knot 14 (see Fig. \ref{big_diagram}\textit{a), b)}). Besides its brightness, this feature is also interesting because of its distance from the main outflowing complex and ALMA sources; $\sim2\arcmin$; 1.2 pc. Assuming a constant velocity of 100 $\mathrm{km\:s^{-1}}$ away from its driving core, which is a typical value found by previous studies of such knots \citep[see][and references therein]{fedriani18,fedriani20}, and a distance of 2 kpc to AFGL 5180 (see §\ref{sec:introduction}), we estimate this knot to be $\sim10^4$ years old, placing a lower limit on the timescale of star formation in AFGL 5180. A similar analysis shows that the closest knots to their driving sources have dynamical ages of a few hundred years, providing indication that star formation is ongoing in AFGL 5180. Knot 14 is also notably off-axis by $\sim90^{\circ}$ from the main bipolar outflow, so it provides some of the strongest evidence for multiple driving sources in the region.
        
            
        
\subsubsection{S4: The Most Massive Protostar in AFGL 5180}\label{sec:s4_jet}
        
\begin{table*}
\centering
\caption{Knot features. Significance levels are listed in levels of $\sigma$ sampled above the local background, see §\ref{sec:knot_ID}. Knots which can be attributed to multiple cores/clusters are indicated by a slash (/) along with the corresponding values for separation and PA 
for each core/cluster. 
Knots which were not at a high enough $\sigma$ peak in a given image to constitute a detection (or which are unresolvable sub-knots in a given image) are denoted by $\cdots$, and those which are outside the field of view of a given image are denoted by -.}   
\label{Knot_table}             
\begin{tabular}{c c c c c c c c c c}     
\hline\hline       
Feature &R.A.(J2000)&Dec.(J2000)&$\mathrm{H_{2}}$&$\mathrm{H_{2}}$ AO&  [FeII]& Core(s) & PA& Sep. 
            \\ 
            & (J2000) & (J2000) & ($\sigma$) & ($\sigma$) & ($\sigma$) &  & $(^{\circ})$ & (\arcsec) 
            \\
\hline                    
K1 & 06:08:51.71 & 21:37:55.60 & 5 & - & $\cdots$ & S & 236 & 25 
               \\  
K2 & 06:08:52.09 & 21:38:34.75 & 30 & - & 65 & M & 288 & 18 
               \\
K3 & 06:08:52.14 & 21:36:20.69 & 3 & - & - & S & 188 & 110 
               \\
               K4 & 06:08:52.38 & 21:38:35.74 & 25 & - & $\cdots$ & M & 297 & 15 
               \\
               K5 & 06:08:52.80 & 21:37:50.53 & 5 & - & $\cdots$ & S & 197 & 20 
               \\
               K6 & 06:08:52.99 & 21:38:10.57 & $\cdots$ & - & 8 & S & 283 & 3 
               \\
               K7 & 06:08:53.15 & 21:38:15.85 & 19 & - & $\cdots$ & S & 351 & 6 
               \\
               K8 & 06:08:53.16 & 21:38:12.42 & 5 & - & $\cdots$ & S & 343 & 3 
               \\
               K9 & 06:08:53.59 & 21:38:31.67 & $\cdots$ & 3 & $\cdots$ & M & 54 & 5 
               \\
               K10 & 06:08:53.74 & 21:38:14.68 & 56 & 38 & $\cdots$ & S & 56 & 9 
               \\
               K11 & 06:08:53.93 & 21:38:39.20 & $\cdots$ & 5 & $\cdots$ & M & 40 & 13 
               \\
               K12 & 06:08:53.95 & 21:38:14.83 & 13 & 9 & $\cdots$ & S & 64 & 11 
               \\
               K13 & 06:08:54.01 & 21:38:18.99 & $\cdots$ & 5 & 5 & M/S & 137/50 & 14/14 
               \\
               K14A & 06:08:54.02 & 21:36:26.84 & 21 & - & - & M/S & 176/174 & 123/104 
               \\
               K14B & 06:08:54.04 & 21:36:24.78 & 19 & - & - & M/S & 176/174 & 125/106 
               \\
               K14C & 06:08:54.53 & 21:36:50.22 & 11 & - & - & M/S & 170/167 & 100/82 
               \\
               K15A & 06:08:54.06 & 21:38:53.06 & $\cdots$ & - & 29 & M/13/14 & 23/3/357 & 26/19/19 
               \\
               K15B & 06:08:54.06 & 21:38:50.76 & $\cdots$ & - & 11 & M/13/14 & 25/3/356 & 24/17/16 
               \\
               K15C & 06:08:54.15 & 21:38:49.85 & $\cdots$ & - & 8 & M/13/14 & 29/8/1 & 24/16/15 
               \\
               K16A & 06:08:54.18 & 21:38:27.93 & $\cdots$ & $\cdots$ & 500 & 3 & 95 & 12 
               \\
               K16B & 06:08:54.21 & 21:38:28.47 & $\cdots$ & $\cdots$ & 590 & 3 & 92 & 12 
               \\
               K16C & 06:08:54.23 & 21:38:27.45 & $\cdots$ & 15 & $\cdots$ & 3 & 97 & 13 
               \\
               K16D & 06:08:54.29 & 21:38:28.20 & $\cdots$ & 55 & $\cdots$ & 3 & 93 & 13 
               \\
               K16E & 06:08:54.31 & 21:38:28.47 & $\cdots$ & 40 & $\cdots$ & 3 & 92 & 14 
               \\
               K16F & 06:08:54.32 & 21:38:28.66 & $\cdots$ & $\cdots$ & 540 & 3 & 91 & 14 
               \\
               K16G & 06:08:54.33 & 21:38:28.55 & $\cdots$ & 20 & $\cdots$ & 3 & 92 & 14 
               \\
               K16H & 06:08:54.34 & 21:38:30.01 & $\cdots$ & $\cdots$ & 540 & 3 & 86 & 14 
               \\
               K16I & 06:08:54.37 & 21:38:29.99 & $\cdots$ & 25 & $\cdots$ & 3 & 86 & 15 
               \\
               K16J & 06:08:54.37 & 21:38:29.29 & $\cdots$ & 25 & $\cdots$ & 3 & 89 & 15 
               \\
               K16K & 06:08:54.40 & 21:38:30.66 & $\cdots$ & 15 & $\cdots$ & 3 & 83 & 15 
               \\
               K16L & 06:08:54.45 & 21:38:31.01 & $\cdots$ & 15 & $\cdots$ & 3 & 83 & 16 
               \\
               K16M & 06:08:54.46 & 21:38:30.74 & $\cdots$ & 20 & $\cdots$ & 3 & 84 & 16 
               \\
               K17A & 06:08:54.26 & 21:38:35.37 & 11 & 9 & $\cdots$ & M/13/14 & 64/72/63 & 14/4/2 
               \\
               K17B & 06:08:54.28 & 21:38:35.43 & $\cdots$ & 7 & $\cdots$ & M/13/14 & 64/73/65 & 15/4/2 
               \\
               K18 & 06:08:54.55 & 21:38:28.21 & 15 & 5 & 85 & 3 & 92 & 17 
               \\
               K19A & 06:08:56.54 & 21:38:05.02 & 5 & - & $\cdots$ & M/S & 118/96 & 51/47 
               \\
               K19B & 06:08:56.66 & 21:38:01.38 & 9 & - & $\cdots$ & M/S & 121/100 & 54/49 
               \\
               K19C & 06:08:56.75 & 21:38:00.71 & 15 & - & $\cdots$ & M/S & 121/101 & 55/50 
               \\
               K20 & 06:08:57.61 & 21:39:51.75 & 5 & - & - & M/13/14 & 36/33/32 & 102/92/91 
               \\
               K21 & 06:09:01.03 & 21:37:42.79 & 9 & - & - & M/S & 113/104 & 117/112 
               \\
\hline                  
\end{tabular}
\end{table*}
                    
The massive protostellar core S4, with a current envelope mass of $\sim2\:M_{\odot}$ (see Table \ref{core_table}) and a current protostellar mass from SED fitting of $m_*\simeq11.3^{+6.1}_{-4.0}\,M_\odot\:$ (see §\ref{sec:stellar_densities}) (Telkamp et al., in prep.), is likely to be the driver of the prominent east-west outflow, which dominates the NIR emission in the region. The evidence for this is: its position in the complex at the tip of the eastern outflow cavity (see, e.g., Fig. \ref{RGB_LBT_AO} and \ref{big_diagram}\textit{f)}); the fact that it is the brightest core in the AFGL 5180 M complex by a factor of a few (see Table \ref{core_table}); and the only core in the complex associated with a methanol maser, a strong indication of massive star formation activity \citep{breen13}. 
        
        
        
Although there are prominent knots, knots 2 and 4, in the western nebulosity which appear to be associated with S4, the knots are noticeably off-axis, and therefore difficult to confidently attribute to S4, as opposed to any of the other sources in AFGL 5180 M.
        
It is worth noting, however, the bow-like morphology of knots 2 and 4, which seem to point back to the heart of AFGL 5180 M (see Fig. \ref{big_diagram}\textit{d)}). Due to the similar estimated dynamical ages for knots 2 and 4 and the knots 16A-M and 18 (see Table \ref{Knot_table}), it would be difficult to justify the small timescales necessary for S4 to have produced both sets of knots due to precessional or tidal interactions with the nearby cores. However, the possibility of S4 being the driving core behind knots 2 and 4 cannot be entirely ruled out. In the eastern lobe of the outflow, however, which appears to be the blue-shifted lobe based on evidence from previous studies \citep[e.g.,][]{snell88, tamura91}, there are a number of features that can be reliably attributed to S4, notably the knot 16 complex and knot 18 (see Fig. \ref{big_diagram}\textit{f)}). Proper motions have been observed in these knots that provide additional strong evidence of their origin from S4 (Fedriani et al. in prep). 

            
An image with all identified knots in this complex is shown in Figure \ref{knot16}. There are several very bright knots traced by H$_2$ and [FeII] identified in both the LUCI-1 SOUL AO and HST [FeII] images in the eastern outflow powered by S4. In particular, the high angular resolution of the AO data provides a high level of detail of a large H$_2$ structure in the eastern outflow. Notably, this H$_2$ structure is spatially overlapping with multiple bright [FeII] knots, many of which are significant up to $>500\sigma$ (see Table \ref{Knot_table}). These [FeII] knots are noteworthy for being significantly brighter than the rest of the knots in the entire AFGL 5180 complex, indicating outflowing activity from S4 within the last couple thousand years producing strong shocks. It is noteworthy that the [FeII] emission is more concentrated at the center of knot 16, while the H$_2$ extends across the entire length of knot 16, particularly in the wings of the knot. Additionally, despite some overlapping emission, there are regions where only [FeII] is detected. This is likely indicative of a strong shock where the H$_2$ emission cannot persist, and only the [FeII] emission remains. This is a good example of the often-referenced ``onion-like'' structure in protostellar jets \citep[see][and references therein]{bally16}.
        

\subsubsection{YSO Surface Number Density Analysis}\label{sec:Nstar}

\begin{figure*}[t]
\centering
\includegraphics[width=0.47\textwidth]{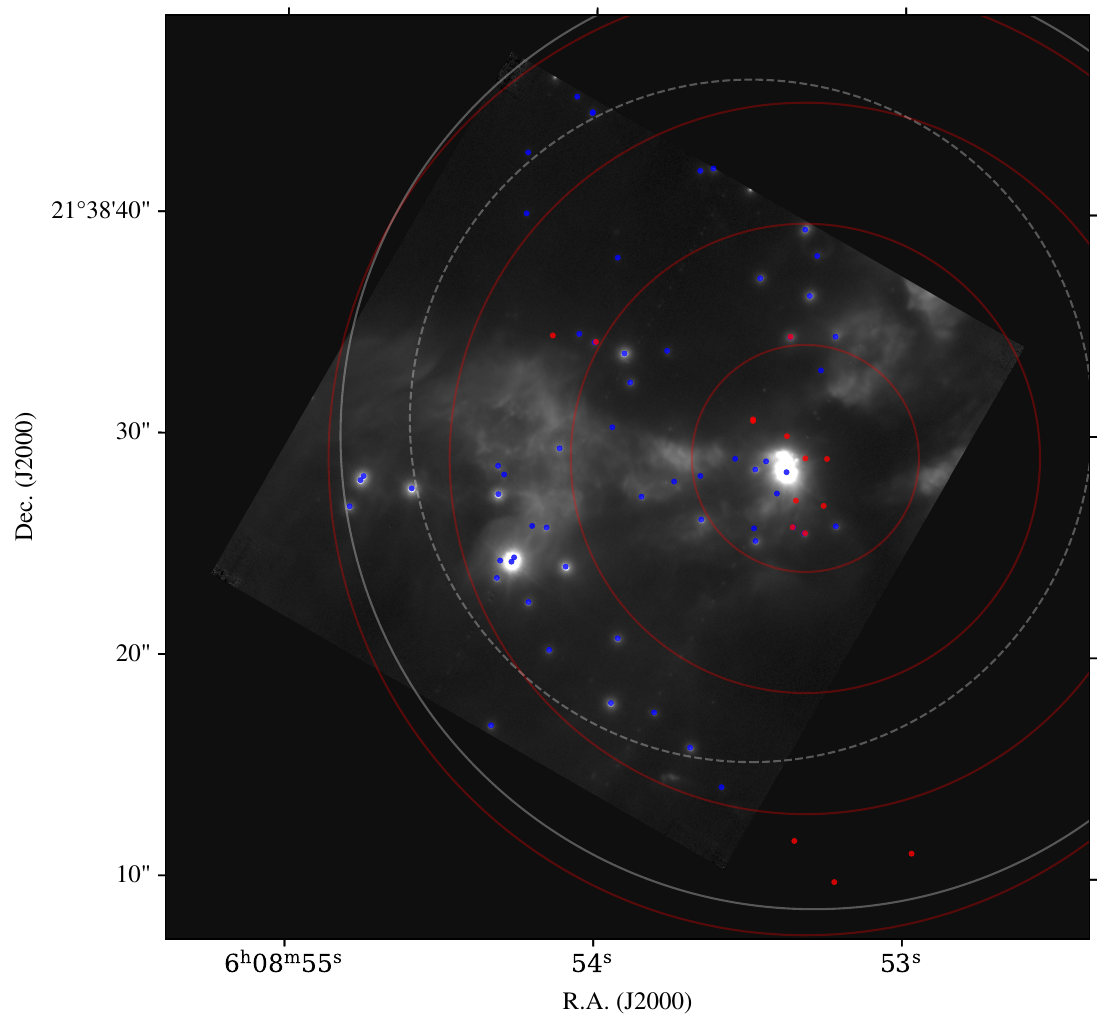}
\includegraphics[width=0.49\textwidth]{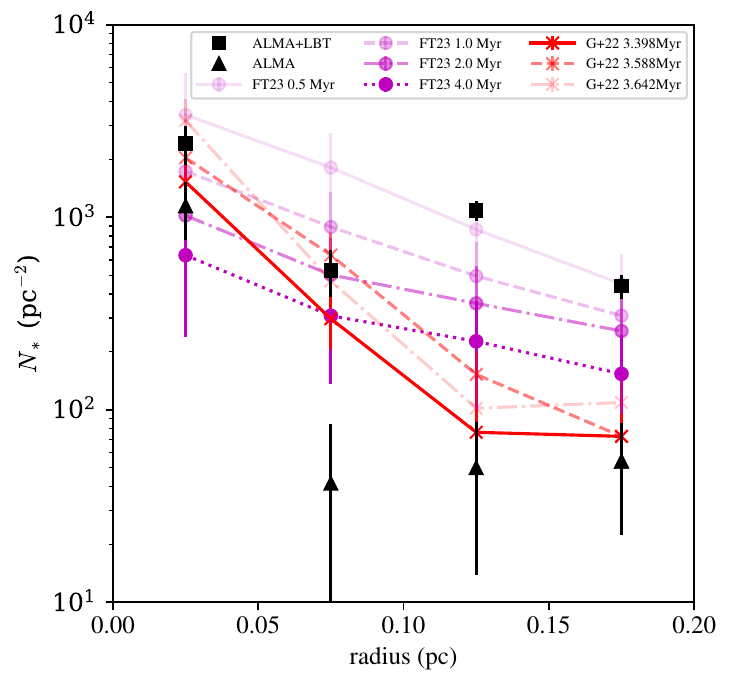}
        
\caption{\label{fig:stellar_density} 
Massive protostar companion YSO surface number density analysis.
\textit{Left:} K-band LUCI-1 SOUL AO image of the central region of AFGL~5180. NIR sources are shown as blue dots and ALMA sources as red dots. ALMA FOVs for Bands 6 and 7 are shown as white solid and dotted circles, respectively. Annuli in which stellar surface densities have been sampled are shown as red circles. 
\textit{Right:} Radial surface number densities derived from the NIR + ALMA (black squares) and ALMA-only (black triangles). Poisson uncertainties are indicated by the error bars.
Data from the STARFORGE simulation \citep{grudic22} is shown by the red lines and symbols, as labeled (see text). 
Turbulent clump model of protocluster formation from \citet{farias23} is shown by the purple lines and symbols, as labeled (see text).
}
\end{figure*}

A major open question in massive star formation concerns the importance of interactions between high-mass protostars and their lower-mass counterparts during the formation process. As discussed in \S\ref{sec:introduction}, there are two main massive star formation models under active investigation: Core Accretion \citep[e.g.,][]{mckee03} and Competitive Accretion \citep[e.g.,][]{bonnell01,wang10,grudic22}. 
A major difference between these models is that Competitive Accretion predicts that high-mass stars necessarily form surrounded by a concentrated cluster of lower-mass YSOs. On the other hand, Core Accretion models do not require such a condition, i.e., massive star formation can occur in both isolated and clustered environments. Therefore, observational studies such as this one can place key constraints on massive star formation theories, through direct measurement of the distribution of YSOs around high-mass protostars.

Here we investigate the distribution of the surface number density of YSOs (including protostars) around the massive protostar S4. Recall that this source has an (inner) envelope mass of $\sim2\,M_\odot$ (see Table\,\ref{core_table}), and an SED fitting-derived current stellar mass of $11.3^{+6.1}_{-4.0}\,M_\odot$ and surrounding clump mass surface density of $\Sigma_\mathrm{cl}=0.182^{+0.249}_{-0.105}\:{\rm g\:cm}^{-2}$ (Telkamp et al. in prep.). 
To estimate the YSO surface number density we make use of the NIR K-band LUCI-1 SOUL AO data, chosen because of its high sensitivity and angular resolution. We also complement the NIR-identified sources with the cores identified in the ALMA Band 6 and Band 7 images (see \S\ref{sec:cores} and Table \ref{core_table}). We note that the mass sensitivity of the ALMA observations is about $M_{\rm mm}\sim 0.01\:M_\odot$.

For K-band source identification we use the algorithm \textit{DAOFIND} \citep{stetson87} implemented in the function \textit{DAOStarFinder} in photutils \citep{photutils}. The input parameters are 3 pixels for the \textit{fwhm} and $5\sigma$ for the \textit{threshold}, and the remaining parameters are kept as default. We mask the edges of the image to avoid false detections, and then perform a visual inspection to make sure there are no spurious detections. See Figure\,\ref{fig:stellar_density}\textit{a} for a depiction of the identified NIR and ALMA sources.

To determine the minimum mass of K-band AO detected YSOs, the image was flux calibrated following the standard method of using isolated 2MASS \citep{skrutskie06} sources in the AO FOV. The minimum K-band magnitude retrieved was about $17.5$\,mag. The estimated clump mass surface density of $0.18\:{\rm g\:cm}^{-2}$ corresponds to a visual extinction of $A_V\simeq 40\:$mag and a K-band extinction of $A_K\simeq 4.6\:$mag. However, the number of observed YSO candidates around the massive protostar is larger with respect to the outer regions. This suggests that the typical extinction to a YSO in the region is expected to be smaller. First, for YSOs clustered around the massive protostar, the average extinction will be about half of the total through the clump, because the massive protostar is typically assumed to be at the center of the clump. Thus, in this case, only half the material of the total clump mass surface density is along the line of sight from us to the massive protostar. Second, the impact of the protostellar outflows from the massive protostar is to clear away a significant part of the clump material. Thus we have considered a range of extinction values from $A_V\simeq 10$ to 20~mag.
For a fiducial value of $A_V= 15$\,mag, a source distance of 2\,kpc, and considering $\sim3$\,Myr isochrones from \citet{baraffe98}, 
we obtain a mass limit of $\sim0.1\,M_\odot$. If one assumed A$_V$ = 40 mag, then the K-band observations would be sensitive to $\sim0.7 M_{\odot}$.


The radial profile of YSO surface number density is shown in Fig.~\ref{fig:stellar_density}\textit{b}, evaluated in annuli of width 0.05~pc. The surface number density was derived by counting the number of YSO candidates within a given annulus. Note that most of the annuli have only partial coverage with the AO FOV, so their $N_*$ measurement is based only on the area that is covered by this FOV. 
Within the central 0.05~pc, $N_*\simeq 2\times 10^3\:{\rm pc}^{-2}$. It declines to about $4\times 10^2\:{\rm pc}^{-2}$ by a radius of about 0.2~pc.


We compare the observed $N_*$ radial profile to two different theoretical models.
First, the STARFORGE simulation of \citet{grudic22}, which is an example of a model in which a massive protostar forms via Competitive Accretion. The simulation models the collapse of a giant molecular cloud (GMC) with a mass of $20,000\,M_\odot$ and an initial size of 10~pc into a young stellar cluster with a variety of feedback physics, reaching an overall SFE of $8.5\%$ at 9 Myr. Individual protostars are identified as sink particles that accrete material and grow in mass. 
For this analysis, we retrieved the simulation snapshots 695, 734, and 745 (from the StarforgeFullPhysics repository) with an age range of $3.398-3.642\,$Myr and a mass range of $8-16\,M_\odot$ for the most massive sink particle, which is a representative value of the current stellar mass of S4 via SED fitting. At this age, the SFE of the simulation is about 2\%, corresponding to a total stellar mass of $\sim400\,M_\odot$.
Given our NIR sensitivity to YSOs, we only consider sink particles with masses $\geq0.1\:M_\odot$. The radial profile of the surface number density of these sink particles around the most massive protostar is shown in Fig.~\ref{fig:stellar_density}\textit{b}. We have also verified that these density profiles are typical of the broader STARFORGE parameter study of different cloud properties, although there is some scatter of about 0.3\,dex \citep{guszejnov22}.







Second, we measured the $N_*(r)$ profile around the most massive star (constrained to be in the range 10 to $20\:M_\odot$) in the protocluster N-body simulations of \citet{farias23}. These involve gradual formation of stars, drawn from a \citet{kroupa01} IMF, from a parental clump described by the Turbulent Clump Model of \citet{mckee03}. Note, these models involve initial locations of star formation that follow a power law distribution in 3D of $r^{-1.5}$, but no primordial mass segregation. However, subsequent mass segregation can occur due to dynamical interactions in the protocluster. We only considered models with fiducial star formation efficiencies per free-fall time of $\epsilon_{\rm ff}=0.03$. Again, we only counted stars with masses $\geq0.1\:M_\odot$. We found that the models that best fit the observed data are the those with initial clump masses of $M_{\rm cl}=300\,M_\odot$ and initial surrounding cloud mass surface density of $\Sigma_{\rm cloud} = 0.1\:{\rm g\:cm}^{-2}$. We examined 200 simulations with these parameters, from which about a hundred of them contained a most massive star with a mass between 10 and 20~$M_\odot$ that were then used in the analysis. As shown in Fig.~~\ref{fig:stellar_density}\textit{b}, reasonable matches to the observed profile are obtained at $t=0.5-2\:$Myr. At ages of 0.5, 1 and 2 Myr these simulations have a SFE of 7, 13 and 27\%, respectively, representing total stellar masses of about 20, 40 and~80 $M_\odot$. 




Both sets of theoretical models have values of $N_*\sim 10^3\:{\rm pc}^{-2}$ on scales within $r\sim 0.1\:$pc. In the case of the STARFORGE simulation, it is interesting that this level is achieved without fine tuning. However, we note that the radial profile of $N_*$ in these models is somewhat steeper than that of the observational data, i.e., in the region from 0.1 to 0.2~pc the simulation predicts about $10\times$ lower values of $N_*$ than derived from the NIR imaging.

In the case of the Turbulent Clump / Core Accretion model, which is a more general model, the overall normalization is free to be adjusted by choosing different values of $M_{\rm cl}$ and $\Sigma_{\rm cl}$. For a given case, the number density profile initially increases in its amplitude (during the first $\sim 0.5\:$Myr) as more stars are formed. However, it later decreases as the cluster expands. Such models have profiles that are relatively shallow and so are a better match to the data out to 0.2~pc, but a worse match to the profile within 0.1~pc.

The following caveats should be considered. We have made quite simple estimates of the level of completeness of the observed YSOs. The source counts are mostly dominated by the K-band identified YSOs and these are affected by the dust extinction in the region and therefore the A$_V$ value, which clearly shows spatial variation. Our analysis has also assumed that the level of foreground field star contamination in the region is small compared to the outer value of $N_*\sim 4\times 10^2\:{\rm pc}^{-2}$. Any significant foreground contamination would lead to a steepening of the radial profile. Finally, we note that there are significant Poisson uncertainties in the derived $N_*$ measurements given the number of sources present.




In summary, we see that the observed radial profile of $N_*$ is similar to that predicted by both a Competitive Accretion type model \citep{grudic22} and a model in which stars form independently within a given turbulent clump, which is one possible scenario for a Core Accretion model \citep{farias23}. We notice that the Competitive Accretion model has a moderately steeper radial profile of $N_*$ than the observed region when considered out to 0.2~pc, but is a good match in the inner 0.1~pc. The flatter overall profile of the data out to 0.2~pc is a better match to the Turbulent Clump / Core Accretion model. Other points to note are that the Competitive Accretion model requires about 3~Myr to have formed its massive protostar, while the Turbulent Clump / Core Accretion model can achieve a reasonable match in a shorter period of time. Finally, we note that a significant part of the steeper inner $N_*$ profile is due to the presence of a relatively significant number of ALMA-identified sources within 0.05~pc. However, the nature of these sources is somewhat uncertain. We return to this point in \S\ref{sec:stellar_densities}.

\section{Discussion}\label{sec:Discussion}

\subsection{Implications for Outflow Launching}

In \S\ref{sec:Results}, we have shown that the AFGL 5180 complex contains a considerable number of jet knots revealed by NIR imaging, particularly in the form of the shock tracers H$_2$ and [FeII], and which are plausibly being powered by several cores identified from ALMA Bands 6 and 7 data, S1-14. Despite the richness and detail of this data set, there are a number of open questions which arise from these results.
    
In particular, apart from the main E-W outflow features that appear to be driven by the primary massive protostar S4, it has been difficult to associate the other outflow knots with particular protostellar sources.
This is primarily due to the close proximity of many of the sources, which precludes the ability to determine precisely the specific driving source of each knot. Still, in many cases we can nonetheless differentiate between knots generated by the the two main sites of star formation, i.e., AFGL 5180 M and S, with a few knots attributable to other cores (Table \ref{Knot_table}). In order to better characterize each core and its respective outflows with greater clarity, proper motion studies are needed.
    
Due to the wide distribution of knots, which seem to emanate in a variety of directions from the central complex, it is worth considering the possibility that this region is characterized by an 'explosive outflow'. A handful of examples of such outflows have been proposed \citep[see, e.g.,][]{zapata09, zapata17, zapata20, bally20,bally22}, with the driving mechanism thought to involve protostellar collisions that lead to impulsive injection of energy into surrounding gas, which then expands in a quasi-isotropic manner. However, with at least fifteen mm-continuum cores identified with ALMA (\S\ref{sec:cores}) and with a large fraction of the outflow knots associated with the main E-W flow, we conclude that a scenario of multiple outflows driven by multiple sources is more plausible.

\subsection{Implications for Massive Star Formation}
\label{sec:stellar_densities}
    
The combined analysis of NIR and mm wavelength data of AFGL~5180 indicate the presence of many lower-mass YSOs, including protostars, around the main massive protostar. The observations are sensitive to masses down to $\sim 0.1\:M_\odot$. The overall projected number density of these sources is $N_*\sim 10^3\:{\rm pc}^{-2}$ on scales of $\sim 0.1\:$pc. There is an enhanced level of $N_*$ by factors of a few on scales of $\lesssim 0.05\:$pc, with a large part of this driven by the presence of ALMA-detected mm sources.

While such a clustering of sources is a prediction of Competitive Accretion models, we have found that simple protocluster models in which stars form independently of each other from a turbulent clump can also match the observed $N_*$ radial profile around the massive protostar. Furthermore, examining the masses of the ALMA-detected mm sources surrounding the massive protostar, S4, we see that apart from S9 (MM2), they are all of very low mass compared to the primary massive protostar. In addition, there is limited evidence that these mm sources are themselves protostars driving outflows, rather than simply being overdensities within the primary protostellar core. Deep cm radio continuum observations will be helpful in establishing the protostellar nature of these mm sources. Until their nature is more firmly established, it is difficult to draw firm conclusions about whether Core Accretion or Competitive Accretion is occurring in this region.

As assessment of YSO density around a massive protostar has been made in a number of other regions. \citet{costasilva22} examined NIR spectro-imaging data of the region around IRAS 18264-1152 with central $m_*\sim 4$ to $8\:M_\odot$ from SED fitting and identified 14 YSO candidates in a 0.0256~$\rm pc^2$ region (i.e., on scales of $r\sim 0.08\:$pc) to yield an estimate of $N_* \sim 500\:{\rm pc}^{-2}$. This level of YSO surface number density is quite similar to that we have found in AFGL~5180. On the other hand, \citet{law22} presented evidence for a massive protostar G28.20-0.05 with $m_*\sim 40\:M_\odot$ that has very limited evidence for any protostellar companions in mm emission on scales from $r=0.1$ to 0.4~pc. Extending the analysis methods of our study in a uniform manner to a larger number of massive protostars is needed to obtain the data to allow more general conclusions to be made about the connection of massive star formation to star cluster formation.

\subsection{Placing AFGL 5180 into Context: Triggered Star Formation?}

\begin{figure*}[t]
\centering
\includegraphics[width=1\textwidth,height=1\textheight,keepaspectratio]{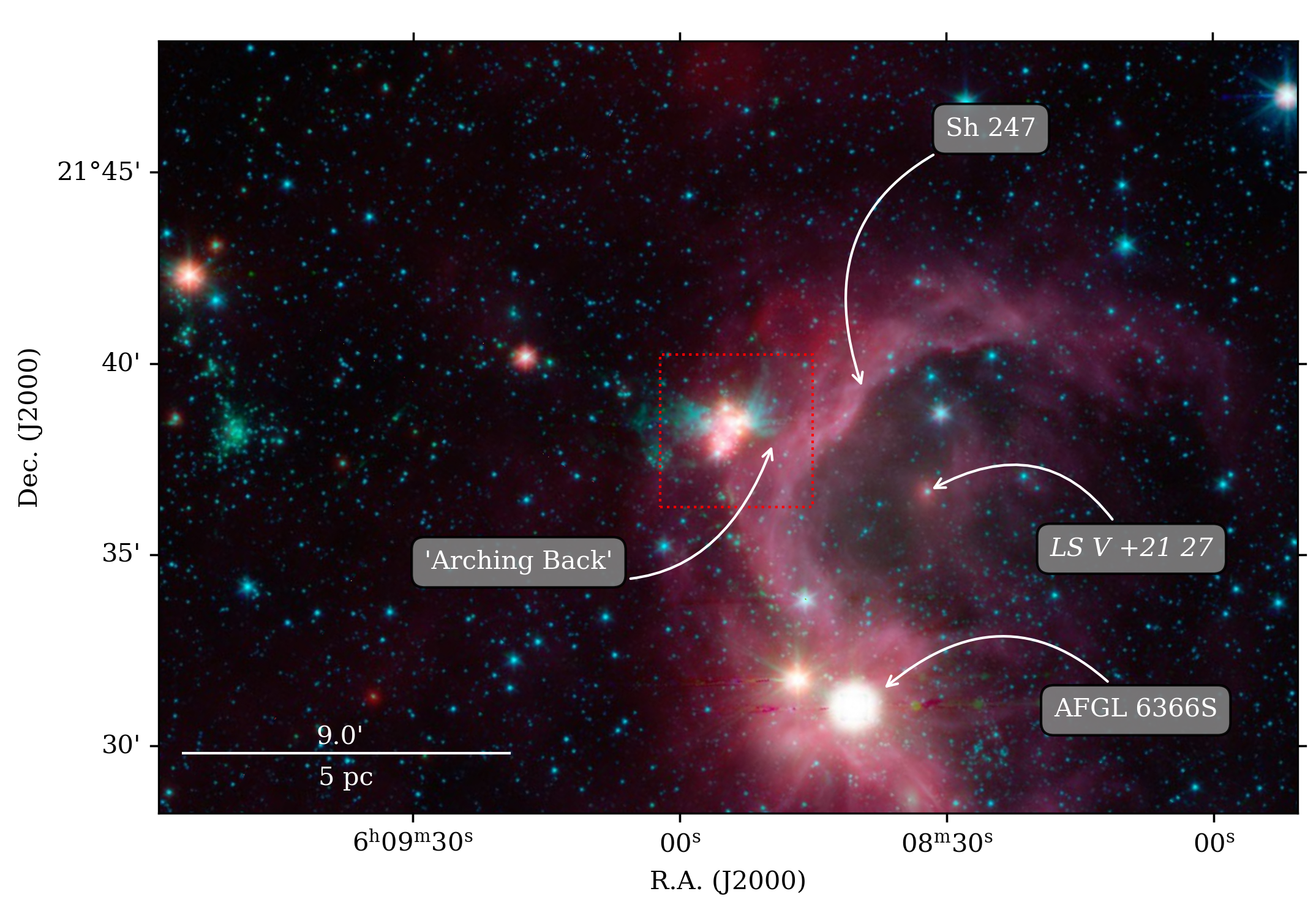}
\caption{\label{glimpse} Spitzer GLIMPSE image of the wider region surrounding AFGL 5180, taken from the Aladin Sky Atlas. 10 $\rm \mu m$ data from the Wide-field Infrared Survey Explorer (WISE) has been used in place of missing IRAC 8.0 $\rm \mu m$ data in red, Spitzer IRAC 4.5 $\rm \mu m$ is shown in green, and IRAC 3.6 $\rm \mu m$ in blue. The FOV of the LBT seeing-limited observations (see Fig. \ref{RGB_LBT_limited}) is shown as the red box.}
\end{figure*}

Figure \ref{glimpse} shows a wide-field RGB of the larger Gemini OB1 association, in which the AFGL 5180 complex is located \citep{carpenter95a, carpenter95b,zucker20}; red represents 10 $\rm \mu m$ data from the Wide-field Infrared Survey Explorer (WISE) mission, green represents Spitzer IRAC 4.5 $\rm \mu m$ data, and blue represents Spitzer IRAC 3.6 $\rm \mu m$ data. A consensus is yet to be reached about whether star formation in AFGL 5180 and the wider Gemini OB1 association is the result of triggered star formation from swept-up gas shells \citep{carpenter95a,carpenter95b}, or whether it is primarily the result of a cloud-cloud collision, as suggested by \citet{vasyunina10,shimoikura13,maity23} on the basis of the morphology of the blueshifted and redshifted gas flows in the region. 
        
However, inspection of the wider region around AFGL~5180 presented in Fig.~\ref{glimpse} indicates that it appears to be spatially coincident with the H\RomanNumeralCaps{2} region Sh 247 \citep[in red in Fig.~\ref{glimpse};][]{sharpless59}, a pressurized region driven by the energetic photons from the massive O9.5 type star \textit{LS V +21 27} \citep{romanlopes19}. Also present on this dust ridge is AFGL~6366S, another massive star-forming complex associated with a Class II 6.7 GHz methanol maser \citep[][]{maity23}. This morphology would seem to suggest that star formation has been triggered in AFGL 5180 and AFGL 6366S via an H\RomanNumeralCaps{2} region from a massive star, a scenario which is supported by studies of other massive star-forming regions \citep[e.g.,][]{fukui18,cosentino20,fukui19}. In Fig.~\ref{glimpse} an ``arching back'' morphology at the interface between AFGL~5180 and the H\RomanNumeralCaps{2} region can even be seen, which appears to be due to feedback caused by the prominent protostellar outflow pushing back against the expansion of the H\RomanNumeralCaps{2} region. Indeed, asymmetric bubble expansion is typically considered a signature of interaction between the expanding shell (mostly supernova remnants) and the surrounding material \citep[see, e.g.,][for a recent review]{slane16}. This would provide evidence for the spatial coincidence between AFGL~5180 and Sh~247, which is key to determining if the latter could have triggered star formation in the former. 

AFGL~6366S, on the other hand, appears to be more evolved, lacking any outflow structure and instead displaying what appears to be an ionizing effect of its own, instanced by the more extended and bright red emission around it. This would seem to indicate that AFGL 6366S is a more evolved region which has begun to ionize its surrounding environment and generate its own H\RomanNumeralCaps{2} region. Under this interpretation, in this region we are witnessing multi-generational massive star formation, with the feedback generated by the massive star \textit{LS V +21 27} and its H\RomanNumeralCaps{2} region Sh~247 potentially triggering nascent star formation in AFGL~5180 and AFGL~6366S, the latter of which may have begun generating ionizing photons from massive stars and its own H\RomanNumeralCaps{2} region in turn. For these reasons, we favor the latter interpretation of triggered star formation in the particular region of AFGL~5180 over the cloud-cloud collision theory, although it remains a possibility that a cloud collision could have initiated the global star formation process in the region.
        
Also noticeable in Fig. \ref{glimpse} is the prominent extended bipolar outflow in green (IRAC 4.5 $\rm \mu m$ emission); this bipolar outflow corresponds neatly with that captured in this study (see Figs.~\ref{RGB_LBT_limited} and \ref{RGB_HST}). This structure represents a clear example of an ``Extended Green Object (EGO)'', also known as ``green fuzzy'', that, to our knowledge, has not been previously identified in catalogs of EGOs from the GLIMPSE survey, likely because of this region's distance from the Galactic Center, where most surveys of EGOs have focused \citep{cyganowski08,chen13}. The corroboration of the IRAC 4.5 $\rm \mu m$ ``green'' emission, which is known to trace shocked H$_2$ emission due to containing H$_2$ emission lines, most prominently the $\textit{v} = 0-0 S(9)$ line \citep{cyganowski08}, with the shocked H$_2$ emission seen in the 2.12 $\rm \mu m$ LBT data, confirms the presence of prominent shocked H$_2$ emission tracing outflows in this region. This green emission has also been associated with CO (1–0) ro-vibrational lines that stretch from $4.3-5.2\,\mathrm{\mu m}$ in the HH211 system \citep[][and references therein]{ray23}, likely contributing some of the emission seen around AFGL~5180, in which a CO outflow has been detected \citep{snell88}. However, we believe that this green emission is at least partly associated with H$_2$ for the AFGL5180 outflows as mentioned above. Indeed, excess 4.5 $\rm \mu m$ emission is also seen overlapping with the southern bow-shock seen in the LBT data (knot 14) and previously identified in \citet{davis98} (see the bottom of the red square in Fig. \ref{glimpse}), corroborating the detection of shocked H$_2$ emission in that feature at multiple wavelengths as well. Nonetheless, follow-up observations with telescopes such as the James Webb Space Telescope (JWST) are required to discern the nature of these green fuzzy objects. Following the naming conventions for EGOs established by \citep{cyganowski08}, we name the EGO associated with AFGL 5180 EGO G188.95+0.92.
        
In addition to outflows, EGOs are also associated with methanol masers and massive star formation, in particular with massive stars still contained within their infalling envelopes \citep{cyganowski08}, all of which correspond with the star formation conditions of AFGL~5180. Thus, this finding of an EGO associated with the AFGL 5180 complex strengthens the association of EGOs with massive star formation, shocked H$_2$ and outflow emission, and Class II 6.7 GHz Methanol maser emission.

\section{Conclusions}\label{sec:Conclusions}

In this paper, we have presented high-resolution NIR data towards the massive star-forming region AFGL~5180, which reveals the presence of $\sim$40 outflow knots and a similar number of YSO candidates.
With additional analysis of archival ALMA Band 6 \& Band 7 data, we determine the locations of several protostellar YSO candidates.  

We have carried out an analysis to attempt to match outflow knots to driving sources. We conclude that a significant part of the outflow activity can be associated with a E-W bipolar outflow from the primary massive protostar in AFGL~5180, with many prominent knots in the east detected in both HST [FeII] and high-resolution LUCI-1 SOUL AO H$_2$ observations. The E-W structure is also clearly seen as an ``extended green object'' in Spitzer IRAC images, i.e., with enhanced 4.5~$\rm \mu m$ emission. The massive protostar, which we identify as the S4 mm source, is estimated to have a current protostellar mass of $\sim 11\:M_\odot$ from SED fitting and is associated with strong Class II 6.7 GHz methanol maser emission. In addition to the primary massive protostar, there are several other protostellar sources in the region, especially from AFGL 5180 S, that appear to be driving independent outflows in a variety of directions.


From our census of YSOs, we have derived the YSO surface number density, $N_*$, i.e., by combining the ALMA-detected mm sources with the NIR K-band detected sources and with an estimated sensitivity of the latter down to $\sim 0.1\:M_\odot$. From this analysis we find central values of $N_*\sim {\rm few}\times 10^3\:{\rm pc}^{-2}$ inside 0.1~pc and with a radially decreasing profile out to 0.2~pc. The overall projected number densities inside 0.1~pc are consistent with models of both Competitive Accretion and Core Accretion (in the latter case with random sampling of cores/stars from a turbulent clump with an assumed power law density profile). The shape of the observed $N_*(r)$ profile is good match to the Competitive Accretion model inside 0.1~pc, but is shallower on larger scales, where it is better match to the Core Accretion from Turbulent Clump model. Additional follow-up observations are needed to confirm the protostellar nature of the ALMA-detected sources in the inner 0.05~pc to better test predictions of these models.

Analysis of the larger cloud structure reveals that AFGL 5180 sits on a dust ridge powered by the massive O9.5 type star \textit{LS V +21}, consistent with pressure-triggered star formation the region. However, indications have been also put forward which rather support a cloud-cloud collision as the primary driver of star formation in the wider region.

\begin{acknowledgements}
    The authors thank the anonymous referee for their insightful comments which helped to improve the quality of the manuscript. The authors would like to acknowledge fruitful discussion with John Bally. S.C. acknowledges support from the Virginia Space Grant Consortium (VSGC) Undergraduate Research Scholarship Program. R.F. acknowledges support from the grants Juan de la Cierva FJC2021-046802-I, PID2020-114461GB-I00 and CEX2021-001131-S funded by MCIN/AEI/ 10.13039/501100011033 and by ``European Union NextGenerationEU/PRTR''. J.C.T. acknowledges support from ERC grant MSTAR, VR grant 2017-04522, and NSF grant 1910675. A.C.G. acknowledges support from PRIN-MUR 2022 20228JPA3A “The path to star and planet formation in the JWST era (PATH)” and by INAF-GoG 2022 “NIR-dark Accretion Outbursts in Massive Young stellar objects (NAOMY)”. J.P.F. acknowledges support from NASA grant 80NSSC20K0507 and NSF Career award 1748571. G.C. acknowledges support from the Swedish Research Council (VR Grant; Project: 2021-05589).
\end{acknowledgements}

\bibliographystyle{aa}
\bibliography{masterbib.bib}

\begin{appendix}
    
    \section{Knot Feature Significance Levels}\label{sec:Knot_significance}
        \begin{figure*}[!htb]
            \centering
            \includegraphics[width=0.40\textwidth]{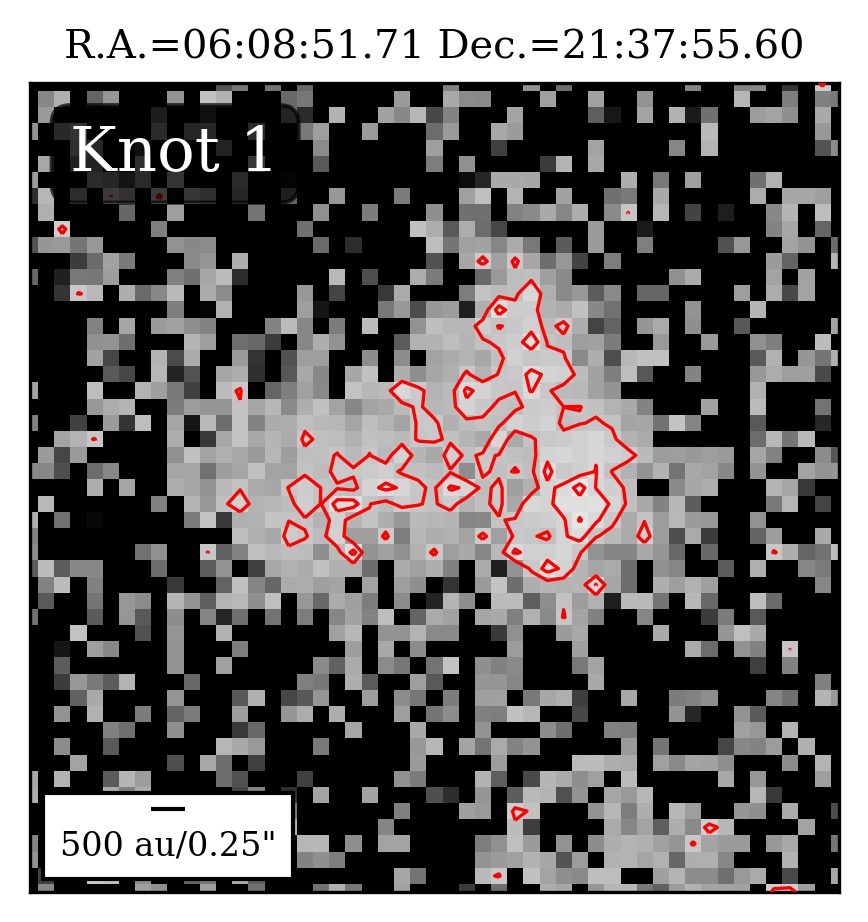}
            \includegraphics[width=0.40\textwidth]{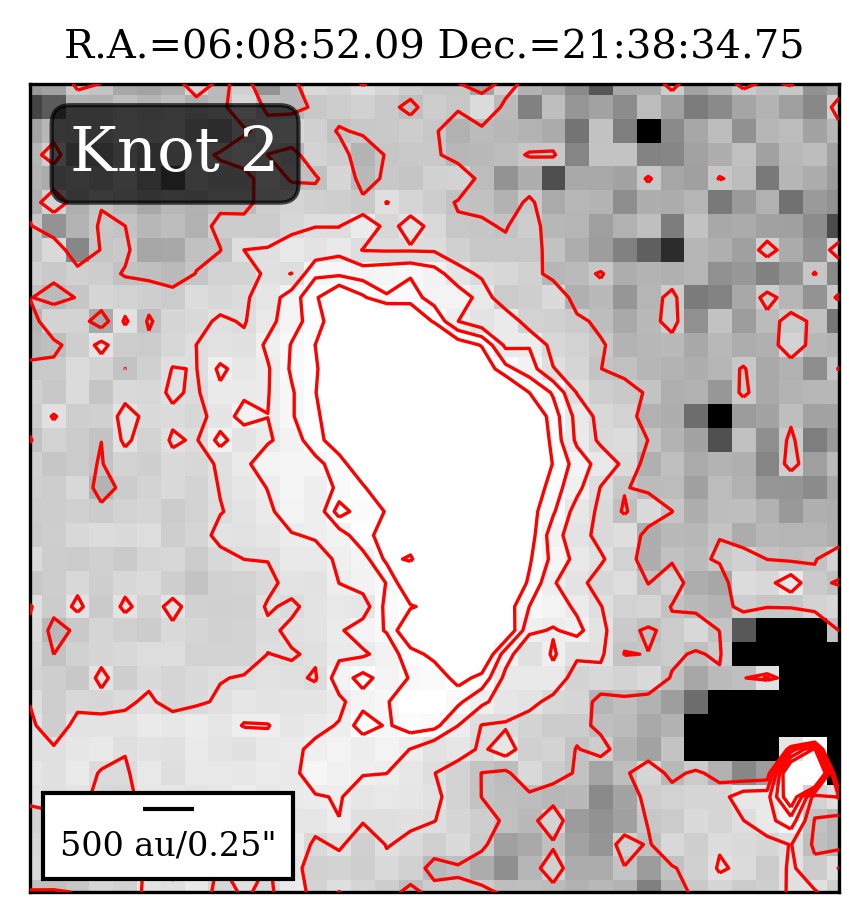}
            \includegraphics[width=0.40\textwidth]{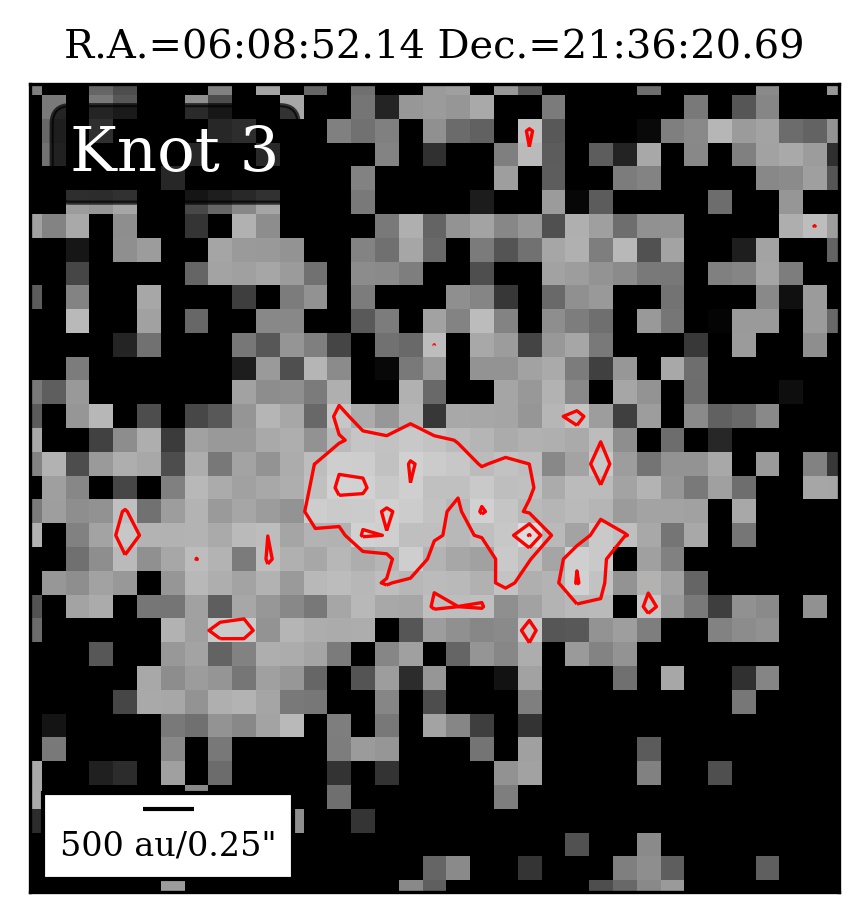}
            \includegraphics[width=0.40\textwidth]{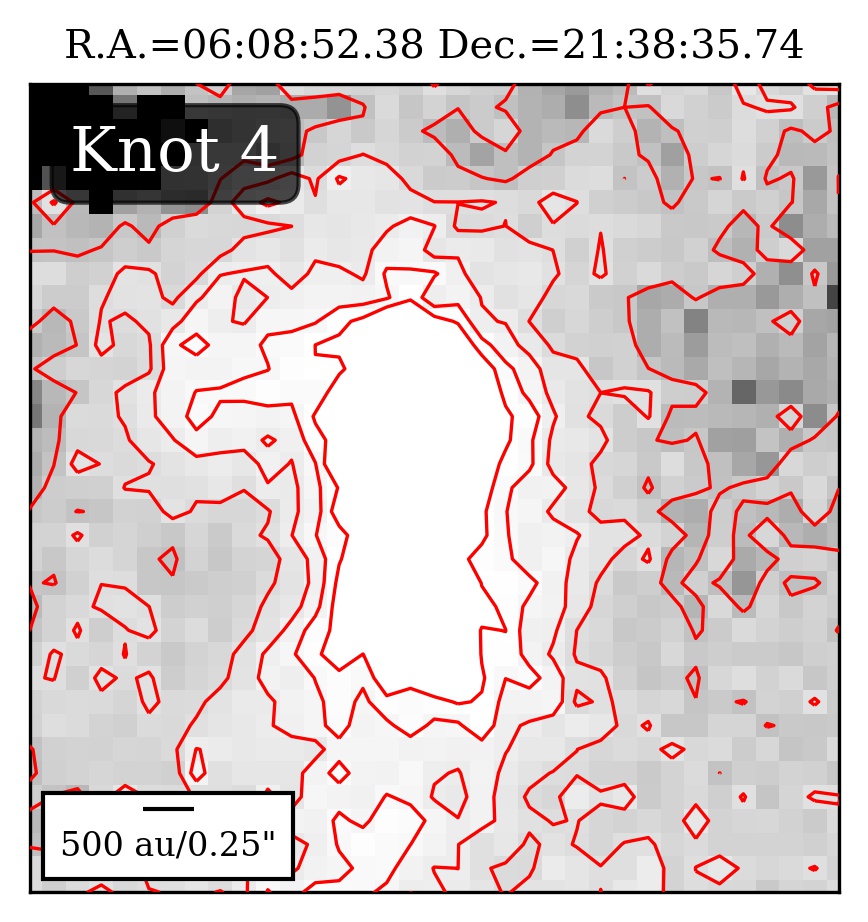}
            \includegraphics[width=0.40\textwidth]{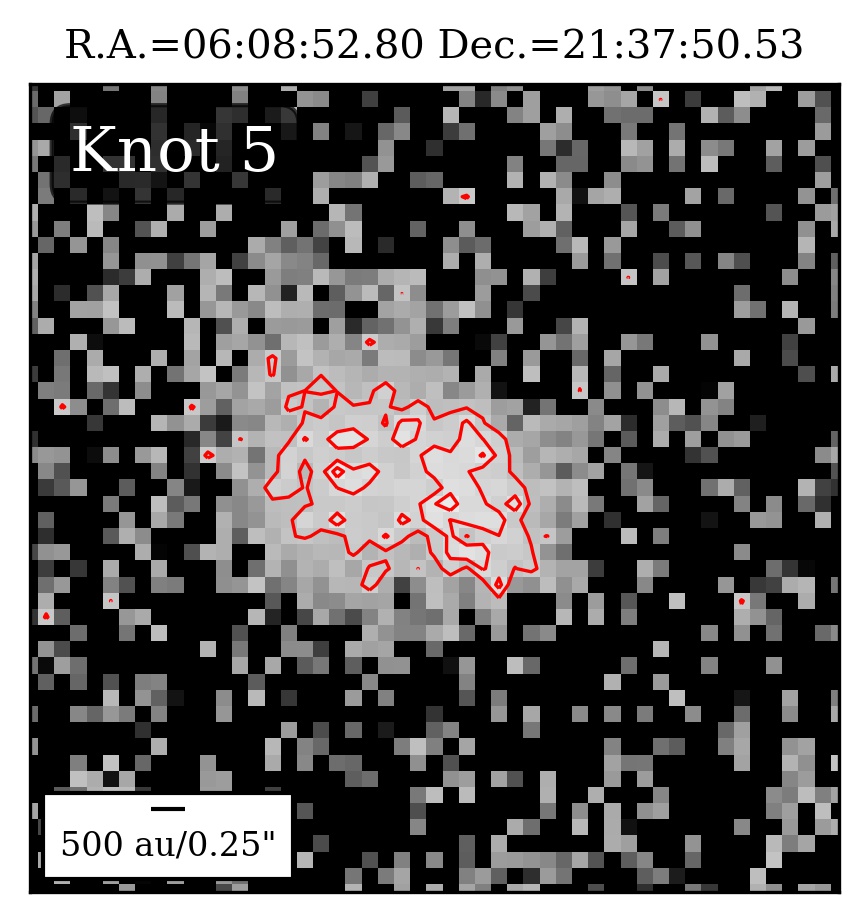}
            \includegraphics[width=0.40\textwidth]{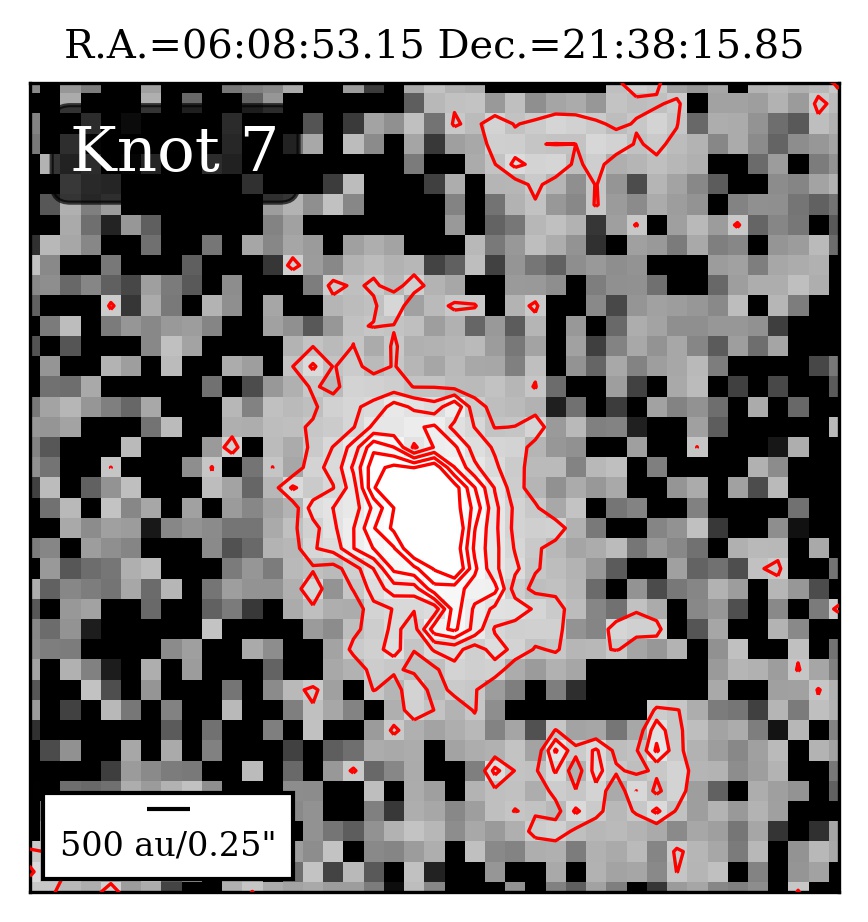}
            \caption{\label{seeing-limited knots}Significance level contour maps of all knot features identified in the LBT seeing-limited continuum-subtracted H$_2$ image and compiled in Table \ref{Knot_table}. The contour levels shown represent 3 to 15$\mathrm{\sigma}$ in steps of 2$\sigma$ above the local background. The central coordinates of each knot determined by 2D Gaussian fitting are given on the top of each panel. A physical scalebar of 500 au is given in the bottom-left corner of each panel. North is up and east is to the left in all panels.}
        \end{figure*}
        \renewcommand{\thefigure}{A\arabic{figure}}
        \addtocounter{figure}{-1}
        \begin{figure*}[!htb]
            \centering
            \includegraphics[width=0.40\textwidth]{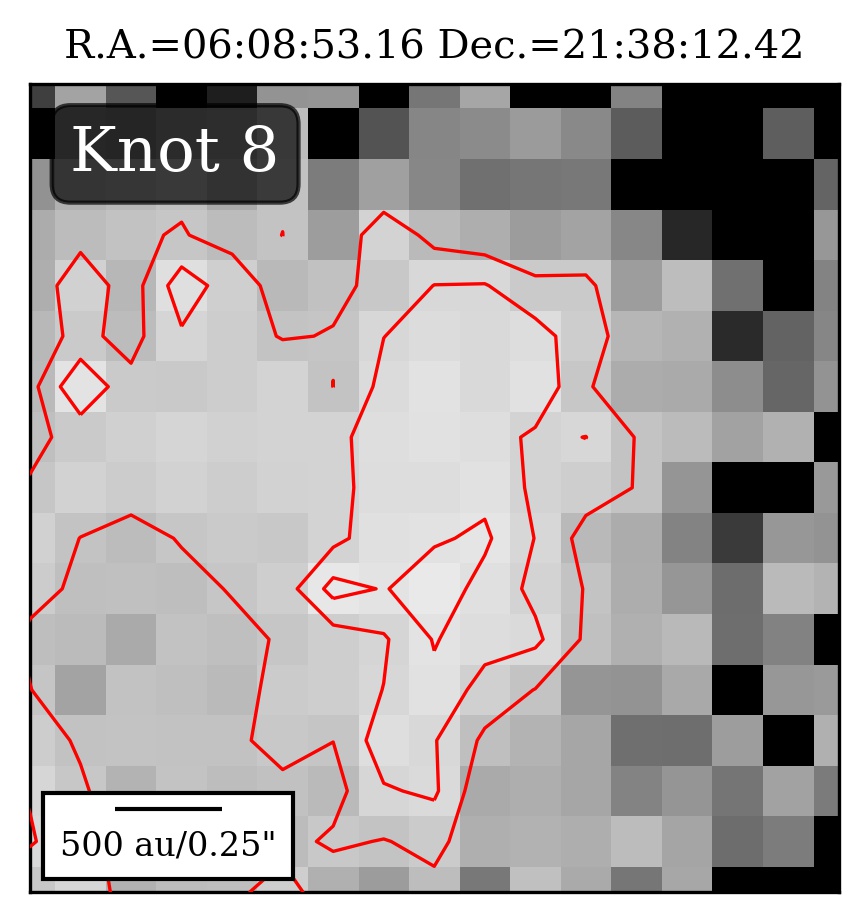}
            \includegraphics[width=0.40\textwidth]{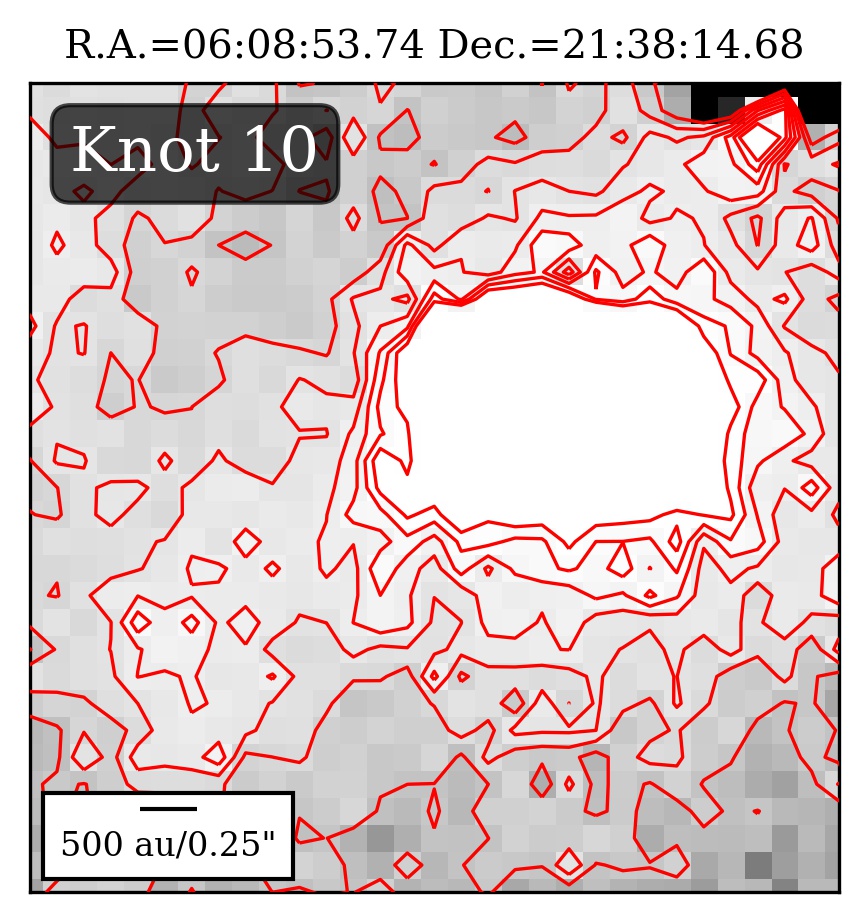}
            \includegraphics[width=0.40\textwidth]{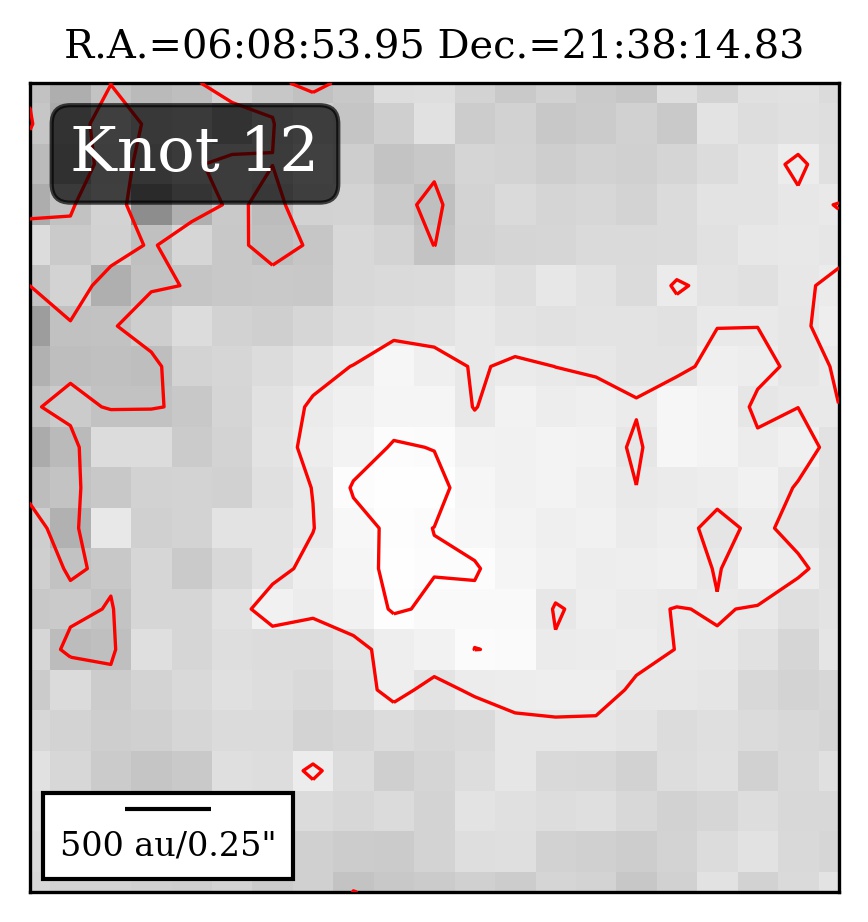}
            \includegraphics[width=0.40\textwidth]{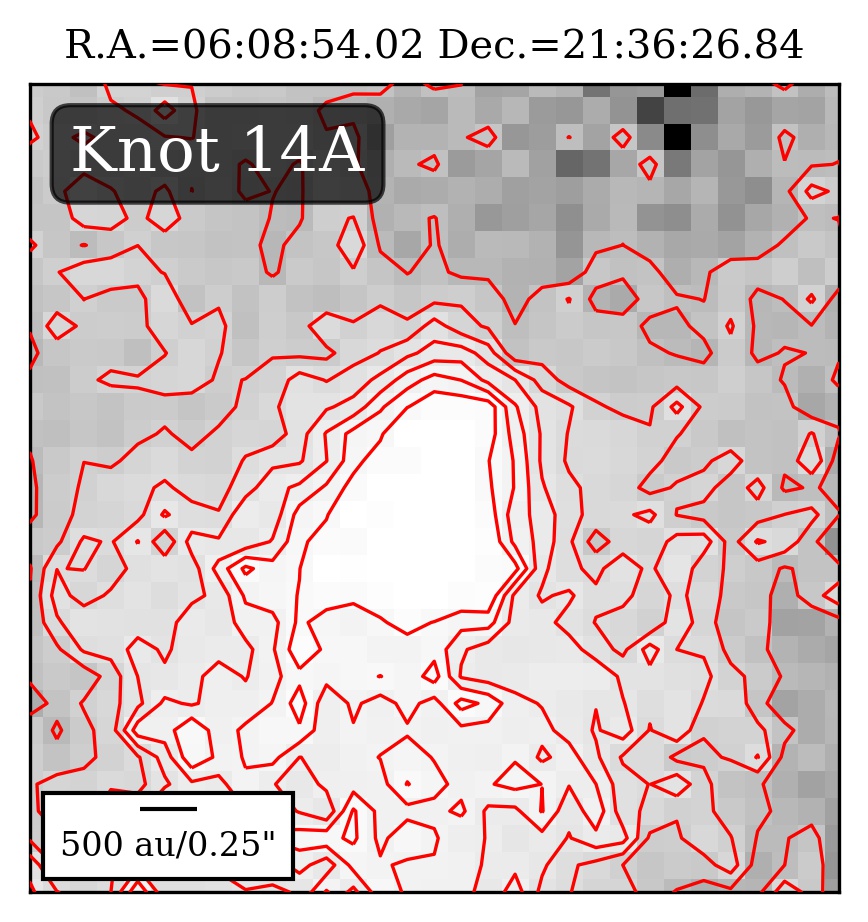}
            \includegraphics[width=0.40\textwidth]{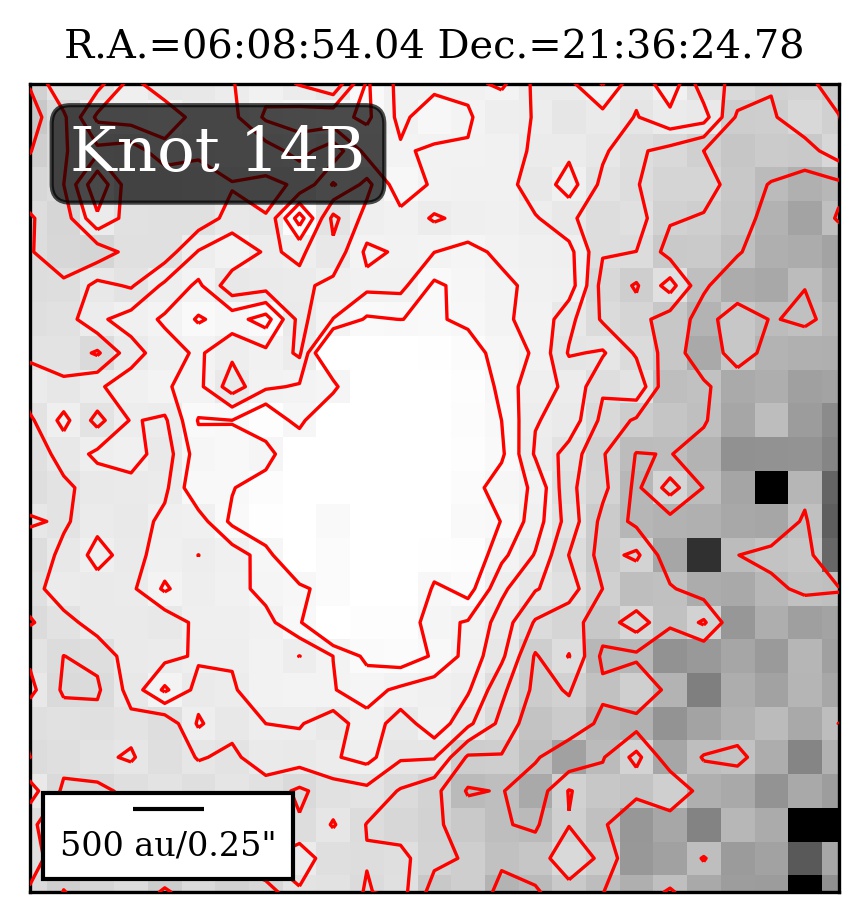}
            \includegraphics[width=0.40\textwidth]{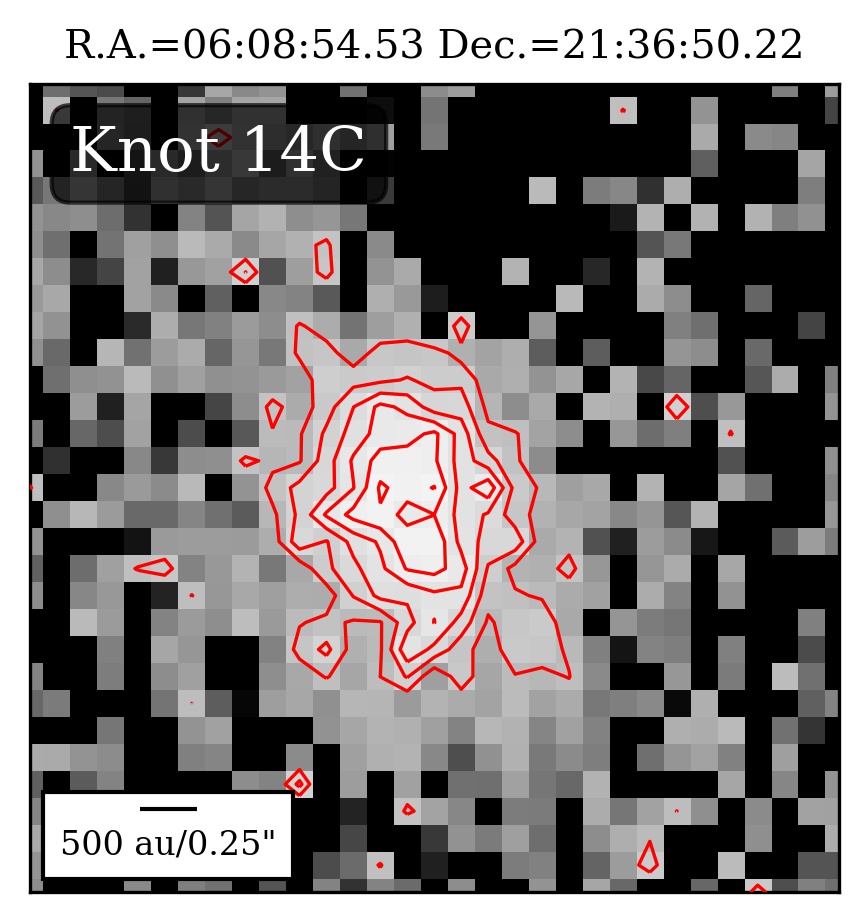}
            \caption{cont.}
        \end{figure*}
    
        \renewcommand{\thefigure}{A\arabic{figure}}
        \addtocounter{figure}{-1}
        \begin{figure*}[!htb]
            \centering
            \includegraphics[width=0.40\textwidth]{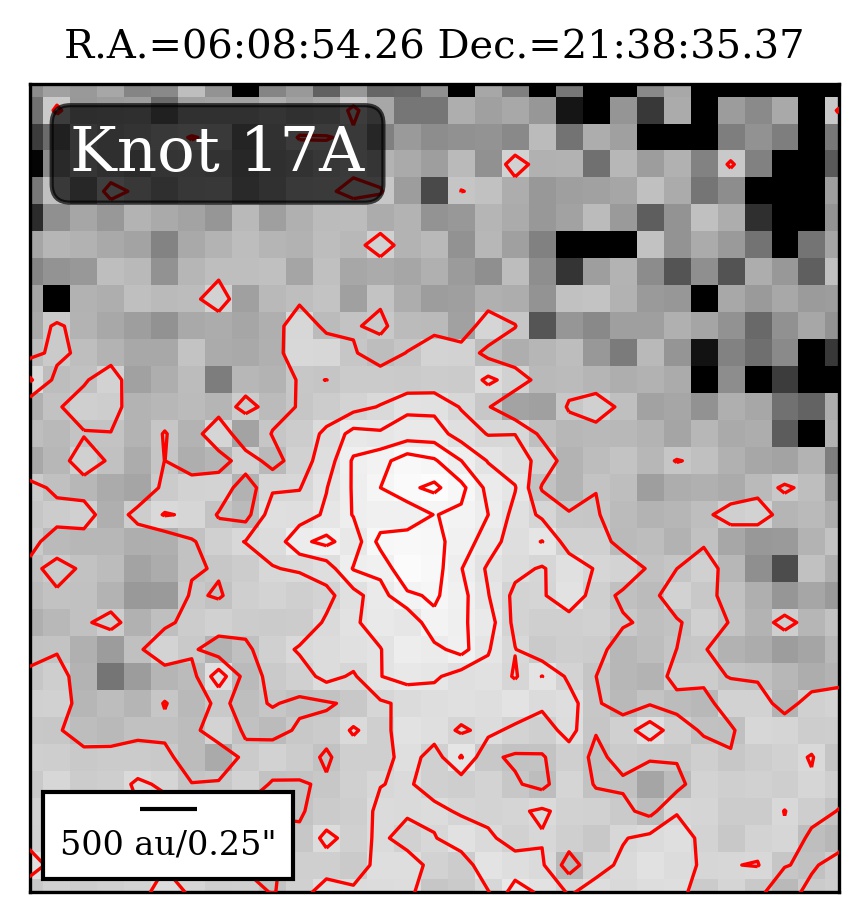}
            \includegraphics[width=0.40\textwidth]{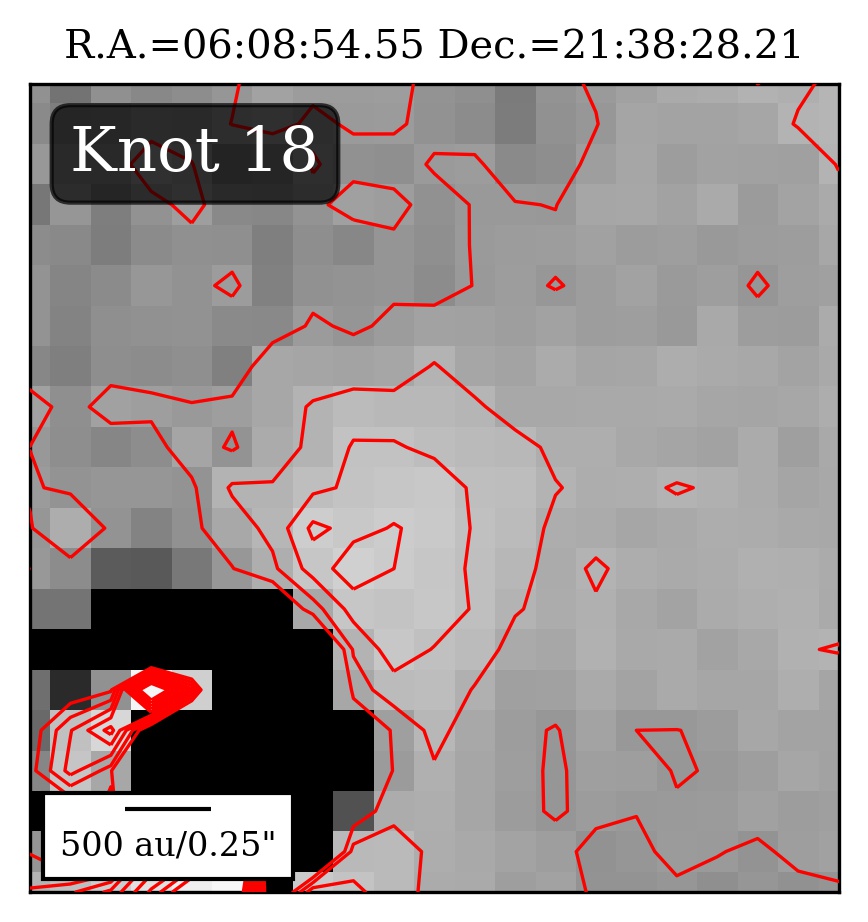}
            \includegraphics[width=0.40\textwidth]{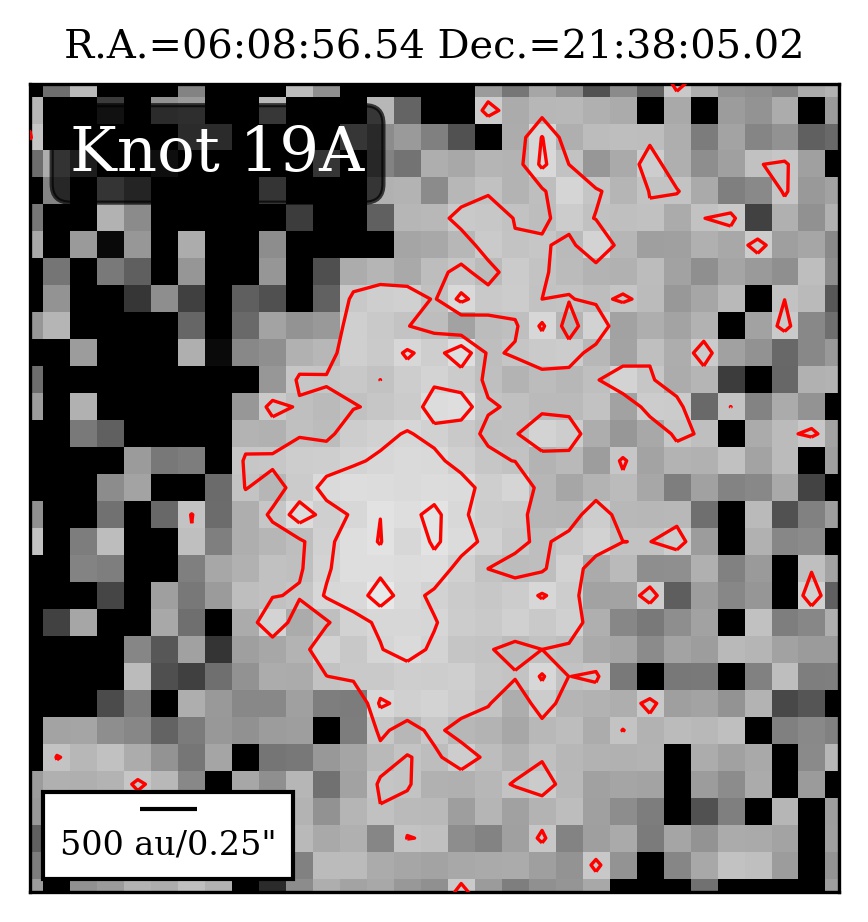}
            \includegraphics[width=0.40\textwidth]{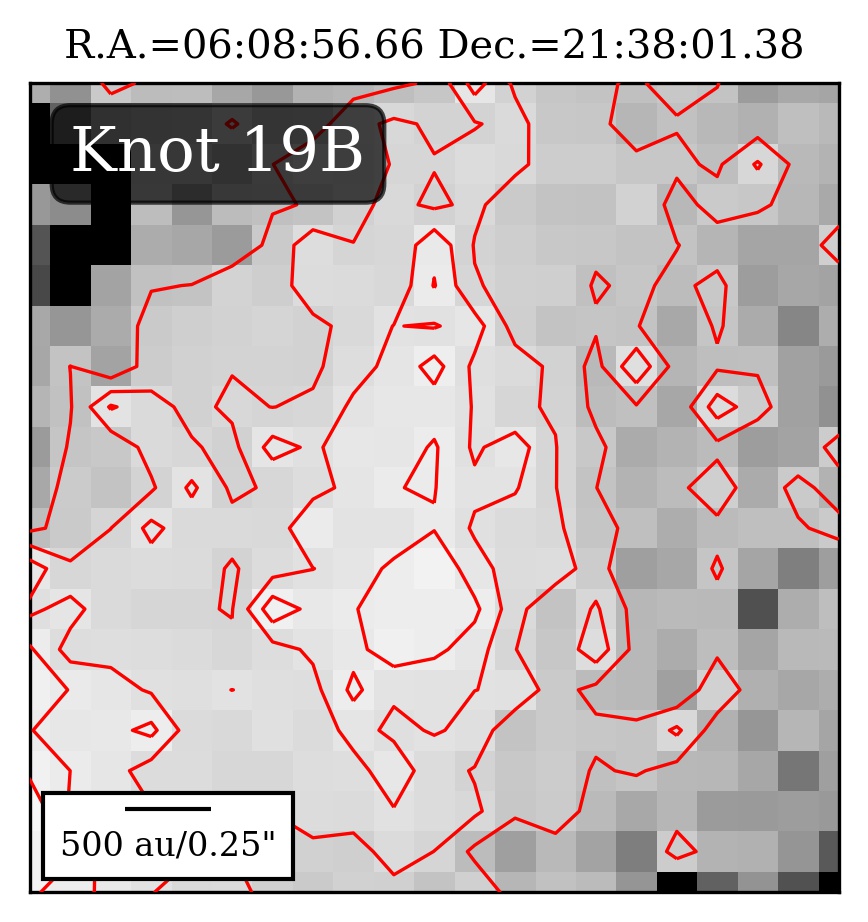}
            \includegraphics[width=0.40\textwidth]{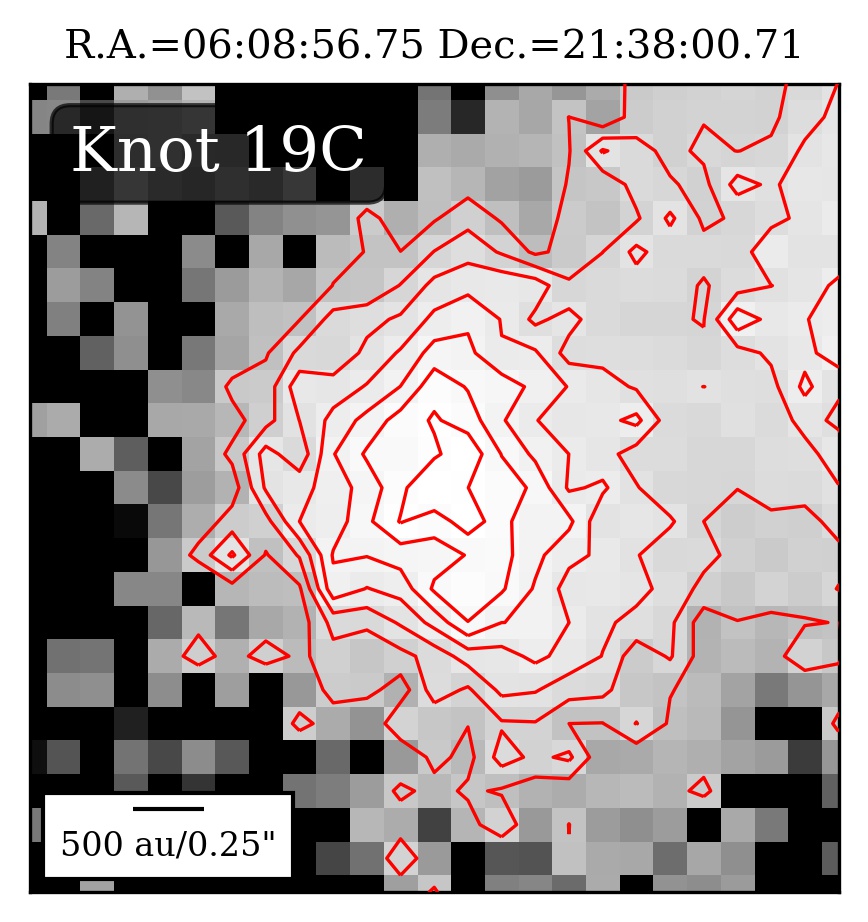}
            \includegraphics[width=0.40\textwidth]{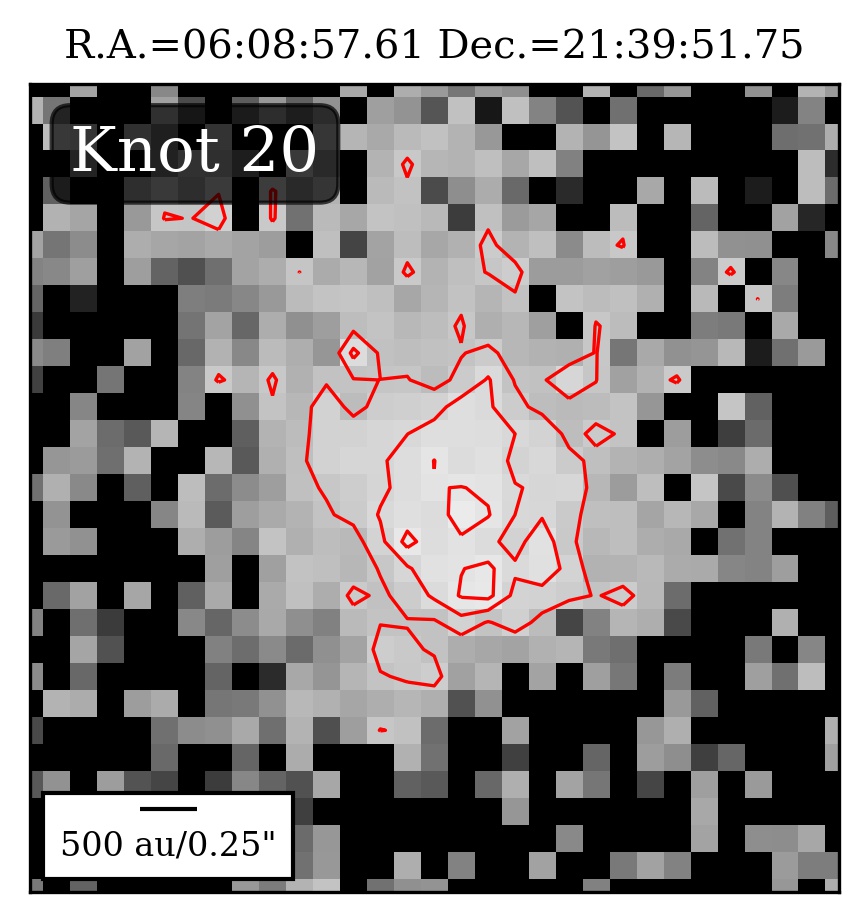}
            \caption{cont. Note that for knot 18, which is part of the S4 Jet as discussed in §\ref{sec:s4_jet}, the contour levels represent 5 to 45$\mathrm{\sigma}$ in steps of 5$\sigma$ above the local background instead, due to its brighter nature; also note the presence of a residual from a continuum-subtracted star in the lower left corner of the panel.}
        \end{figure*}
    
        \renewcommand{\thefigure}{A\arabic{figure}}
        \addtocounter{figure}{-1}
        \begin{figure*}[!htb]
            \centering
            \includegraphics[width=0.40\textwidth]{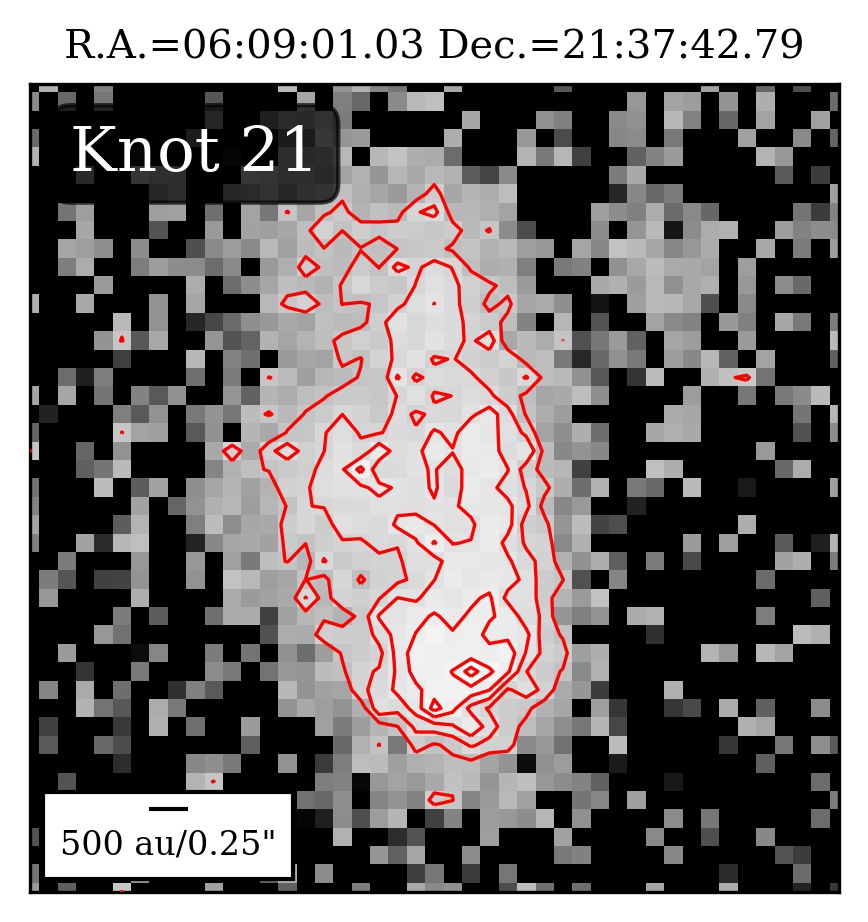}
            
            \caption{cont.}
        \end{figure*}
        
        \begin{figure*}[!htb]
                \centering
                \includegraphics[width=0.40\textwidth]{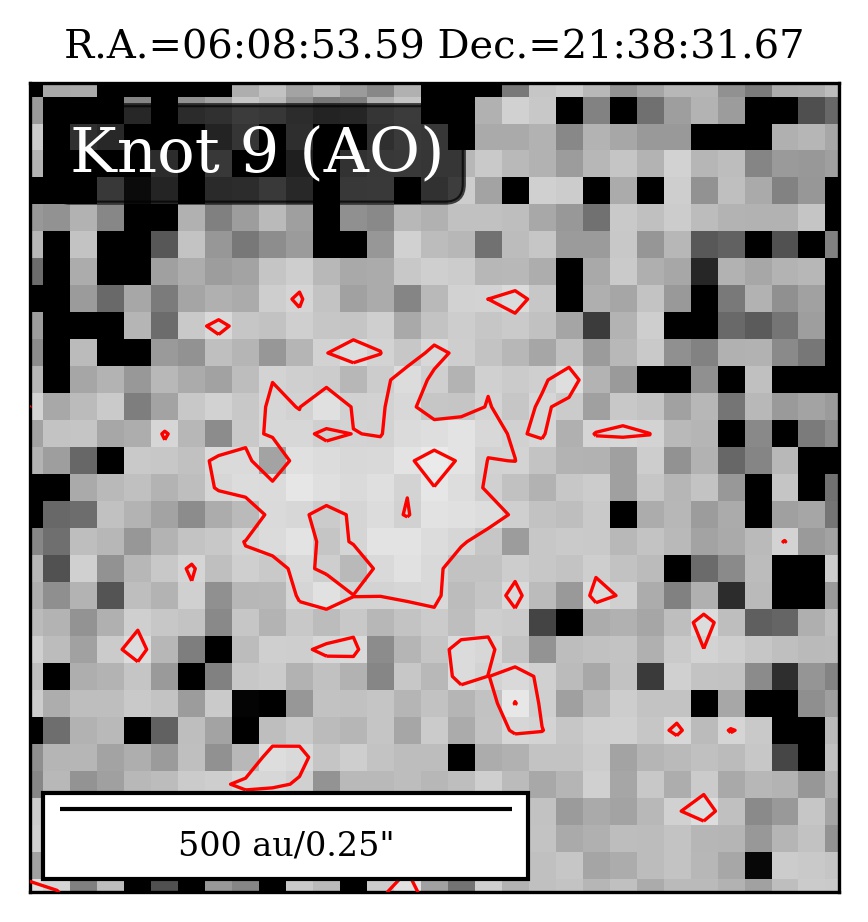}
                \includegraphics[width=0.40\textwidth]{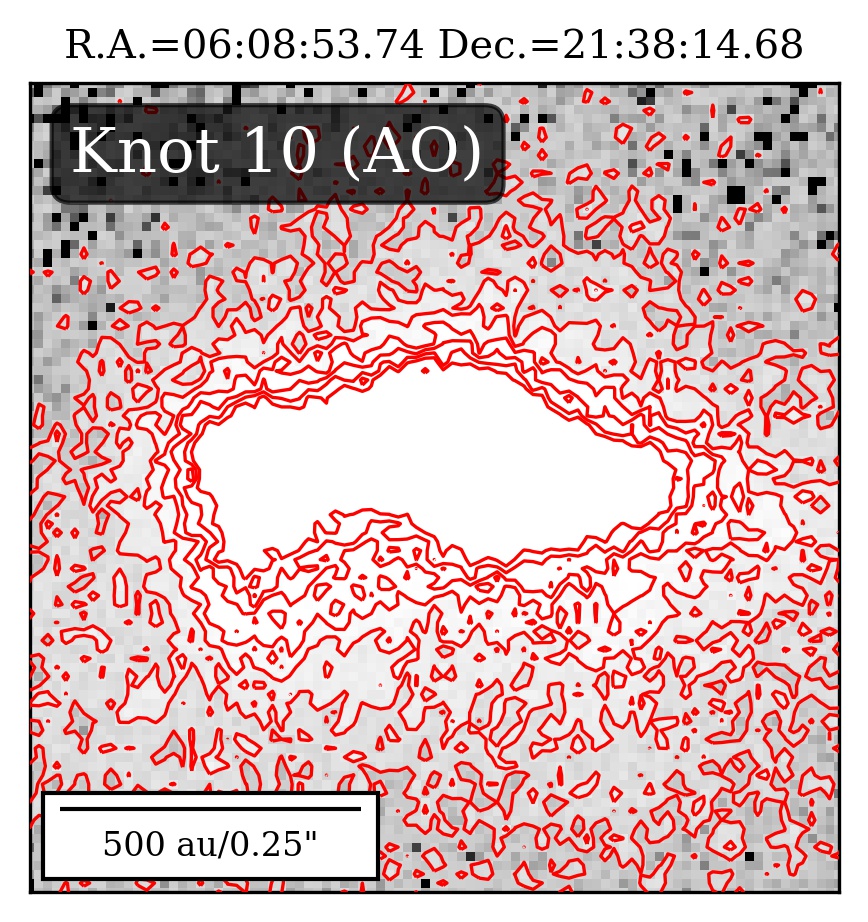}
                \includegraphics[width=0.40\textwidth]{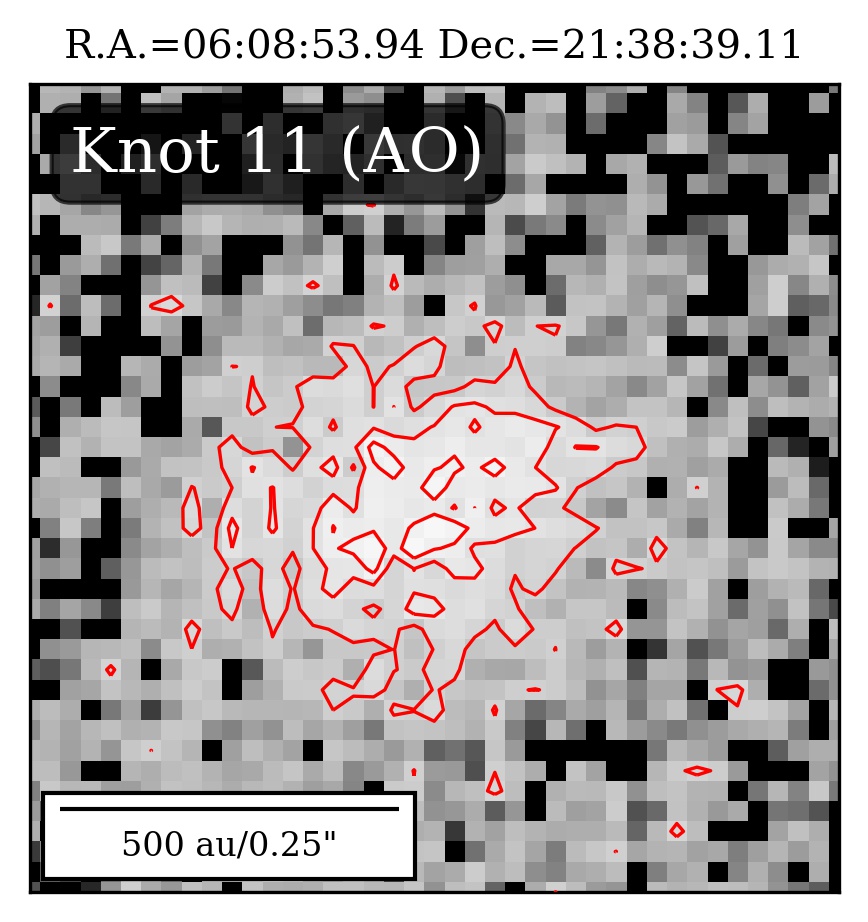}
                \includegraphics[width=0.40\textwidth]{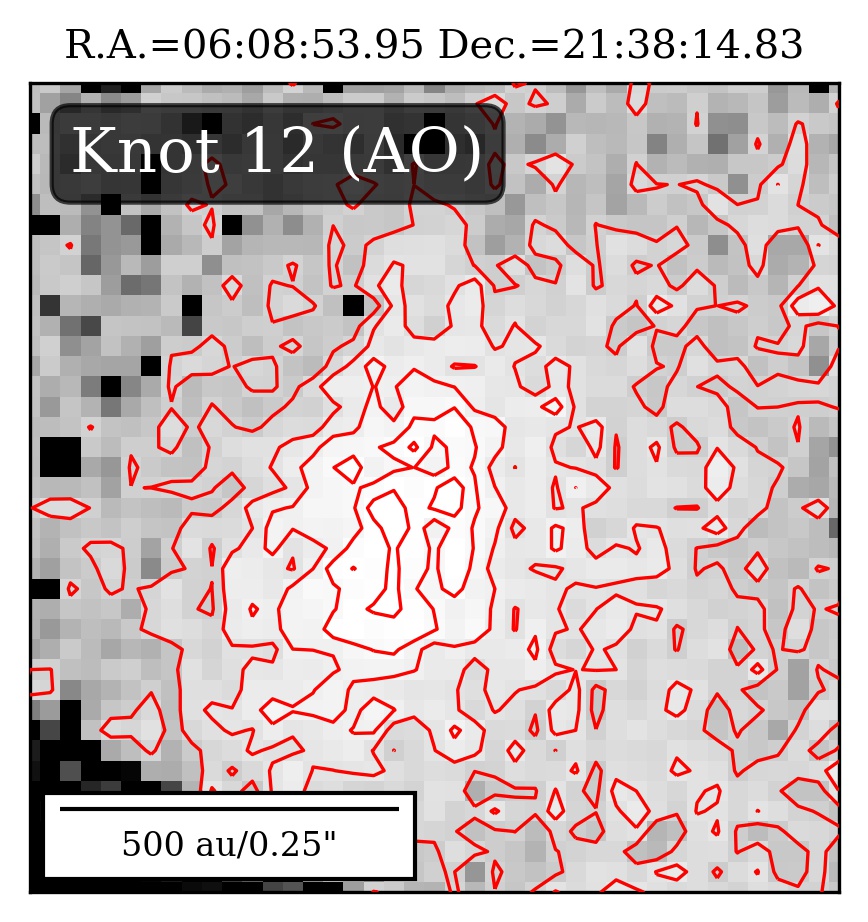}
                \includegraphics[width=0.40\textwidth]{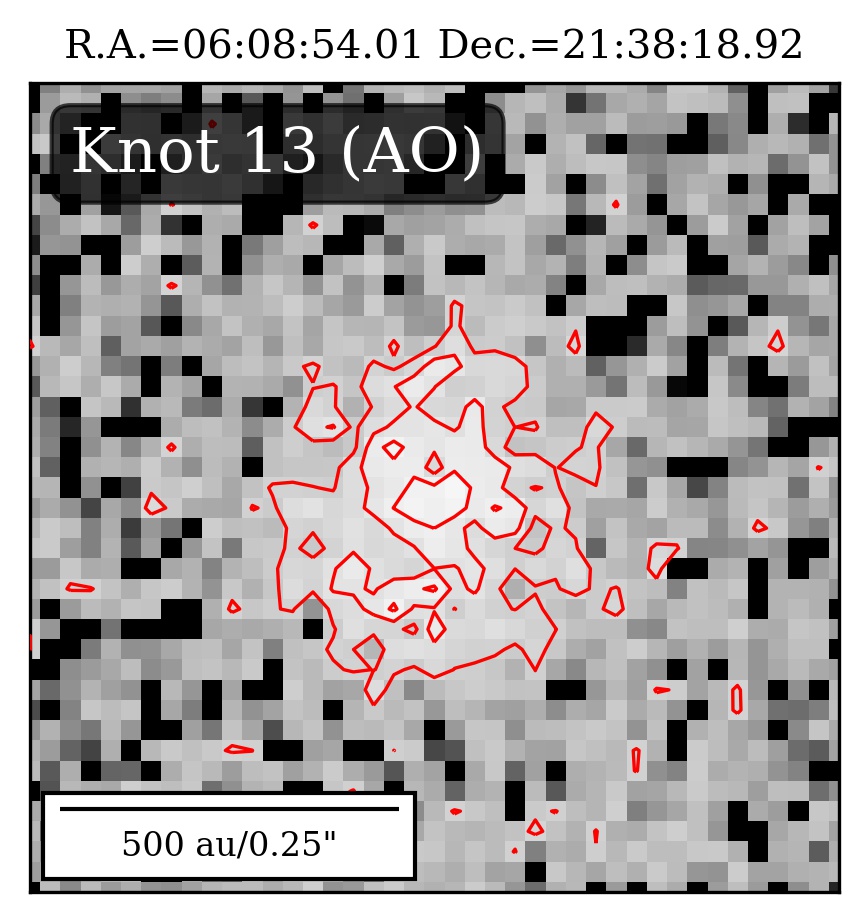}
                \includegraphics[width=0.40\textwidth]{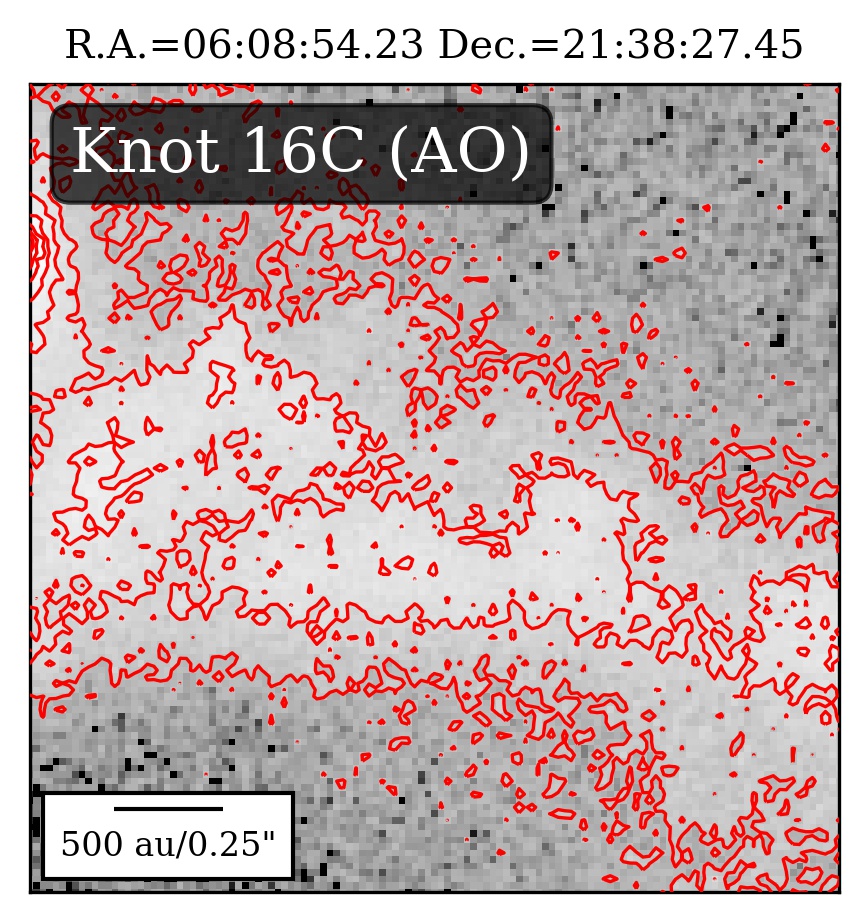}
        
          \caption{\label{AO_knot_sig}Significance level contour maps of all knot features identified in the LUCI-1 SOUL AO continuum-subtracted H$_2$ image and compiled in Table \ref{Knot_table}. The contour levels shown represent 3 to 15$\mathrm{\sigma}$ in steps of 2$\sigma$ above the local background. Note that for knot 16C, which is part of the S4 Jet as discussed in §\ref{sec:s4_jet}, the contour levels represent 5 to 45$\mathrm{\sigma}$ in steps of 5$\sigma$ above the local background instead. A physical scalebar of 500 au is given in the bottom-left corner of each panel. North is up and east is to the left in all panels.}
        \end{figure*}
    
        \renewcommand{\thefigure}{A\arabic{figure}}
        \addtocounter{figure}{-1}
    
        \begin{figure*}[!htb]
                \centering
                \includegraphics[width=0.40\textwidth]{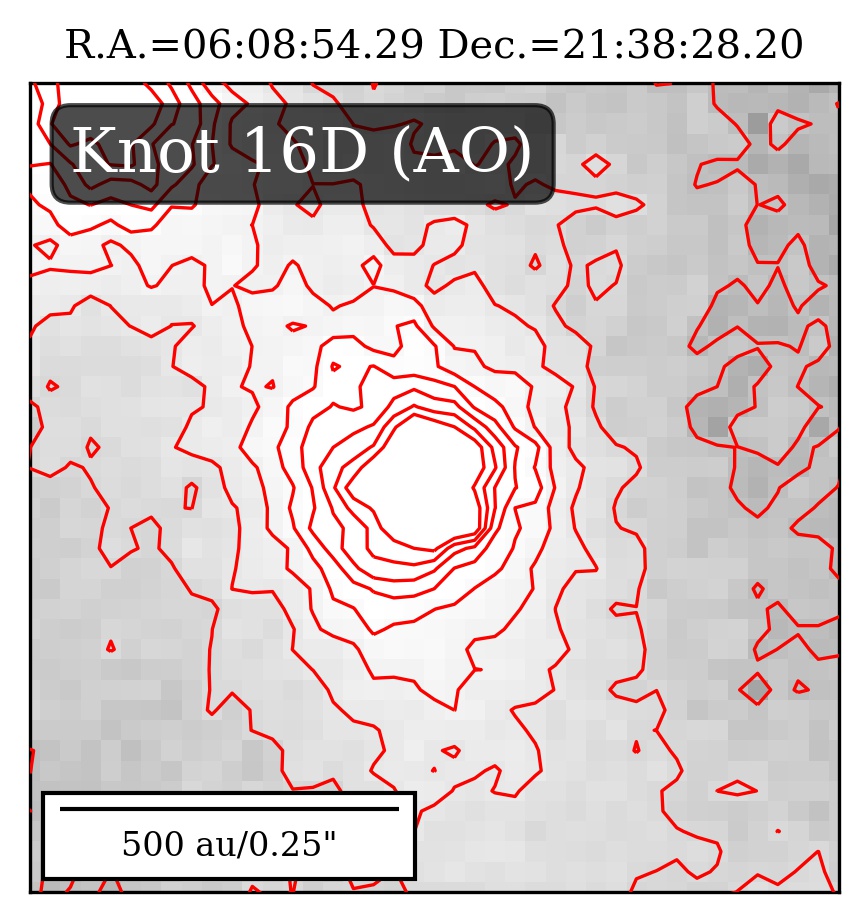}
                \includegraphics[width=0.40\textwidth]{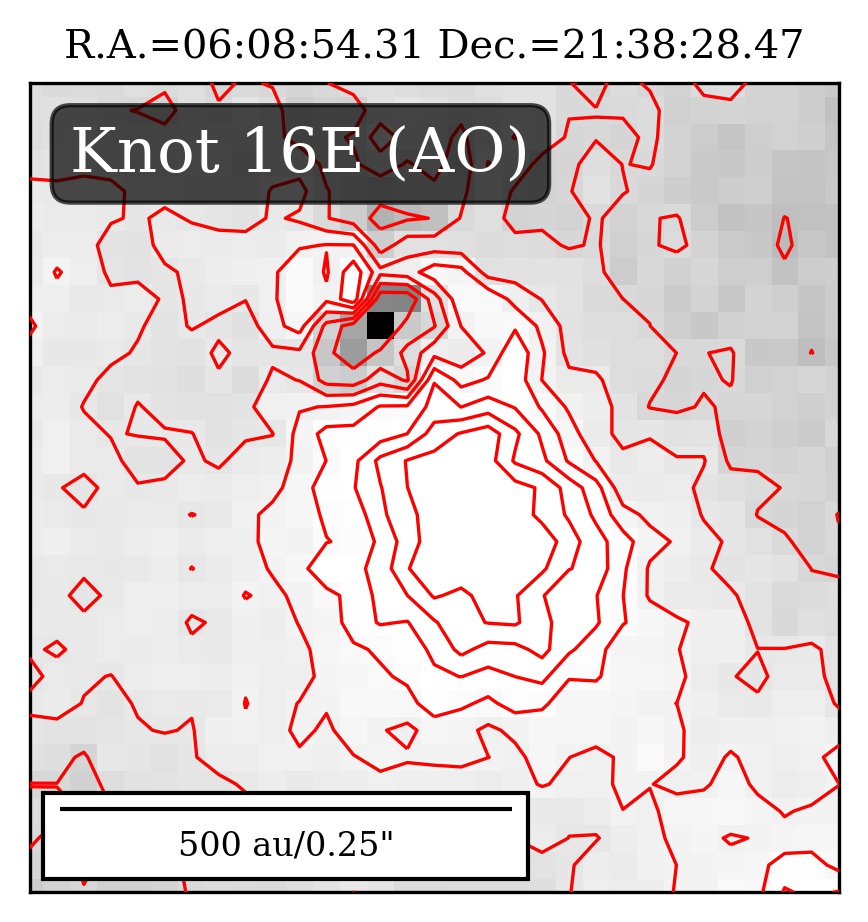}
                \includegraphics[width=0.40\textwidth]{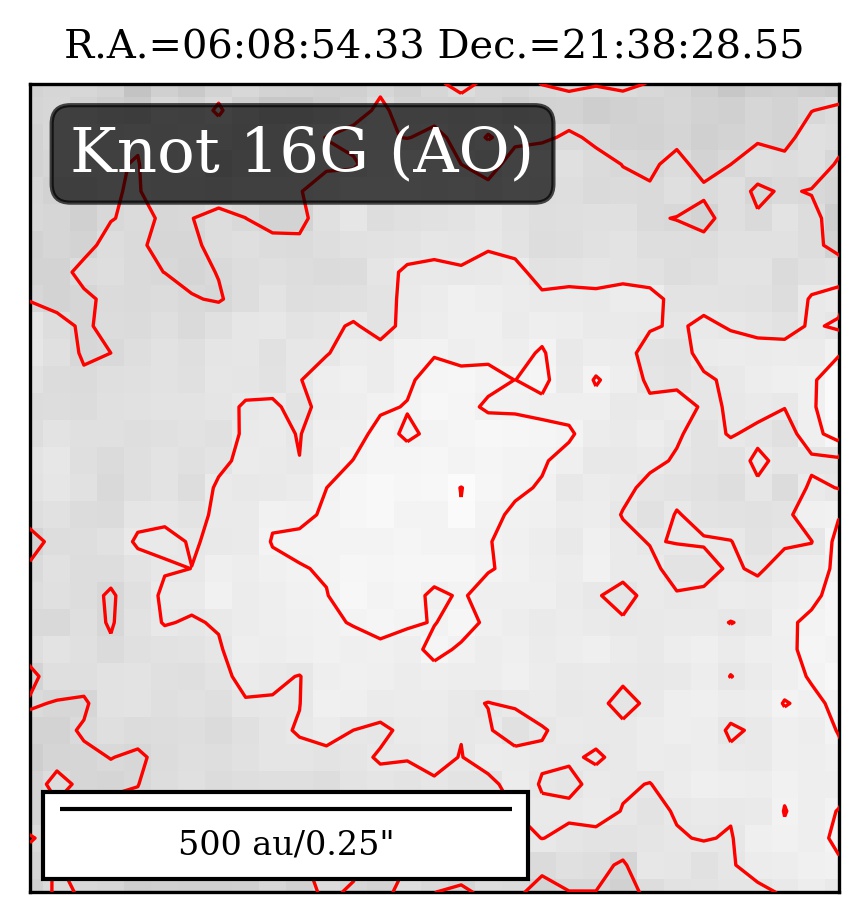}
                \includegraphics[width=0.40\textwidth]{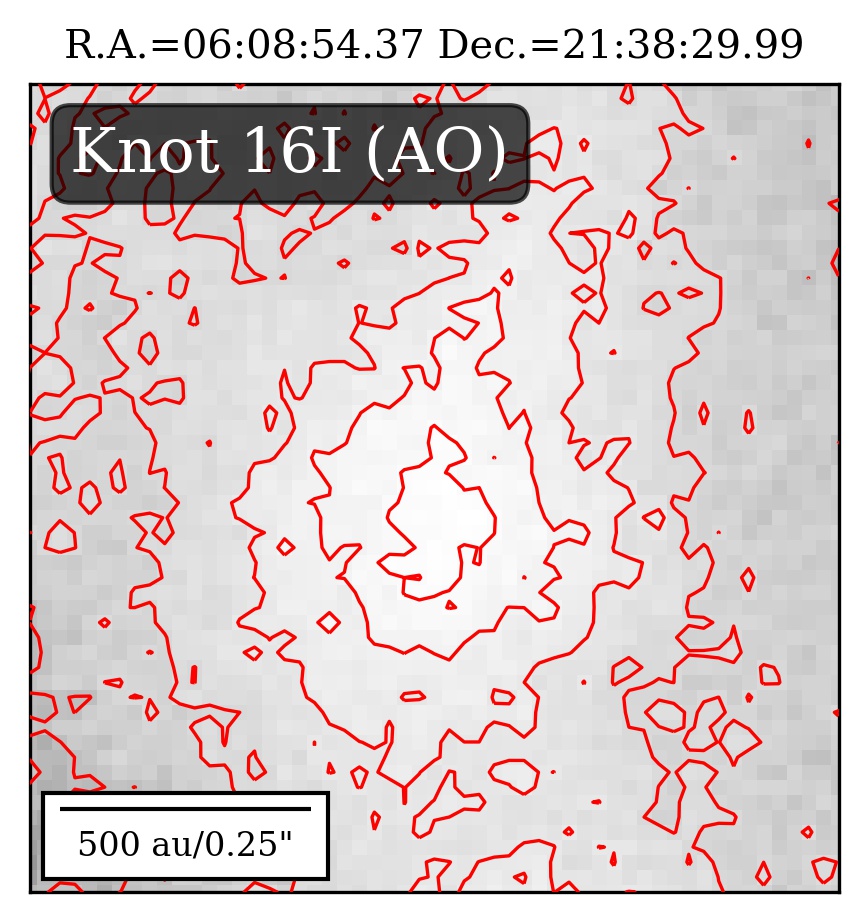}
                \includegraphics[width=0.40\textwidth]{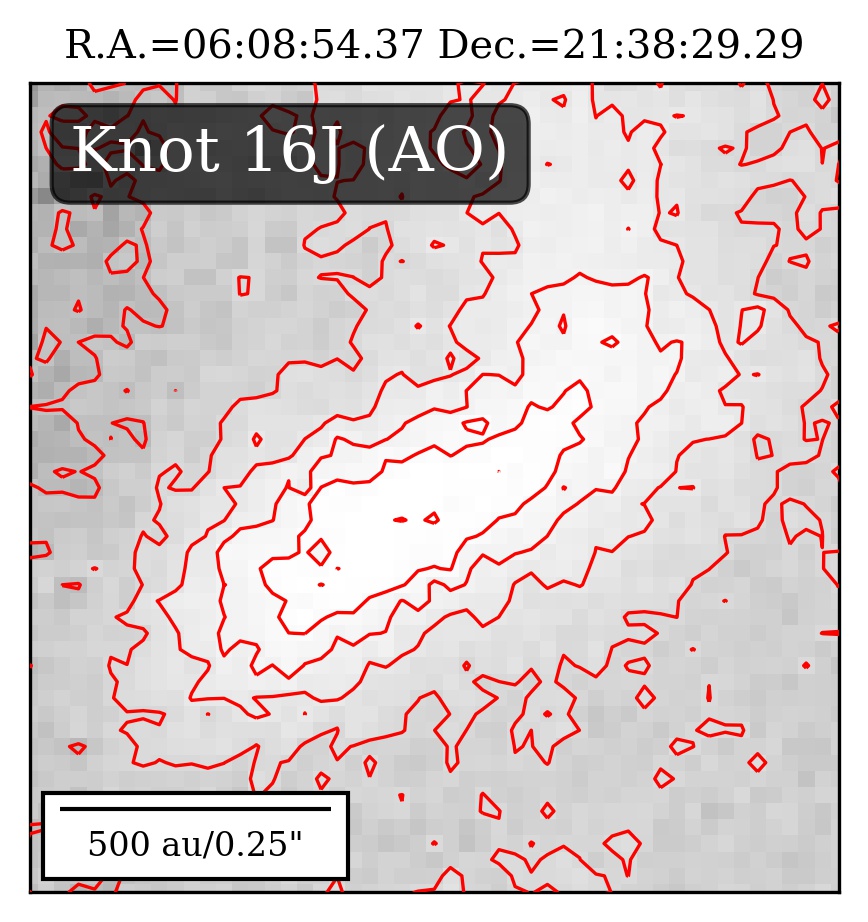}
                \includegraphics[width=0.40\textwidth]{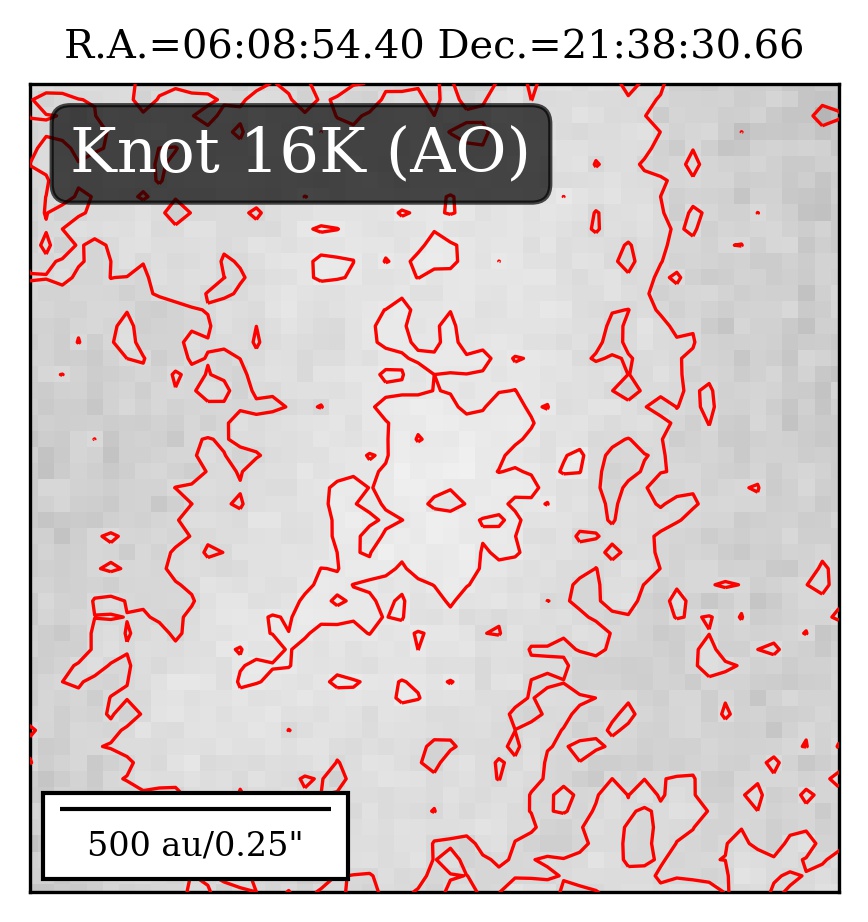}
        
          \caption{cont. Note that for each of these knots, which are part of the S4 Jet as discussed in §\ref{sec:s4_jet}, the contour levels represent 5 to 45$\mathrm{\sigma}$ in steps of 5$\sigma$ above the local background. For knot 16E, note the presence of a residual from a continuum-subtracted star in the upper left corner of the panel.}
        \end{figure*}
    
        \renewcommand{\thefigure}{A\arabic{figure}}
        \addtocounter{figure}{-1}
    
        \begin{figure*}[!htb]
                \centering
                \includegraphics[width=0.40\textwidth]{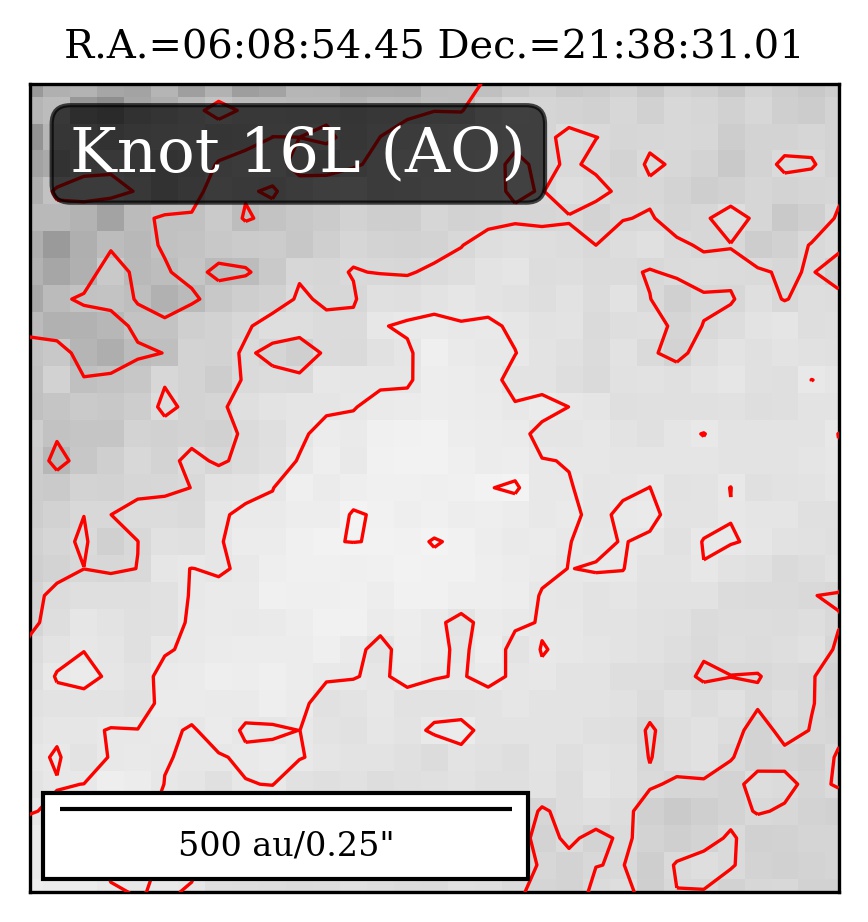}
                \includegraphics[width=0.40\textwidth]{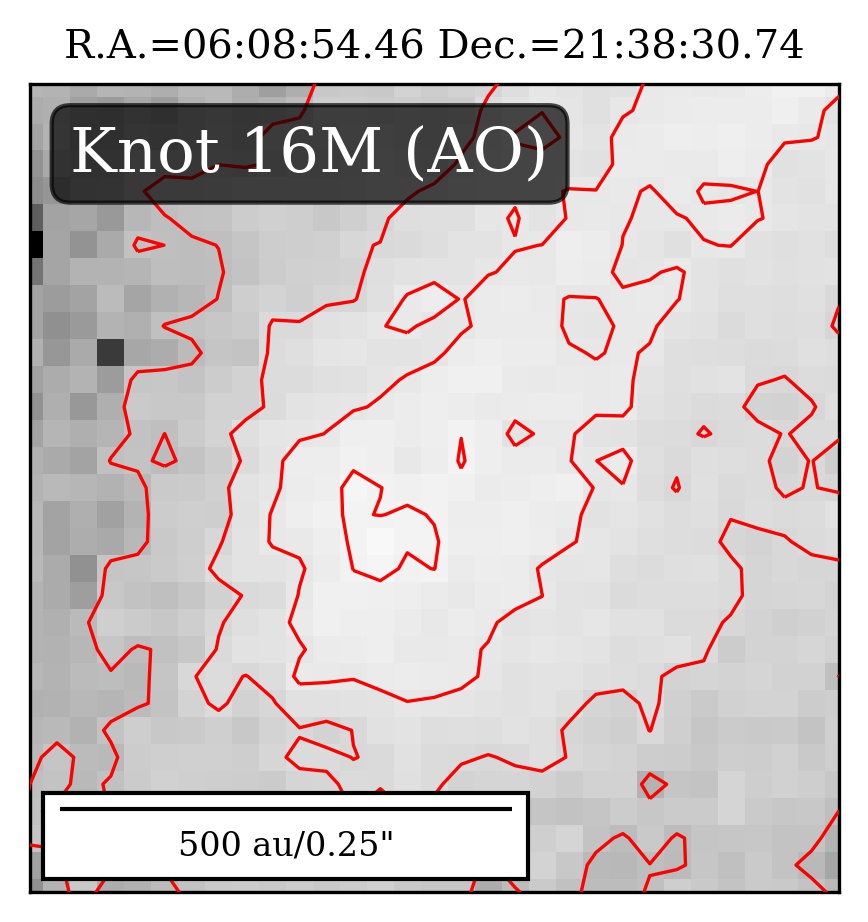}
                \includegraphics[width=0.40\textwidth]{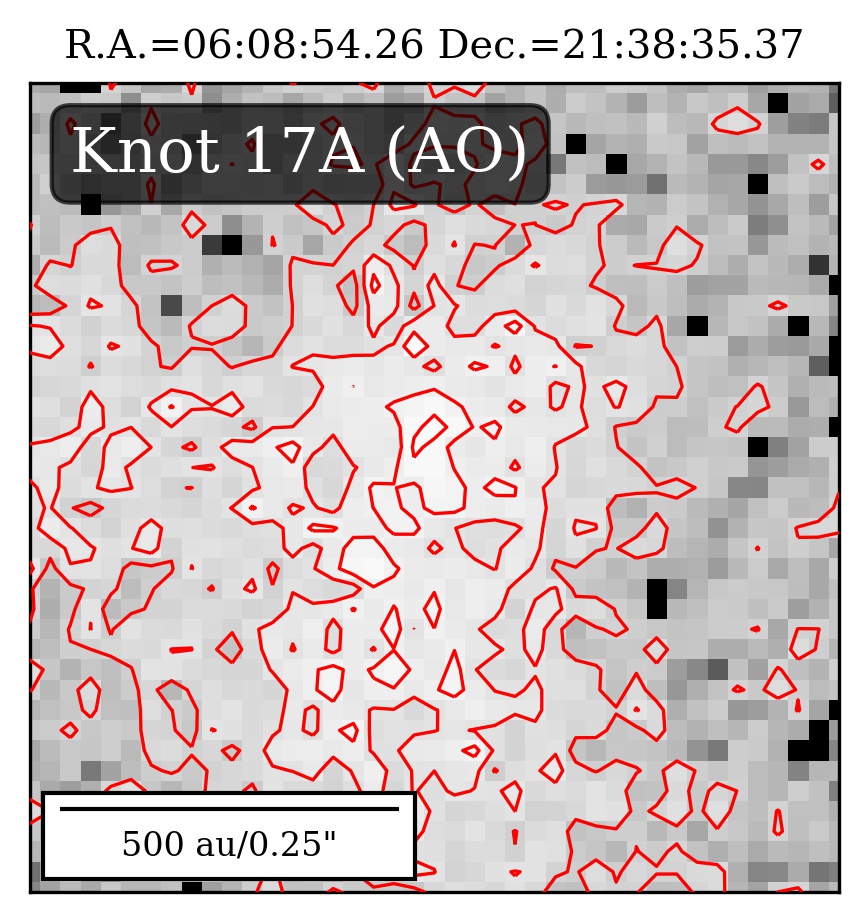}
                \includegraphics[width=0.40\textwidth]{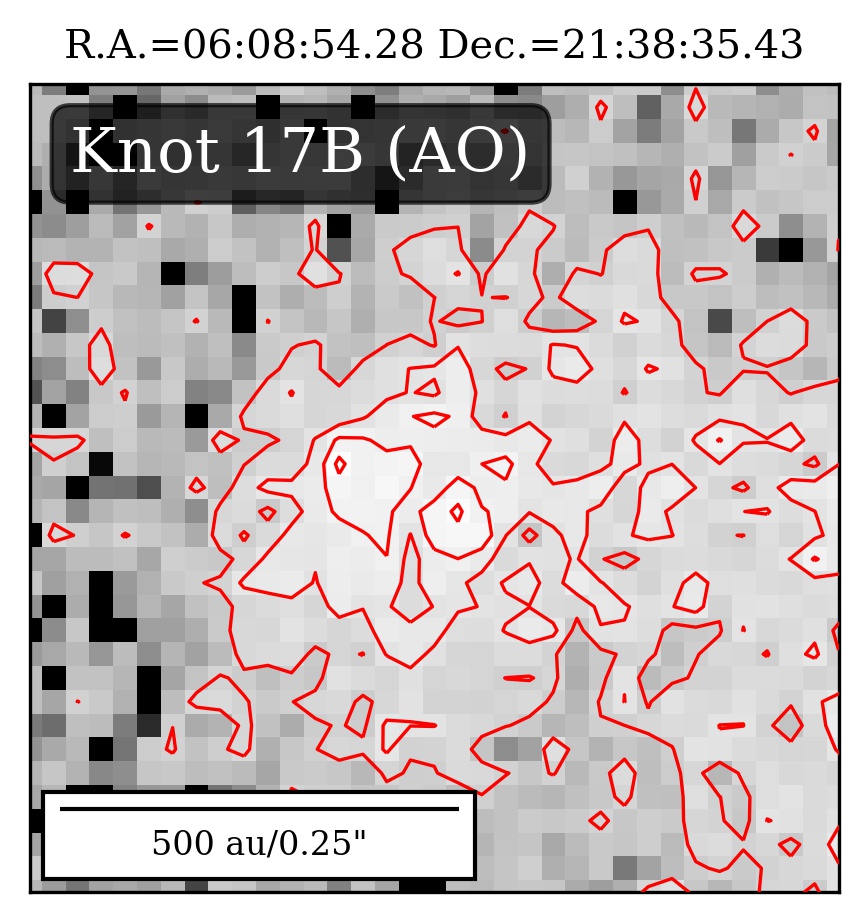}
                \includegraphics[width=0.40\textwidth]{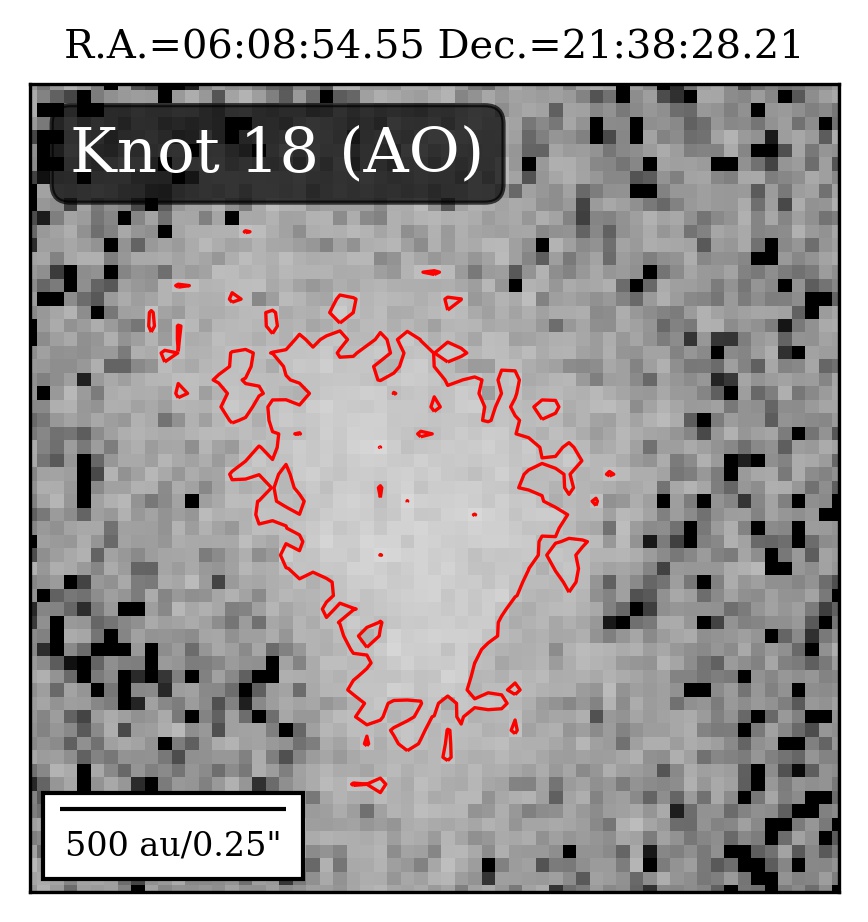}
        
          \caption{cont. Note that for knots 16L, 16M, and 18, which are part of the S4 Jet as discussed in §\ref{sec:s4_jet}, the contour levels represent 5 to 45$\mathrm{\sigma}$ in steps of 5$\sigma$ above the local background.}
        \end{figure*}
    
        \begin{figure*}[!htb]
                \centering
                \includegraphics[width=0.40\textwidth]{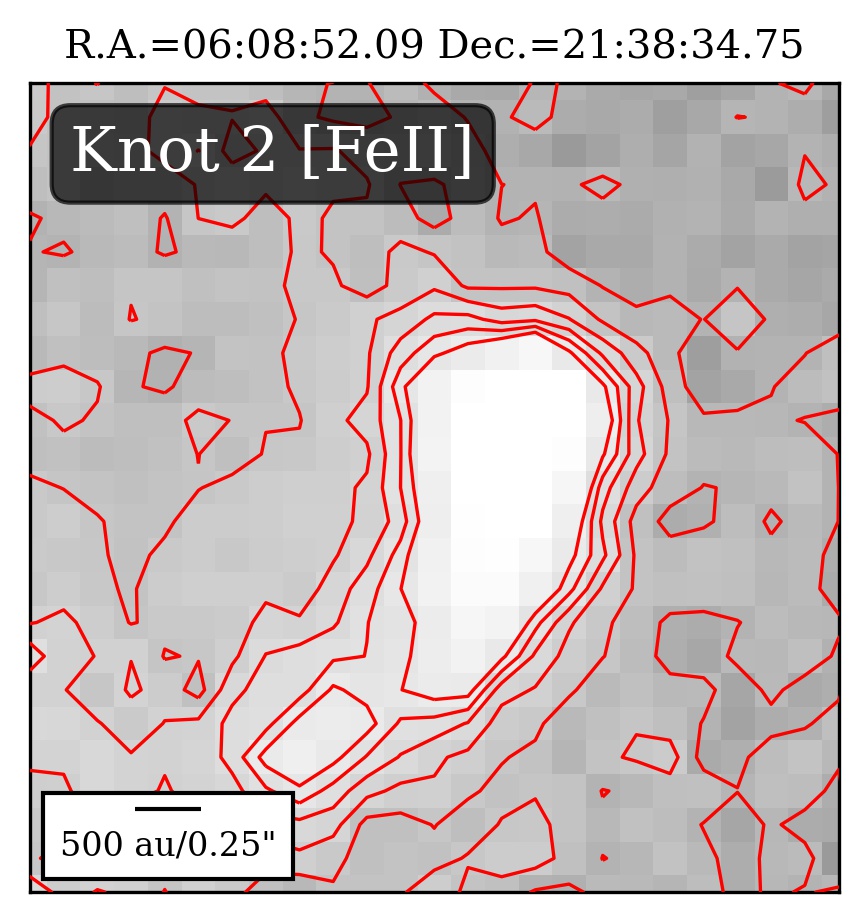}
                \includegraphics[width=0.40\textwidth]{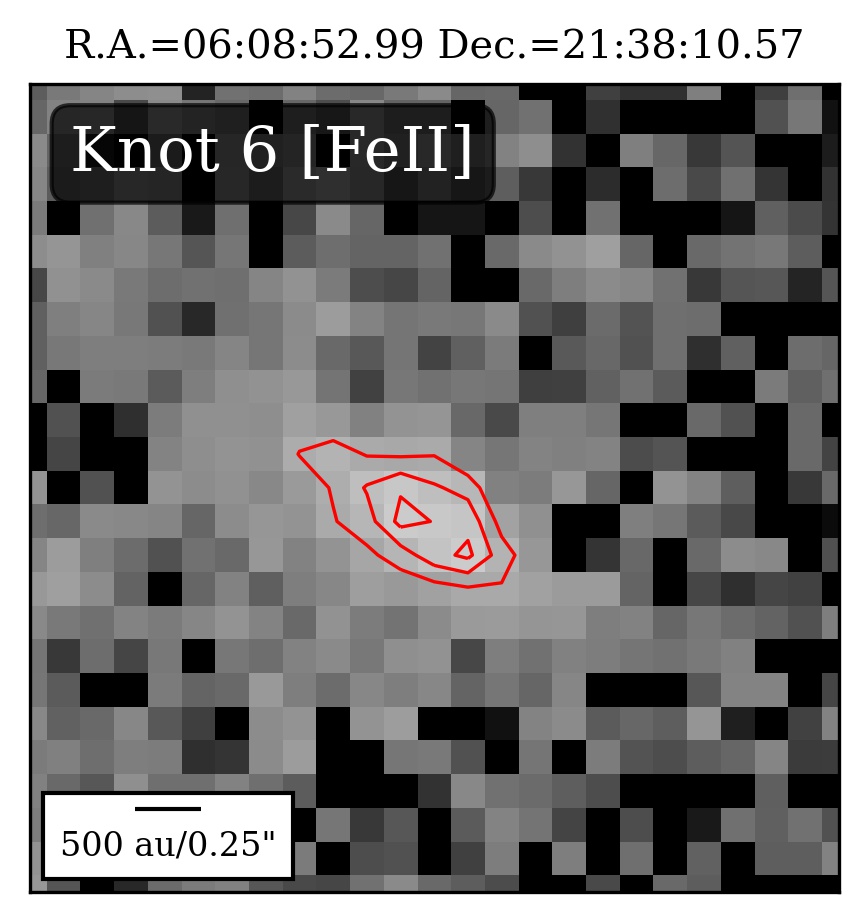}
                \includegraphics[width=0.40\textwidth]{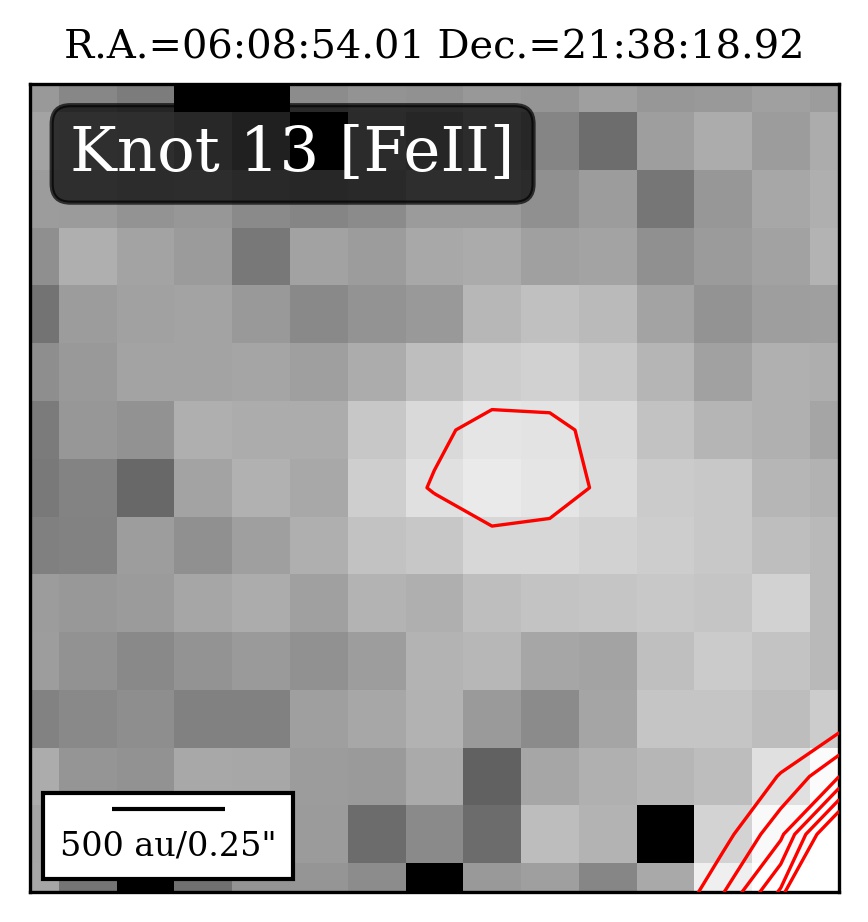}
                \includegraphics[width=0.40\textwidth]{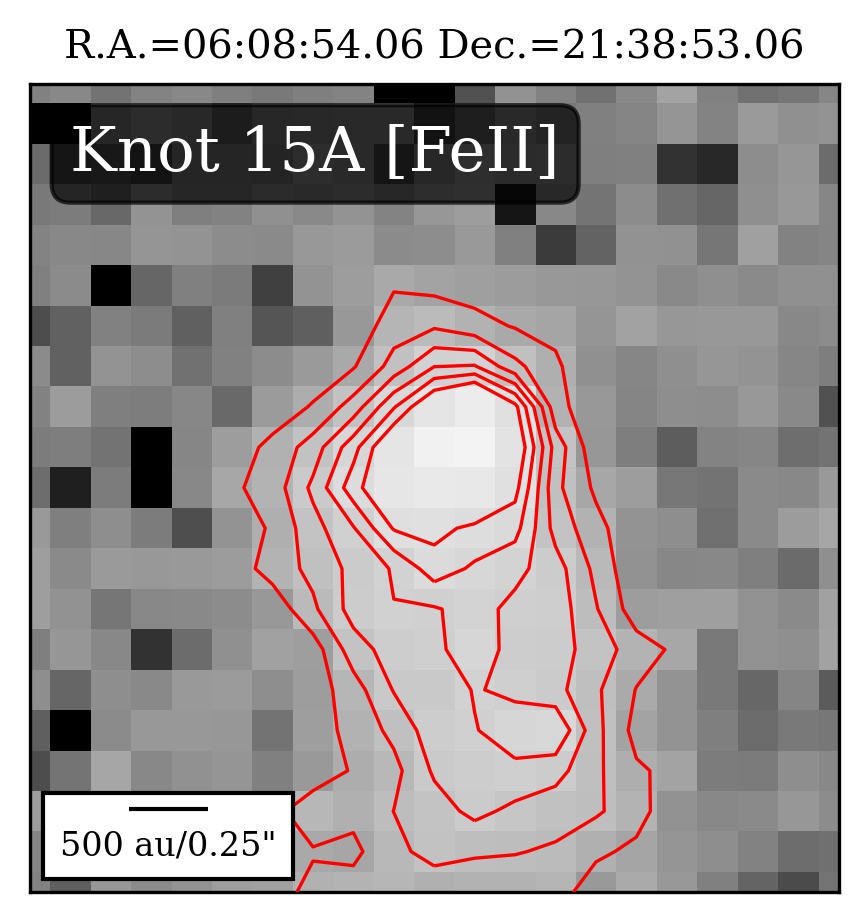}
                \includegraphics[width=0.40\textwidth]{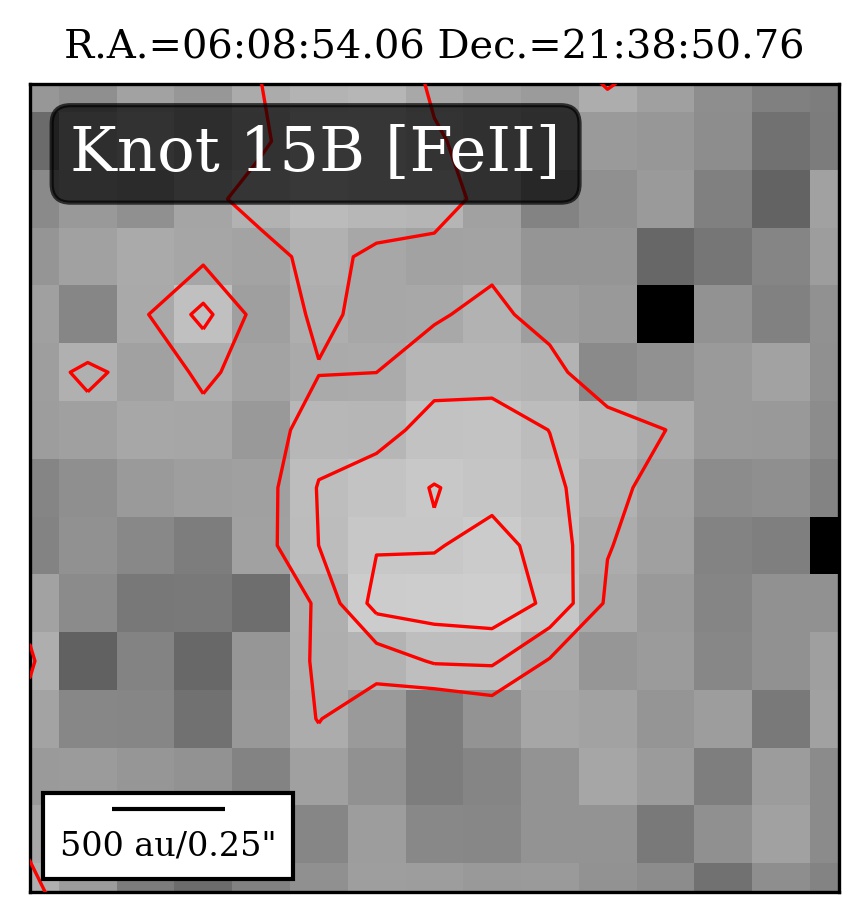}
                \includegraphics[width=0.40\textwidth]{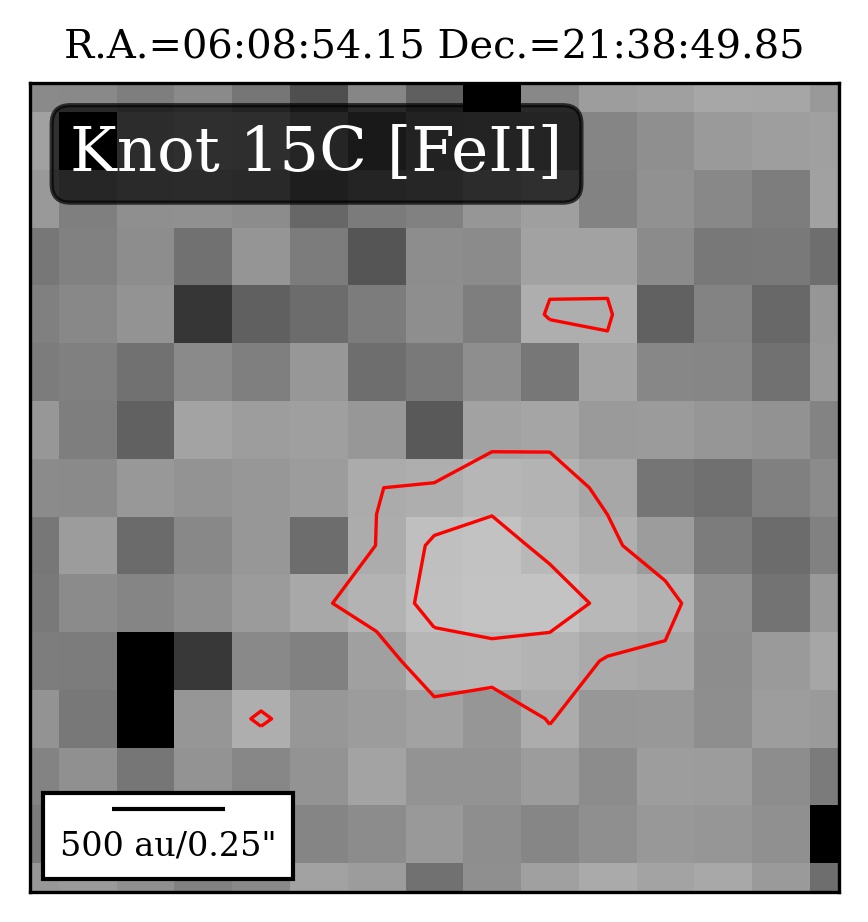}
          \caption{\label{FEII_knot_sig}Significance level contour maps of all knot features identified in the HST [FeII] continuum-subtracted image and compiled in Table \ref{Knot_table}. The contour levels shown represent 5 to 20 $\mathrm{\sigma}$ in steps of 3$\sigma$ above the local background. A physical scalebar of 500 au is given in the bottom-left corner of each panel. North is up and east is to the left in all panels.}

        \end{figure*}
    
        \renewcommand{\thefigure}{A\arabic{figure}}
        \addtocounter{figure}{-1}
    
        \begin{figure*}[!htb]
                \centering
                \includegraphics[width=0.40\textwidth]{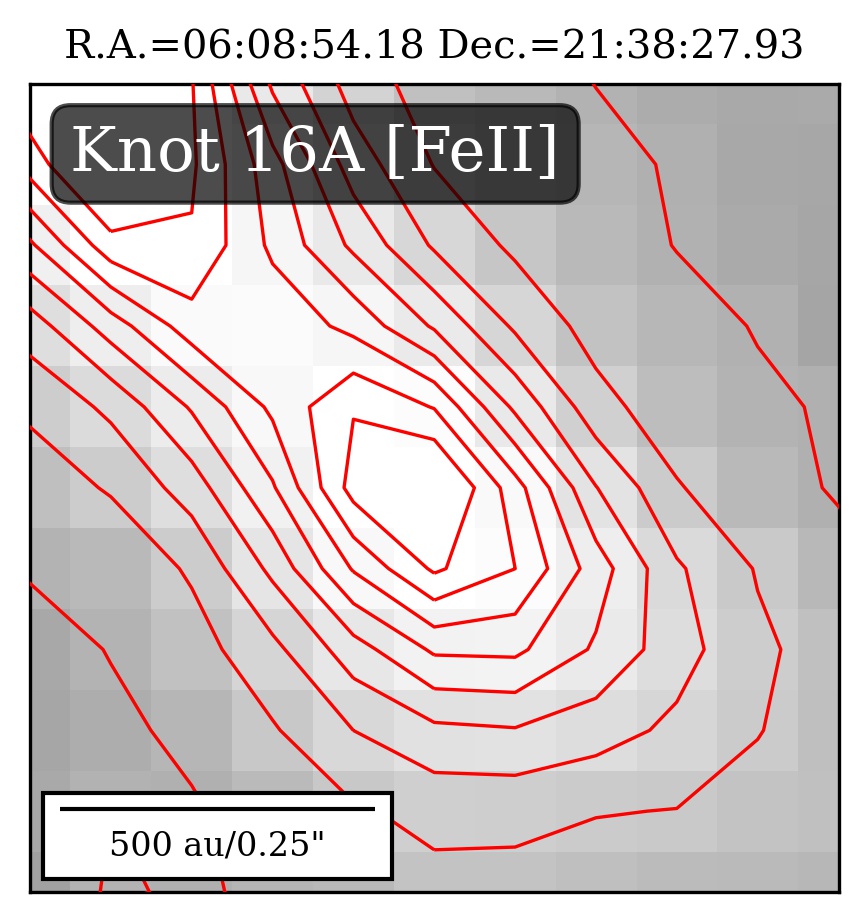}
                \includegraphics[width=0.40\textwidth]{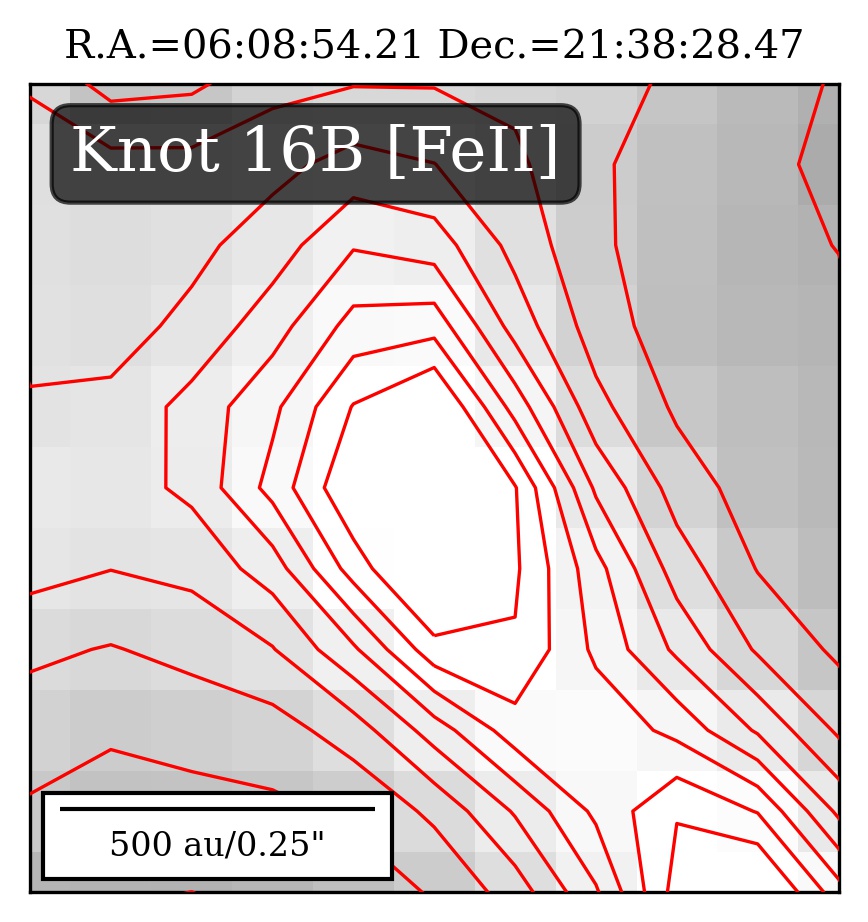}
                \includegraphics[width=0.40\textwidth]{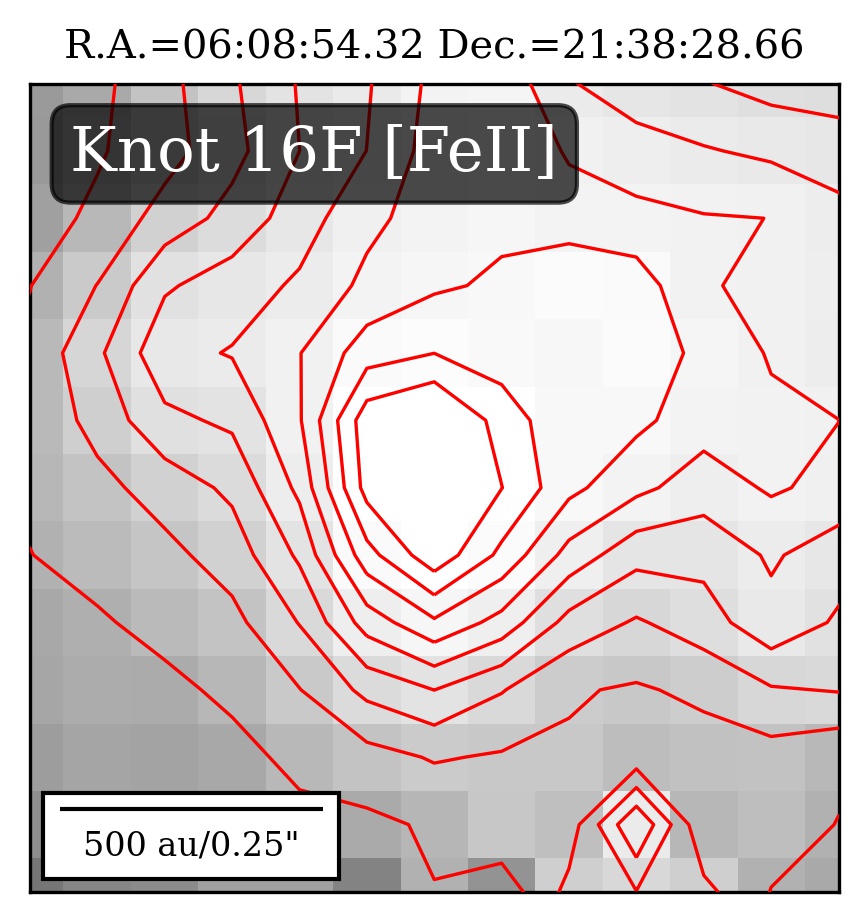}
                \includegraphics[width=0.40\textwidth]{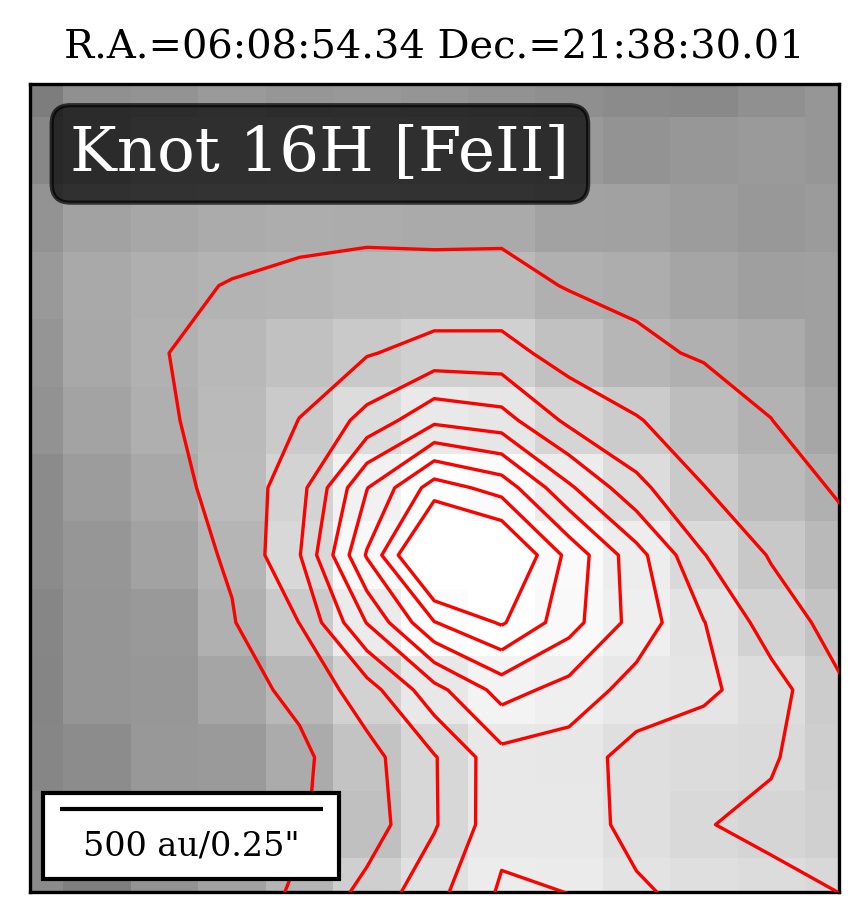}
                \includegraphics[width=0.40\textwidth]{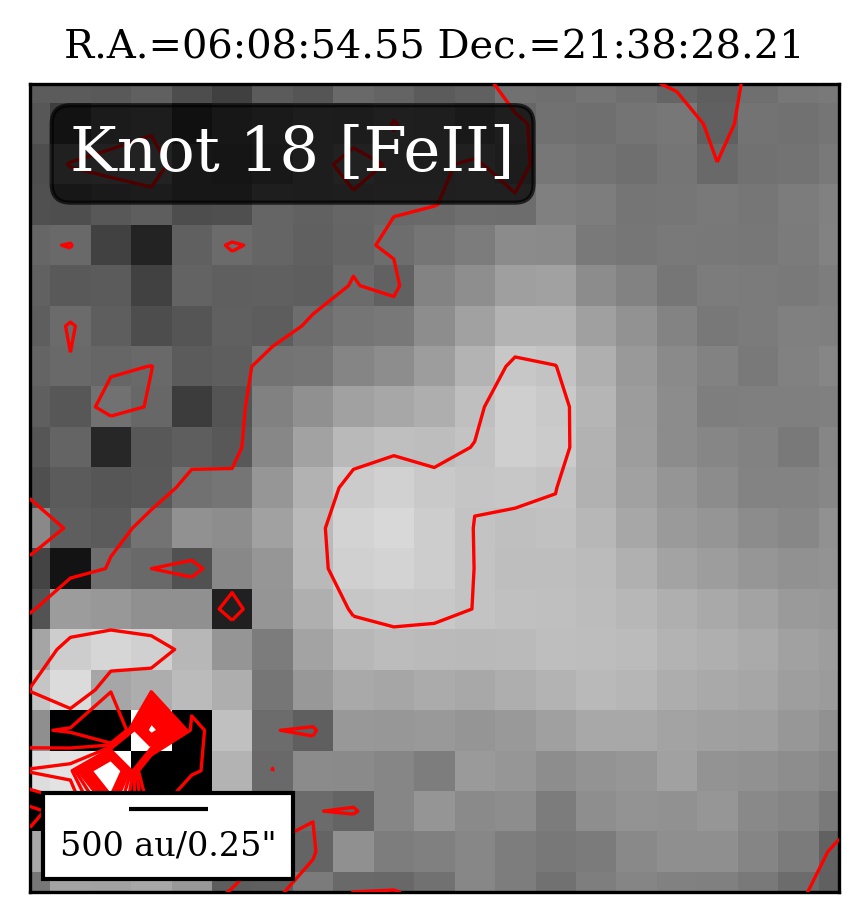}
            \caption{cont. Note that for each of these knots, which are part of the S4 Jet as discussed in §\ref{sec:s4_jet}, the contour levels represent 5 to 500$\mathrm{\sigma}$ in steps of 55 $\sigma$ above the local background.}
        \end{figure*}
    \section{Demonstration of Knot-Source Attribution}\label{sec:knot_attribution}
        \renewcommand{\thefigure}{B\arabic{figure}}
        \begin{figure*}[t]
            \centering
            \includegraphics[width=1\textwidth,height=1\textheight,keepaspectratio]{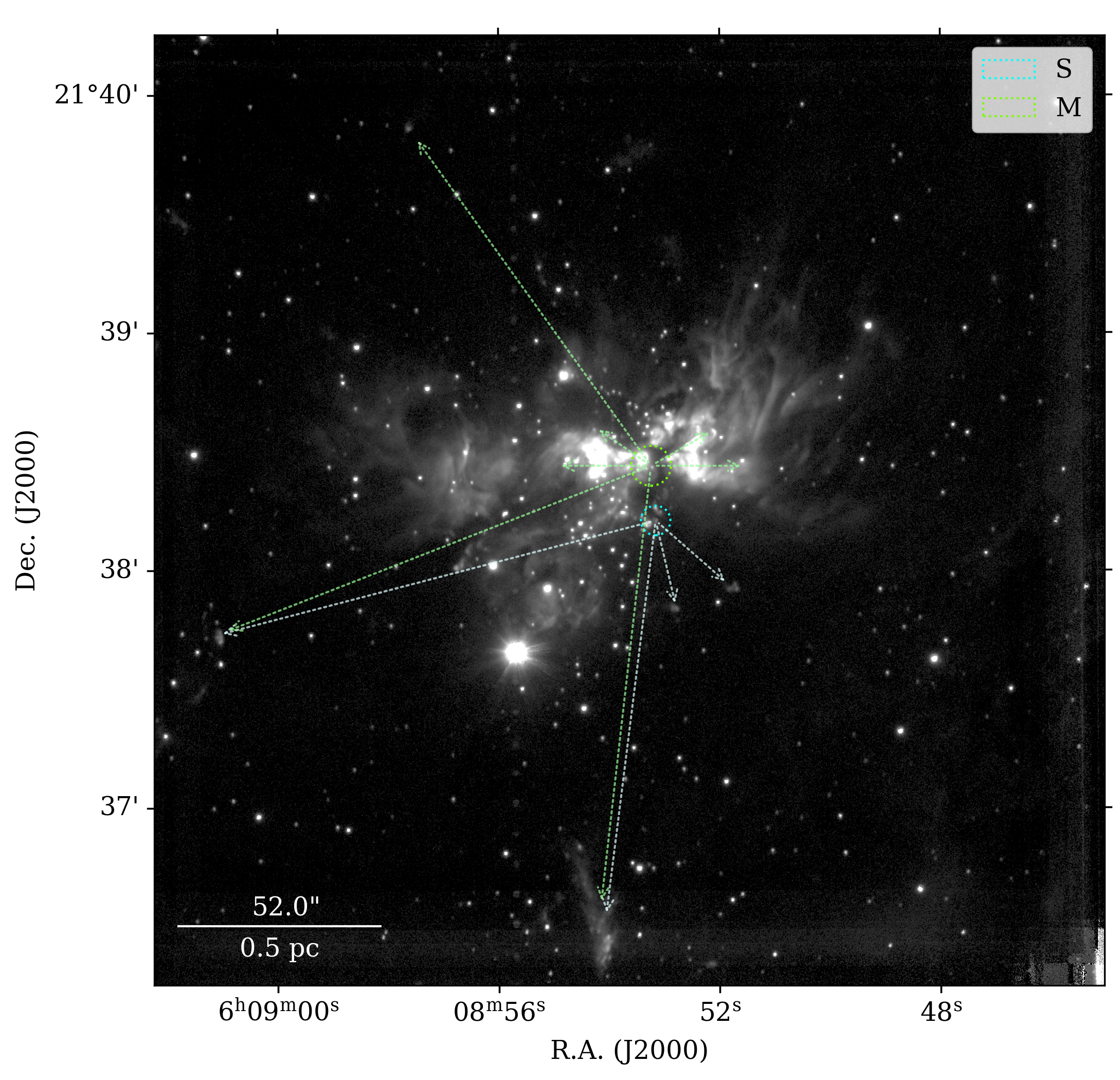}
            \caption{\label{knot_attribution} Diagram demonstrating the attribution of knots to potential driving sources/clusters (see \S\ref{sec:knot_ID}). AFGL 5180 M and S and their potentially associated jets are shown as the circles and arrows in green and blue, respectively.}
        \end{figure*}
\end{appendix}

%
%
\end{document}